\documentclass[aps,prd,superscriptaddress,nofootinbib,tighten,preprint]{revtex4}
\pdfoutput=1
\usepackage[utf8]{inputenc}
\usepackage[english]{babel}
\usepackage{amsmath}
\usepackage{subfigure}
\usepackage{color}
\usepackage{amsfonts}
\usepackage{amssymb}
\usepackage{graphicx}
\newcommand{\be}{\begin{equation}}
\newcommand{\ee}{\end{equation}}
\newcommand{\bea}{\begin{eqnarray}}
\newcommand{\eea}{\end{eqnarray}}

\newcommand{\ubl}{{\rm U}(1)_{\rm B-L}}
\newcommand{\mD}{\mathcal{M_{D}}}
\newcommand{\mR}{\mathcal{M_{R}}}
\newcommand{\mzbl}{M_{Z_{BL}}}

\newcommand{\vbl}{v_{BL}}
\newcommand{\gbl}{g_{BL}}
\newcommand{\nbl}{n_{BL}}
\newcommand{\zbl}{Z_{BL}}
\newcommand{\dm}{\phi_{DM}}
\newcommand{\mdm}{M_{DM}}
\newcommand{\ldh}{\lambda_{Dh}}
\newcommand{\ldH}{\lambda_{DH}}
\newcommand{\smgauge}{{\rm SU}(3)_{\rm c}\times{\rm SU}(2)_{\rm L}
\times {\rm U}(1)_{\rm Y}}
\newcommand{\nn}{\nonumber}

\catcode`@=12
\def\la{\mathrel{\mathchoice {\vcenter{\offinterlineskip\halign{\hfil
$\displaystyle##$\hfil\cr<\cr\sim\cr}}}
{\vcenter{\offinterlineskip\halign{\hfil$\textstyle##$\hfil\cr<\cr\sim\cr}}}
{\vcenter{\offinterlineskip\halign{\hfil$\scriptstyle##$\hfil\cr<\cr\sim\cr}}}
{\vcenter{\offinterlineskip\halign{\hfil$\scriptscriptstyle##$\hfil\cr<\cr\sim
\cr}}}}}
\def\ga{\mathrel{\mathchoice {\vcenter{\offinterlineskip\halign{\hfil
$\displaystyle##$\hfil\cr>\cr\sim\cr}}}
{\vcenter{\offinterlineskip\halign{\hfil$\textstyle##$\hfil\cr>\cr\sim\cr}}}
{\vcenter{\offinterlineskip\halign{\hfil$\scriptstyle##$\hfil\cr>\cr\sim\cr}}}
{\vcenter{\offinterlineskip\halign{\hfil$\scriptscriptstyle##$\hfil\cr>\cr\sim
\cr}}}}}
\begin{document}
\title{Neutrino Mass, Leptogenesis and FIMP Dark Matter
in a ${\rm U}(1)_{\rm B-L}$ Model}
\author{Anirban Biswas}
\email{anirbanbiswas@hri.res.in}
\affiliation{Harish-Chandra Research Institute, Chhatnag Road,
Jhunsi, Allahabad 211 019, India}
\affiliation{Homi Bhabha National Institute,
Training School Complex, Anushaktinagar, Mumbai - 400094, India}
\author{Sandhya Choubey}
\email{sandhya@hri.res.in}
\affiliation{Harish-Chandra Research Institute, Chhatnag Road,
Jhunsi, Allahabad 211 019, India}
\affiliation{Homi Bhabha National Institute,
Training School Complex, Anushaktinagar, Mumbai - 400094, India}
\affiliation{Department of Theoretical Physics, School of
Engineering Sciences, KTH Royal Institute of Technology, AlbaNova
University Center, 106 91 Stockholm, Sweden}
\author{Sarif Khan}
\email{sarifkhan@hri.res.in}
\affiliation{Harish-Chandra Research Institute, Chhatnag Road,
Jhunsi, Allahabad 211 019, India}
\affiliation{Homi Bhabha National Institute,
Training School Complex, Anushaktinagar, Mumbai - 400094, India}
\begin{abstract}

The Standard Model (SM) is inadequate to explain the origin of
tiny neutrino masses, the dark matter (DM) relic abundance and also
the baryon asymmetry of the Universe. In this work to address
all the three puzzles, we extend the SM by a local U$(1)_{\rm B-L}$
gauge symmetry, three right-handed (RH) neutrinos for the cancellation
of gauge anomalies and two complex scalars having nonzero
U$(1)_{\rm B-L}$ charges. All the newly added particles
become massive after the breaking of U$(1)_{\rm B-L}$
symmetry by the vacuum expectation value (VEV) of one of the
scalar fields $\phi_H$. The other scalar field $\dm$,
which does not have any VEV, becomes automatically
stable and can be a viable DM candidate.
Neutrino masses are generated using
Type-I seesaw mechanism while the required lepton asymmetry
to reproduce the observed baryon asymmetry, can be attained
from the CP violating out of equilibrium decays of RH neutrinos
in TeV scale. More importantly within this framework,
we have studied in detail the production of DM via
freeze-in mechanism considering all possible annihilation
and decay processes. Finally, we find a
situation when DM is dominantly produced from the annihilation
of RH neutrinos, which are at the same time also responsible
for neutrino mass generation and leptogenesis.
\end{abstract}
\maketitle
\section{Introduction}
\label{intro}
The presence of non-zero neutrino mass and mixing has been confirmed by
observing neutrino flavour oscillations \cite{Fukuda:1998mi, Ahmad:2002jz}
among its different flavours. Neutrino experiments
have measured the three intergenerational mixing angles
($\theta_{12}, \theta_{23}, \theta_{13}$) and the
two mass square differences ($\Delta m_{21}^{2}$ and $\Delta m_{32}^{2}$) 
\footnote{$\Delta m_{ij}^{2} = m_{i}^{2}
- m_{j}^{2}\,\,\,\, {\rm and}\,\,\,\, \Delta m_{atm}^{2} =
m_{3}^{2} - \frac{m_{1}^{2} + m_{2}^{2}}{2}$} 
with an unprecedented
accuracy \cite{Eguchi:2002dm, An:2015nua, RENO:2015ksa, Abe:2014bwa,
Abe:2015awa, Salzgeber:2015gua, Adamson:2016tbq, Adamson:2016xxw}.
Neutrinos are massless in the Standard Model (SM) of
particle physics because in SM there is no right
handed (RH) counterpart of the left handed (LH)
neutrinos. To generate tiny neutrino masses
and their intergenerational mixing angles,
as suggested by different experiments, we have
to think of some new interactions and/or new particles
beyond the Standard Model (BSM). 
Moreover, there are still some unsolved problems 
in the neutrino sector. For example, we do not know
the exact octant of the atmospheric mixing angle
$\theta_{23}$ i.e. whether it lies in the lower octant
($\theta_{23} < 45^{\circ}$) or in the higher octant
($\theta_{23} > 45^{\circ}$), the exact sign of $\Delta
m^2_{32}$ which is related to the mass hierarchy between
$m_2$ and $m_3$ (for the normal hierarchy (NH) $\Delta m^2_{32}>0$ while
for the inverted hierarchy (IH) $\Delta m^2_{32}<0$) and
also about the Dirac CP phase $\delta$, responsible for
the CP violation in the leptonic sector. Recently,
T2K and No$\nu$A experiments have reported their preliminary result which
predicts that the value of Dirac CP phase is
around $\delta_{CP} \sim 270^{\circ}$ \cite{xyz}. Besides these,
we do not know whether the neutrinos are Dirac fermion or
Majorana fermion. Observation of neutrino less double
$\beta$ decay \cite{Gando:2012zm, Agostini:2013mzu, Asakura:2014lma,
Albert:2014awa, xyz1, KamLAND-Zen:2016pfg} will confirm the
Majorana nature of neutrinos and might also provide important information
about the Majorana phases which could be the other source
of CP violation in the leptonic sector, if the SM neutrinos
are Majorana fermions. 

Besides these unsolved problems in the neutrino sector,
another well known puzzle in recent times is the presence
of dark matter (DM) in the Universe. Many indirect evidence suggests the
existence of DM. Among the most compelling evidence of DM 
are the observed flatness of rotation curves of spiral
galaxies \cite{Sofue:2000jx}, gravitational lensing
\cite{Bartelmann:1999yn}, the observed spatial
offset between DM and visible matter in the collision of
two galaxy clusters (e.g. Bullet cluster \cite{Clowe:2003tk},
Abell cluster \cite{Biviano:1996bg, Kahlhoefer:2013dca}) etc. The latter
also imposes an upper bound on the ratio between self interaction
and mass of DM particles,  which is $\frac{\sigma_{DM}}{M_{DM}}
\la 1\,{\rm barn}/{\rm GeV}$ \cite{Harvey:2015hha}. Moreover, 
satellite borne experiments like WMAP \cite{Hinshaw:2012aka}
and Planck \cite{Ade:2015xua} have made a precise measurement
of the amount of dark matter present in the Universe from
the cosmic microwave background (CMB) anisotropy \cite{Ade:2015xua} 
and the current measured value of this parameter lies in the range  
$0.1172\leq\Omega_{\rm DM} h^2\leq0.1226$ at 67\%
C.L \cite{Ade:2015xua}.    

Despite the compelling observational
evidence for DM due to its gravitational
interactions, our knowledge about
its particle nature is very limited.\,\,The
only thing we know about the DM is
that it is very weakly interacting and electromagnetically
blind.\,\,The SM of particle physics does not have any fundamental
particle which can play the role of a cold dark matter (CDM),
consequently a BSM scenario containing new fundamental
stable particle(s) is required. There are earth
based ongoing DM direct detection experiments,
namely Xenon-1T \cite{Aprile:2012zx}, LUX \cite{Akerib:2015rjg},
CDMS \cite{Ahmed:2010wy, Agnese:2014aze}
amongst others, which have been trying
to detect the Weakly Interacting Massive Particle (WIMP)
\cite{Gondolo:1990dk, Jungman:1995df, ArkaniHamed:2008qn}
type DM by measuring recoil energies of the detector nuclei
scattered by the WIMPs. However, no convincing
DM signal has been found yet and hence the $M_{DM}-\sigma_{\rm SI}$
plane for a WIMP type DM is now getting severely
constrained. Therefore, invoking particle DM models outside the WIMP 
paradigm seems to be pertinent at this stage \cite{Arcadi:2017kky}. 
In the present work we study one of the possible
alternatives of WIMP, namely, the Feebly Interacting Massive
Particle (FIMP) \cite{Hall:2009bx, Yaguna:2011qn, Molinaro:2014lfa,
Biswas:2015sva, Merle:2015oja, Shakya:2015xnx, Biswas:2016bfo,
Konig:2016dzg, Biswas:2016iyh, Biswas:2016yjr}. 
A major difference between the WIMP and FIMP scenarios is that while 
in the former the DM particle is in thermal equilibrium with the plasma 
in the early Universe and freezes-out when the Hubble expansion rate becomes 
larger than its annihilation cross section, in 
the FIMP case the DM is never in thermal equilibrium
with the cosmic soup. This is mainly ensured by its
extremely weak couplings to other particles in the thermal bath.
Therefore, the number density of the FIMP is negligible in the
early Universe and increases when the FIMP is subsequently
produced by the decays and annihilations of other particles to
which it is coupled (very feebly). This process is generally known as 
freeze-in \cite{Hall:2009bx}.

In addition to the above two unsolved problems, another long standing
enigma is the presence of more baryons over anti baryons in
the Universe, which is known as the baryon asymmetry
or the matter-antimatter asymmetry in the Universe. The baryon
asymmetry observed in the Universe is expressed by a quantity
$Y_{B}=\dfrac{\eta_{B}\,n_{\gamma}}{\rm s}$, where
$\eta_{B}=n_{B}-n_{\bar{B}}$ is the excess in the number density
for baryon over anti-baryon while $n_{\gamma}$ and ${\rm s}$
are the photon number density and the entropy density of the
Universe, respectively. At the present epoch, $\eta_{B}=
\left(5.8-6.6 \right) \times 10^{-10}$ at 95\% C.L. \cite{Olive:2016xmw}
while at $T\sim 2.73$ K, the photon density
$n_{\gamma}=410.7\,\,{\rm cm}^{-3}$ \cite{Olive:2016xmw} and
the entropy density ${\rm s}=2891.2\,\,{\rm cm}^{-3}$ \cite{Olive:2016xmw}
(in natural unit with Boltzmann constant $K_{B}=1$). Therefore,
the observed baryon asymmetry at the present 
Universe is $Y_{B}=\left(8.24-9.38\right)\times 10^{-10}$, which 
although small, is sufficient to produce the $\sim 5\%$
energy density (visible matter) of the Universe.
To generate baryon asymmetry in the Universe from a
matter-antimatter symmetric state, one has to satisfy three
necessary conditions, known as the Sakharov's
conditions \cite{Sakharov:1967dj}. These are i) baryon number (B) violation,
ii) C and CP violation and iii) departure from thermal equilibrium.
Since the baryon number (B) is an accidental symmetry of
the SM (i.e. all SM interactions are B conserving)
and also the observed CP violation in quark sector
is too small to generate the requited baryon asymmetry,
hence like the previous cases, here also one has to look for
some additional BSM interactions which by satisfying
the Sakharov's conditions can generate the observed
baryon asymmetry in an initially matter-antimatter
symmetric Universe.       

In this work, we will try to address all of the three
above mentioned issues. The non-observation of 
any BSM signal at LHC implies the
concreteness of the SM. However to address all
the three problems, we need to extend the particles
list and/or gauge group of SM because as already
mentioned, SM is unable to explain either of them.
In our model, we have extended the SM gauge group
$\smgauge$ by a local $\ubl$ gauge group. 
The ${\rm B-L}$ extension of SM \cite{Mohapatra:1980qe,
Georgi:1981pg, Wetterich:1981bx, Lindner:2011it}
has been studied earlier in the context of dark matter
phenomenology \cite{Okada:2010wd, Okada:2012sg,
Basso:2012ti, Basak:2013cga, Sanchez-Vega:2014rka, Basak:2014sza,
Guo:2015lxa, Rodejohann:2015lca, Okada:2016gsh, Patra:2016ofq,
Okada:2016tci} and
baryogenesis in the early Universe in Refs. \cite{Buchmuller:1992qc,
Buchmuller:1996pa, Dulaney:2010dj}. Since we have
imposed a local U(1) symmetry, consequently an extra
gauge boson ($\zbl$) will arise.
To cancel the anomaly due to this extra gauge boson
we need to introduce three right-handed (RH) neutrinos
($N_{i}, i=1$, $2$, $3$) to make the model anomaly free.
Apart from the three RH neutrinos, we have also introduced
two SM gauge singlet scalars namely $\phi_{H}$ and
$\phi_{DM}$, both of them are charged under
the proposed $\ubl$ gauge group.
The $\ubl$ symmetry is spontaneously broken
when the scalar field $\phi_{H}$ takes a nonzero
vacuum expectation value (VEV) and thereby
generates the masses for the three RH neutrinos
as well as the extra neutral gauge boson $\zbl$,
whose mass terms are forbidden initially due
to the $\ubl$ invariance of the Lagrangian. The other
scalar $\phi_{DM}$ does not acquire any VEV 
and by choosing appropriate ${\rm B-L}$
charge $\dm$ becomes naturally stable 
and therefore, can serve as a viable dark matter candidate.
As mention above, anomaly cancellation requires the
introduction of three RH neutrinos in the present model. 
Therefore we can easily generate the neutrino
masses by the Type-I seesaw mechanism after
{\rm B-L} symmetry is broken. Diagonalising
the light neutrino mass matrix ($m_{\nu}$, for detail see Section
\ref{neutrino-sec}), we determine the
allowed parameter space by satisfying the 
$3\sigma$ bounds on the mass square differences
($\Delta m_{12}^{2}$, $\Delta m_{atm}^{2}$),
the mixing angles ($\theta_{12}, \theta_{13},
\theta_{23}$) \cite{Capozzi:2016rtj}
and also the cosmological bound on
the sum of three light neutrinos masses \cite{Ade:2015xua}.
We also determine the effective mass $m_{\beta\beta}$ which is relevant 
for neutrino-less double beta decay 
and compare it against the current bound on $m_{\beta \beta}$
from GERDA phase I experiment \cite{Agostini:2013mzu}.

Next, we explain the possible origin of the baryon asymmetry
at the present epoch from an initially matter-antimatter
symmetric Universe via leptogenesis. We first generate the lepton
asymmetry (or ${\rm B-L}$ asymmetry, $Y_{B-L}$) from the
out of equilibrium, CP violating decays of RH neutrinos.
The lepton asymmetry thus produced has been converted into the
baryon asymmetry by the ($B+L$) violating sphaleron
processes which are effective before and during electroweak phase
transition \cite{Manton:1983nd, Klinkhamer:1984di,
Kuzmin:1985mm}. When the sphaleron processes
are in thermal equilibrium ($10^{12}\,{\rm GeV}
\,\la T \la 10^2$ GeV, $T$ being the temperature of the Universe),
the conversion rate is given by \cite{Khlebnikov:1988sr}
\begin{eqnarray}
Y_{B} = - \frac{8 N_{f} + 4 N_{\phi_{h}}}
{22 N_{f} + 13 N_{\phi_{h}}} Y_{B-L}
\end{eqnarray}
where $N_{f}= 3$ and $N_{\phi_{h}}= 1$, are the number
of fermionic generations and number of Higgs doublet in
the model, respectively.

Finally, in order to address the dark matter issue,
we consider the singlet scalar $\dm$ as a DM candidate.
Since the couplings of this scalar to the rest of the particles of 
the model are free parameters, they could take any value. 
Depending on the value of these couplings, we could consider 
$\dm$ as a WIMP or a FIMP. Detailed study on the
WIMP type scalar DM in the present $\ubl$ framework
has been done in Refs. \cite{Guo:2015lxa, Rodejohann:2015lca,
Biswas:2016ewm}. In most of the earlier works,
it has been shown that the WIMP
relic density is mainly satisfied around the resonance
regions of the mediator particles. Moreover, the WIMP
parameter space has now become severely constrained due to
non-observation of any ``real'' signal in various
direct detection experiments. Thus, as discussed
earlier, in this situation the study of scalar DM
other than WIMP is worthwhile.      
Therefore in this work, we consider the scalar field
$\dm$ as a FIMP candidate which, depending on its mass,
is dominantly produced from the decays of heavy bosonic
particles such as $h_1$, $h_2$, $\zbl$
and also from the annihilations of bosonic as well as
fermionic degrees of freedom present in the model
(e.g. $N_i$, $\zbl$, $h_i$ etc.).
In particular, in Ref. \cite{Biswas:2016yjr},
we have also studied a SM singlet scalar as the
FIMP type DM candidate in a $L_{\mu} - L_{\tau}$
gauge extension of the SM. In that work, we have considered
the extra gauge boson mass in MeV range to explain the
muon $(g-2)$ anomaly. Consequently, the production
of a $\mathcal{O}$(GeV) DM from the decay of $Z_{\mu\tau}$ is
forbidden. Additionally, in that model due to the considered
$L_{\mu} - L_{\tau}$ flavour symmetry the neutrino mass matrices
(both light and heavy neutrinos) have particular shape.
On the other hand, in the present work, we have extensively
studied the FIMP DM production mechanism from all
possible decays and annihilations other particles
present in the model. Moreover, we have found that
depending on our DM mass, a sharp correlation exists
among the three puzzles of astroparticle physics
namely neutrino mass generation, leptogenesis and DM.
Furthermore, earlier in Ref.\,\,\cite{Biswas:2016bfo}, 
one of us, along with other collaborators, has studied the freeze-in DM production mechanism
in the framework of U(1)$_{\rm B-L}$ extension
of the SM. However, in that article they have considered
an MeV range RH neutrino as the FIMP DM candidate. Thus,
in the context of DM phenomenology the current work
is vastly different from Ref.\,\,\cite{Biswas:2016bfo}.

In the non-thermal scenario, most of the production
of the FIMP from the decay of a heavy particle occurs
when $T\sim M$, where $M$ is the mass of the
decaying mother particle, which is generally assumed to be in
thermal equilibrium. Therefore, the non-thermality
condition of the FIMP demands that
$\dfrac{\Gamma}{H}<1\bigg|_{T\sim M}$
\cite{Arcadi:2013aba},
which in turn imposes a severe upper bound
on the coupling strengths of the FIMP. 
Thus the non-thermality condition requires
extremely small coupling of $\dm$ with the thermal
bath ($\la 10^{-10}$) and hence, FIMP DM can
easily evade all the existing bounds from DM direct detection
experiments \cite{Aprile:2012zx, Akerib:2015rjg, Ahmed:2010wy}.  

Rest of the paper has been arranged in the following
manner, in Section \ref{model} we discuss
the model in detail. In Section \ref{result} we
present the main results of the paper. 
In particular, we discuss the neutrino phenomenology
in Section \ref{neutrino-sec}, 
baryogenesis  via leptogenesis 
in Section \ref{baryogenesis}
and non-thermal FIMP dark matter $\dm$ production 
in Section \ref{fimp-DM}. Finally in
Section \ref{conclusion} we end with our conclusions.  
\section{Model}
\label{model}
The gauged $\ubl$ extension of SM is one of the most extensively studied
BSM model so far. In this model, the gauge sector of the SM is enhanced
by imposing a local $\ubl$ symmetry to the SM Lagrangian, where B
and L represent the respective baryon and lepton number of a particle.
Therefore, the complete gauged group is $\smgauge\times\ubl$. Since
the $\ubl$ extension of SM is not an anomaly free theory, hence we
need to introduce some chiral fermions to cancel the anomaly. In order
to achieve this, we have considered three extra right handed (RH)
neutrinos to make the proposed ${\rm B-L}$ extension anomaly free. 
Besides the SM particles and three RH neutrinos, we have introduced
two SM gauge singlet scalars $\phi_H$, $\dm$ in the theory with suitable
${\rm B-L}$ charges. One of the scalar fields namely $\phi_H$
breaks the proposed $\ubl$ symmetry spontaneously by acquiring
a nonzero VEV $\vbl$ and thereby generates masses to all the
BSM particles. 
We have chosen the ${\rm B-L}$ charge of $\dm$ in such a
way that the Lagrangian of our model before the $\ubl$
symmetry breaking does not contain any interaction term
involving odd powers of $\dm$. When $\phi_H$ gets a
nonzero VEV, this $\ubl$ symmetry breaks spontaneously into
a remnant $\mathbb{Z}_2$ symmetry under which only $\dm$
becomes odd. The $\mathbb{Z}_2$ invariance
of the Lagrangian will be preserved as long as the parameters
of the Lagrangian are such that the scalar field $\dm$
does not get any VEV. Under this condition, the scalar
field $\dm$ becomes absolutely stable and, in principle,
can serve as a viable dark matter candidate. The respective
SU(2)$_{\rm L}$, U(1)$_{\rm Y}$ and $\ubl$ charges of
all the particles in the present model are listed
in Table \ref{tab1}.
\def\I{i}
\begin{center}
\begin{table}
\begin{tabular}{||c|c|c|c||}
\hline
\hline
\begin{tabular}{c}
    Gauge\\
    Group\\ 
    \hline
    
    SU(2)$_{\rm L}$\\ 
    \hline
    U(1)$_{\rm Y}$\\ 
    \hline
    $\ubl$\\
\end{tabular}
&

\begin{tabular}{c|c|c}
    \multicolumn{3}{c}{Baryon Fields}\\ 
    \hline
    $Q_{L}^{i}=(u_{L}^{i},d_{L}^{i})^{T}$&$u_{R}^{i}$&$d_{R}^{i}$\\ 
    \hline
    $2$&$1$&$1$\\ 
    \hline
    $1/6$&$2/3$&$-1/3$\\ 
    \hline
    $1/3$&$1/3$&$1/3$\\
\end{tabular}
&
\begin{tabular}{c|c|c}
    \multicolumn{3}{c}{Lepton Fields}\\
    \hline
    $L_{L}^{i}=(\nu_{L}^{i},e_{L}^{i})^{T}$ & $e_{R}^{i}$ & $N_{R}^{i}$\\
    \hline
    $2$&$1$&$1$\\
    \hline
    $-1/2$&$-1$&$0$\\
    \hline
    $-1$&$-1$&$-1$\\
\end{tabular}
&
\begin{tabular}{c|c|c}
    \multicolumn{3}{c}{Scalar Fields}\\
    \hline
    $\phi_{h}$&$\phi_{H}$&$\phi_{DM}$\\
    \hline
    $2$&$1$&$1$\\
    \hline
    $1/2$&$0$&$0$\\
    \hline
    $0$&$2$&$n_{BL}$\\
\end{tabular}\\
\hline
\hline
\end{tabular}
\caption{Charges of all particles under various symmetry groups.}
\label{tab1}
\end{table}
\end{center} 
The complete Lagrangian for the model is as follows,
\begin{eqnarray}
\mathcal{L}&=&\mathcal{L}_{SM} + \mathcal{L}_{DM} +
(D_{\mu}\phi_{H})^{\dagger} (D^{\mu}\phi_{H})
-\frac{1}{4} {F_{BL}}_{\mu \nu} {F_{BL}}^{\mu \nu} + 
\frac{i}{2}\bar{N_i}\gamma^{\mu}D_{\mu} N_{i}
-V(\phi_{h},\phi_{H})\nn \\
&&-\sum_{i=1}^3 \frac{y_{N_{i}}}{2}
\phi_{H} \bar{N^{c}_{i}} N_{i}
-\sum_{i,\,j=1}^3 y_{ij}^{\prime} \bar{L_{i}}
\tilde {\phi_{h}} N_{j} +h.c.\,,
\label{lag}
\end{eqnarray}
with $\tilde{\phi_h} = i \sigma_2 \phi^*_h$. The term $\mathcal{L}_{SM}$ and
$\mathcal{L}_{DM}$ represent the SM and dark sector Lagrangian, respectively.
The dark sector Lagrangian $\mathcal{L}_{DM}$ containing all possible
gauge invariant interaction terms of the scalar field $\dm$,
has the following form
 \begin{eqnarray}
\mathcal{L}_{DM} &=& (D^{\mu}\phi_{DM})^{\dagger}
(D_{\mu}\phi_{DM})
- \mu_{DM}^{2} (\phi_{DM}^{\dagger} \phi_{DM}) 
-\lambda_{DM}\,(\phi_{DM}^{\dagger} \phi_{DM})^{2} 
-\lambda_{Dh}\,(\phi_{DM}^{\dagger} \phi_{DM})
(\phi_{h}^{\dagger} \phi_{h})
\nn \\ && 
-\lambda_{DH}\,(\phi_{DM}^{\dagger} \phi_{DM}) (\phi_{H}^{\dagger} \phi_{H})\,,
\label{ldm}
\end{eqnarray}
where the interactions of $\dm$ with $\phi_h$ and $\phi_H$
are proportional to the couplings $\ldh$ and $\ldH$, respectively.
The fourth term in Eq.~(\ref{lag}) represents the kinetic
term for the additional gauge boson $Z_{BL}^{\mu}$ in terms
of field strength tensor ${F_{BL}}_{\mu\nu}$ of the $\ubl$
gauge group. The covariant derivatives involving in the kinetic
energy terms of the BSM scalars and fermions, $\phi_H$, $\dm$
and $N_i$ (Eq.~(\ref{lag})), can be expressed in a generic form
\begin{eqnarray}
D_\mu \psi = (\partial_{\mu} + i\,g_{BL}\,Q_{BL}(\psi)
\,{Z_{BL}}_{\mu})\,\psi \,,
\label{coderi}
\end{eqnarray}
where $\psi = \phi_{DM}, \phi_H$, $N_i$ and $Q_{BL}(\psi)$ represents the
${\rm B-L}$ charge of the corresponding field (listed in Table \ref{tab1}).
The quantity $V(\phi_h, \phi_H)$ in Eq.~(\ref{lag}) contains
the self interaction terms of $\phi_{H}$ and $\phi_h$ as well as
the mutual interaction term between the two scalar fields.
The expression of $V(\phi_h, \phi_H)$ is given by
\begin{eqnarray}
V(\phi_h, \phi_H) = \mu_{H}^{2} \phi_{H}^{\dagger} \phi_{H} 
+ \mu_{h}^{2} \phi_{h}^{\dagger} \phi_{h}
+ \lambda_{H} (\phi_{H}^{\dagger} \phi_{H})^{2}
+ \lambda_{h} (\phi_{h}^{\dagger} \phi_{h})^{2}
+ \lambda_{hH}(\phi_{h}^{\dagger} \phi_{h})
(\phi_{H}^{\dagger} \phi_{H})\,.
\label{int}
\end{eqnarray}
After the symmetry breaking, the SM Higgs doublet $\phi_{h}$
and the BSM scalar $\phi_{H}$ take the following form,
\begin{eqnarray}
\phi_{h}=
\begin{pmatrix}
0 \\
\dfrac{v+H}{\sqrt{2}}
\end{pmatrix}
\,\,\,\,\,\,\,\,\,
\phi_{H}=
\begin{pmatrix}
\dfrac{v_{BL}+H_{BL}}{\sqrt{2}}
\end{pmatrix}\,\,,
\label{phih}
\end{eqnarray}
where $v$ = 246 GeV is the VEV of $\phi_{h}$, which breaks the SM gauge
symmetry into a residual U(1)$_{\rm EM}$ symmetry. The remaining terms
in Eq.~(\ref{lag}) are the Yukawa interaction terms for the
left handed and right handed neutrinos. As mentioned in the beginning
of this section, when the extra scalar field $\phi_H$ gets a
nonzero VEV $\vbl$, the proposed $\ubl$ gauge symmetry breaks
spontaneously. As a results, the Majorana mass terms for the right
handed neutrinos, proportional to the Yukawa couplings $y_{N_i}$,
are generated. In general, for a three generation of right handed
neutrinos, we will have a $3\times 3$ Majorana mass matrix $\mR$ with
all off diagonal terms are present. However, in the present
scenario for calculational simplicity, we have chosen a basis
for the $N_i$ fields with respect to which $\mR$ is diagonal.
The diagonal elements, representing the masses of $N_i$s,
are given by,
\begin{eqnarray}
M_{N_i} &=& \dfrac{y_{N_i}}{\sqrt{2}} v_{BL} \,.
\label{mni}
\end{eqnarray}
Like the three right handed neutrinos, the extra neutral gauge
bosons also becomes massive through the Eq.~(\ref{coderi})
when $\phi_H$ picks up a VEV. The mass term $\zbl$ is given by
\begin{eqnarray}
M_{Z_{BL}} &=& 2\,g_{BL}\,v_{BL}\,.
\label{mzbl}
\end{eqnarray}

When both $\phi_h$ and $\phi_H$ obtain their respective VEVs,
there will be a mass mixing between the states $H$ and $H_{BL}$.
The mass matrix with respect to the basis $H$ and $H_{BL}$
looks like as follows
\begin{eqnarray}
\mathcal{M}^2_{scalar} = \left(\begin{array}{cc}
2\lambda_h v^2 ~~&~~ \lambda_{hH}\,v_{BL}\,v \\
~~&~~\\
\lambda_{hH}\,v_{BL}\,v ~~&~~ 2 \lambda_H v^2_{BL}
\end{array}\right) \,\,.
\label{mass-matrix}
\end{eqnarray}
Rotating the basis states $H$ and $H_{BL}$ by a suitable
angle $\alpha$, we can make the above mass matrix diagonal.
The new basis states ($h_1$ and $h_2$) with respect to which
the mass matrix $\mathcal{M}^2_{scalar}$ becomes diagonal,
are some linear combinations of earlier basis states $H$ and $H_{BL}$.
The new basis states, now representing two the physical states,
are defined as
\begin{eqnarray}
h_{1}&=& H \cos \alpha + H_{BL} \sin \alpha \,, \nn \\
h_{2}&=& - H \sin \alpha + H_{BL} \cos \alpha\,,
\end{eqnarray}
where we denote $h_1$ as the SM-like Higgs boson while
$h_2$ is playing the role of a BSM scalar field. The mixing angle
between $H$ and $H_{BL}$ can be expressed in terms of
the parameters of the Lagrangian (cf. Eq.~(\ref{lag})) as,
\begin{eqnarray}
\tan 2\alpha = \dfrac{\lambda_{hH}\,v_{BL}\,v}
{\lambda_h v^2 - \lambda_H v^2_{BL}}\,.
\label{mix-ang}
\end{eqnarray}
Besides the two physical scalar fields $h_1$ and $h_2$, as mentioned
earlier, there is another scalar field ($\dm$) in the present model,
which can play the role of a dark matter candidate. The masses of these
three physical scalar fields $h_1$, $h_2$ and $\dm$ are give below,
\begin{eqnarray}
M^2_{h_1} &=& \lambda_h v^2 + \lambda_H v_{BL}^2 - 
\sqrt{(\lambda_h v^2 - \lambda_H v_{BL}^2)^2 + (\lambda_{hH}\,v\,v_{BL})^2}
\ ,\nn \\
M^2_{h_2} &=& \lambda_h v^2 + \lambda_H v^2_s + 
\sqrt{(\lambda_h v^2 - \lambda_H v_{BL}^2)^2 + (\lambda_{hH}\,v\,v_{BL})^2}\,,
\nn \\
M^2_{DM} &=& \mu^2_{DM} + \frac{\lambda_{Dh} v^2}{2} +
\frac{\lambda_{DH} v^2_{BL}}{2}\,,
\label{mh1-mh2}
\end{eqnarray}
where $M_{x}$\footnote{Throughout the paper we have kept
the mass ($M_{h_1}$) of SM-like Higgs boson $h_1$ fixed at 125.5 GeV.}
denotes the mass of the corresponding scalar field $x$.  

In this work, we choose $M_{h_2}$, $\mdm$, $\nbl$, $M_{N_i}$, $\mzbl$,
$\gbl$, $\alpha$, $\ldh$, $\ldH$ and $\lambda_{DM}$ as our
independent set of parameters. The 
other parameters in the Lagrangian namely $\lambda_{h}$,
$\lambda_{H}$, $\lambda_{hH}$, $\mu_{\phi_{h}}^{2}$ and $\mu_{\phi_{H}}^{2}$
can be expressed in terms of these variables as follows
\cite{Biswas:2016ewm}.
\begin{eqnarray}
\lambda_{H} &=& \dfrac{M_{h_{1}}^{2} + M_{h_{2}}^{2} +
(M_{h_{2}}^{2} - M_{h_{1}}^{2})\cos 2 \alpha}{4\,v_{BL}^{2}}\,,\nn \\
\lambda_{h}&=& \dfrac{M_{h_{1}}^{2} + M_{h_{2}}^{2} +
(M_{h_{1}}^{2} - M_{h_{2}}^{2})\cos 2 \alpha}{4\,v^{2}}\,,\nn\\
\lambda_{hH} &=& \dfrac{(M_{h{1}}^{2}-M_{h_{2}}^{2})
\cos \alpha \sin \alpha}{v\,v_{BL}}\,,\nn\\
\mu_{\phi_{h}}^{2}&=& - \dfrac{(M_{h_1}^{2}+M_{h_2}^{2})v
+(M_{h_1}^{2}-M_{h_2}^{2})(v \cos 2 \alpha + v_{BL}
\sin 2 \alpha)}{4\,v} \,,\nn\\
\mu_{\phi_{H}}^{2}&=& \dfrac{-(M_{h_1}^{2}+M_{h_2}^{2})v_{BL}
+ (M_{h_1}^{2}-M_{h_2}^{2})(v_{BL}\cos 2 \alpha - v
\sin 2 \alpha)}{4\,v_{BL}}\,,\nn \\
\mu^2_{DM} &=& {M}^2_{DM}- \frac{\lambda_{Dh} v^2}{2} -
\frac{\lambda_{DH} v^2_{BL}}{2}\,,
\label{quatic-vacuum-coeff}
\end{eqnarray}
where $\vbl$ is defined in terms of $\mzbl$ and $\gbl$ in Eq.~(\ref{mzbl}).

As we already know, in the present scenario two of the three scalar
fields namely $\phi_h$ and $\phi_H$ obtain VEVs. On the other hand,
the remaining scalar field $\dm$ does not have any VEV,
which ensures its stability by preserving its $\mathbb{Z}_2$
odd parity. Therefore, the ground state of the system is
($\langle\phi_h\rangle$, $\langle\phi_H\rangle$,
$\langle\dm\rangle$) = ($v$, $\vbl$, 0).
Now, such a ground state (vacuum) will be bounded from
below when the following inequalities are satisfied
simultaneously \cite{Biswas:2016ewm},
\begin{eqnarray}
&&\mu^2_{\phi_h} < 0, \mu^2_{\phi_H} < 0 ,
\mu^2_{DM} > 0\,,\nonumber \\
&&\lambda_h \geq 0, \lambda_H \geq 0, \lambda_{DM} \geq 0,\nonumber \\
&&\lambda_{hH} \geq - 2\sqrt{\lambda_h\,\lambda_H},\nonumber \\
&&\lambda_{Dh} \geq - 2\sqrt{\lambda_h\,\lambda_{DM}},\nonumber \\
&&\lambda_{DH} \geq - 2\sqrt{\lambda_H\,\lambda_{DM}},\nonumber \\
&&\sqrt{\lambda_{hH}+2\sqrt{\lambda_h\,\lambda_H}}\sqrt{\lambda_{Dh}+
2\sqrt{\lambda_h\,\lambda_{DM}}}
\sqrt{\lambda_{DH}+2\sqrt{\lambda_H\,\lambda_{DM}}} \nonumber \\ 
&&+ 2\,\sqrt{\lambda_h \lambda_H \lambda_{DM}} + \lambda_{hH} \sqrt{\lambda_{DM}}
+ \lambda_{Dh} \sqrt{\lambda_H} + \lambda_{DH} \sqrt{\lambda_h} \geq 0 \,\,\,\,.
\label{vsc}
\end{eqnarray}
Besides the lower limits of $\lambda$s as described by the above inequalities,
there are also upper limits on the Yukawa and quartic couplings
arising from the perturbativity condition which demands that
the Yukawa and scalar quartic couplings have to be less than
$\sqrt{4\,\pi}$ ($y<\sqrt{4\,\pi}$) and $4\,\pi$
($\lambda < 4\,\pi$) respectively \cite{Chakrabarty:2015yia}.
\section{Results}
\label{result}
\subsection{Neutrino Masses and Mixing}
\label{neutrino-sec}
As mentioned earlier, the cancellation of both axial
vector anomaly \cite{Adler:1969gk, Bardeen:1969md}
and gravitational gauge anomaly \cite{Delbourgo:1972xb,
Eguchi:1976db}, in $\ubl$ extended SM, requires the
presence of extra chiral fermions. Hence, in the present
model to cancel these anomalies we have introduced three right handed (RH)
neutrinos ($N_{i}$, $i$=1 to 3). The Majorana masses for the RH neutrinos
are generated only after spontaneous breaking of the proposed ${\rm B-L}$
symmetry by the VEV of $\phi_H$. Also in the present scenario, as stated
earlier, we are working in a basis where the Majorana mass matrix for
the three RH neutrinos are diagonal i.e. $\mR = {\rm diag}\,(M_{N_1},
M_{N_2}, M_{N_3})$. The expression for the mass of $i$th RH neutrino
($M_{N_i}$) is given in Eq.~(\ref{mni}). On the other hand, the Dirac mass
terms involving both left chiral and right chiral neutrinos, are originated
when the electroweak symmetry is spontaneously broken by the VEV of
SM Higgs doublet $\phi_h$, giving rise
to a $3\times 3$ complex matrix $\mD$. 
In general, one can take all the elements of matrix $\mD$ as complex
but for calculational simplicity and also keeping in mind
that only three physical phases (one Dirac phase and
two Majorana phases) exist for three light neutrinos
(Majorana type), we have considered only three complex
elements in the lower triangle part of the Dirac mass
matrix $\mD$. However, the results we have presented
later in this section will not change significantly if
we consider all the elements of $\mD$ are complex.
The Dirac mass matrix $\mD$ we assume has the following
structure:
\begin{eqnarray}
\mathcal{M}_{D} = \left(\begin{array}{ccc}
y_{ee} ~~&~~ y_{e\mu}
~~&~~y_{e\tau} \\
~~&~~\\
y_{\mu e} + i\,\tilde{y}_{\mu e} ~~&~~ y_{\mu \mu}
~~&~~ y_{\mu \tau}\\
~~&~~\\
y_{\tau e} + i\,\tilde{y}_{\tau e} ~~&
~~ y_{\tau \mu} + i\,\tilde{y}_{\tau \mu} ~~&~~ y_{\tau \tau} \\
\end{array}\right) \,,
\label{mdcomplex}
\end{eqnarray}   
where $y_{ij} = \dfrac{y_{ij}^{\prime}}{\sqrt{2}} v$ ($i,j = $ e, $\mu$, $\tau$)
and the Yukawa coupling $y_{ij}^{\prime}$ has been defined in Eq.~(\ref{lag}).

Now, with respect to the Majorana basis $\left(\overline{{\nu_\alpha}_L}
~~\overline{({N_{\alpha}}_R)^c}\right)^T$ and $\left(({\nu_\alpha}_L)^c
~~{N_{\alpha}}_R\right)^T$ one can write down the Majorana mass matrix for
both left and right chiral neutrinos using $\mD$ and $\mR$
matrices in the following way,
\begin{eqnarray}
M = \left(\begin{array}{cc}
0 & \mD \\
\mD^{T} & \mR
\end{array}\right) \,\,.
\label{mtot}
\end{eqnarray}
Since $M_{D}$ and $M_{R}$ are both $3\times3$ matrices (for three generations
of neutrinos), the resultant matrix $M$ will be of order $6\times6$
and also it is a complex symmetric matrix which reflects its Majorana
nature. Therefore, after diagonalisation of the matrix $M$, we get
three light and three heavy neutrinos, all of which are Majorana
fermions. If we use the block diagonalisation technique, we can write
the light and heavy neutrino mass matrices in the leading order as,
\begin{eqnarray}
m_{\nu}&\simeq&-\mD\,\mR^{-1} \mD^T\,, 
\label{activemass}\\
m_N &\simeq& \mR\,.
\label{sterilemass}
\end{eqnarray} 
Here $M_{R}$ is a diagonal matrix and the expression
of all the elements of $m_{\nu}$ in terms of the elements of $\mD$
and $\mR$ matrices are given in Appendix \ref{app:mnu}. After
diagonalising ${m_{\nu}}$ matrix we get three light neutrino
masses ($m_i$, $i=1,\,2,\,3$), three mixing angles ($\theta_{12}$,
$\theta_{13}$ and $\theta_{23}$) and one Dirac CP phase $\delta$.

We have used the Jarlskog Invariant $J_{\rm CP}$
\cite{Jarlskog:1985ht} to determine the Dirac CP
phase $\delta$, which is defined as,
\begin{eqnarray}
J_{\rm CP} = \frac{1}{8} \sin 2\theta_{12} \sin 2\theta_{23}
\sin 2\theta_{13} \cos \theta_{13}
\sin \delta.
\label{jcp}
\end{eqnarray}
Moreover, the quantity $J_{\rm CP}$ is related to the elements of
the Hermitian matrix $h=m_{\nu}m^{\dagger}_{\nu}$ in the
following way,
\begin{eqnarray}
J_{CP} = \frac{{\rm Im}\,(h_{13} h_{23} h_{31})}
{\Delta m^{2}_{21}\, \Delta m^{2}_{32}\,
\Delta m^{2}_{31}}
\label{jcpe} 
\end{eqnarray}
where in the numerator ${\rm Im} (X)$ represents the imaginary part of
$X$ while in the denominator, $\Delta m^{2}_{ij} = m^{2}_{i} - m^{2}_{j}$.
Once we determine the quantity $J_{\rm CP}$ (from Eq.~(\ref{jcpe}))
and the intergenerational mixing angles of neutrinos then one can
easily determine the Dirac CP phase using Eq.~(\ref{jcp}).

In the present scenario we have twelve independent parameters coming from the 
Dirac mass matrix. The RH neutrino mass matrix, in principle, should bring in 
three additional parameters. However, as we will discuss in details in 
Section \ref{baryogenesis}, two of the RH neutrino masses is taken 
to be nearly degenerate. In particular, the condition of resonant leptogenesis 
requires that  $M_{N_2}-M_{N_1}=\Gamma_1/2$, where $\Gamma_1$ is the 
tree level decay width of $N_1$ and is seen to be $\sim 10^{-11} $ GeV.
Therefore, for all practical purposes we have $M_{N_1}\simeq M_{N_2}$, and the 
RH neutrino mass matrix only brings in two independent parameters, 
$M_{N_1}$ and $M_{N_3}$. Thus, we have fourteen independent 
parameters which we vary in the following ranges, 
\begin{eqnarray}
\begin{array}{cccccc}
1 \,\,{\rm TeV}& \le & M_{N_1}& \le & 3\,\,{\rm TeV}\,,&\\
M_{N_1}\,\,& < & M_{N_3}& \le & 15\,\,{\rm TeV}\,,&\\
1\,&\le &\dfrac{\sqrt{2}\,\,y_{ij}}{v}\,\times 10^{8}
& \le & 1000\, & (i,j= e,\,\mu,\,\tau, i=j\neq e)\,,\\
1\,&\le &\dfrac{\sqrt{2}\,\,y_{ee}}{v}\,\times 10^{10}
& \le & 100\,,&\\
1\,&\le &\dfrac{\sqrt{2}\,\,\tilde{y}_{ij}}{v}\,\times 10^{8}
& \le & 1000\,&(i= \tau, j=e,\,\mu)\,,\\
1\,&\le &\dfrac{\sqrt{2}\,\,\tilde{y}_{\mu e}}{v}\,\times 10^{9}
& \le & 1000\,.&
\label{para-ranges}
\end{array}
\end{eqnarray}
We try to find the allowed parameter space which satisfy the
following constraints on three mixing angles ($\theta_{ij}$) and
two mass square differences ($\Delta m_{ij}^{2}$), $J_{CP}$
obtained from neutrino oscillation data and the cosmological
bound on the sum of three light neutrino masses. These
experimental/observational results are listed below.
\begin{itemize}
\item Measured values of three mixing angles in $3\sigma$
range \cite{Capozzi:2016rtj}:\\
$30^{\circ}<\,\theta_{12}\,<36.51^{\circ}$,
$37.99^{\circ}<\,\theta_{23}\,<51.71^{\circ}$ and
$7.82^{\circ}<\,\theta_{13}\,<9.02^{\circ}$.

\item Allowed values of two mass squared differences in $3\sigma$ range
\cite{Capozzi:2016rtj}:\\ $6.93<\dfrac{\Delta m^2_{21}}
{10^{-5}}\,{\text{eV}^2} < 7.97$ and $2.37<\dfrac{\Delta m^2_{31}}
{10^{-3}}\,{\text{eV}^2} < 2.63$ in $3\sigma$ range. 

\item Above mentioned values of the neutrino oscillation parameters
also put an upper bound on the absolute value of $J_{\rm CP}$
from Eq.~(\ref{jcp}), which is $|J_{\rm CP}|\leq\,0.039$.

\item Cosmological upper bound on the sum of three light
neutrino masses i.e. $\sum_i m_{i} < 0.23$ eV at $2\sigma$
C.L. \cite{Ade:2015xua}.
\end{itemize}
While it is possible to obtain both normal hierarchy (NH)
($m_1<m_2<m_3$) and inverted hierarchy (IH) ($m_3<m_1<m_2$)
in this scenario, we show our results only for NH for brevity. 
Similar results can be obtained for IH.
\begin{figure}[h!]
\centering
\includegraphics[angle=0,height=7.50cm,width=8.5cm]{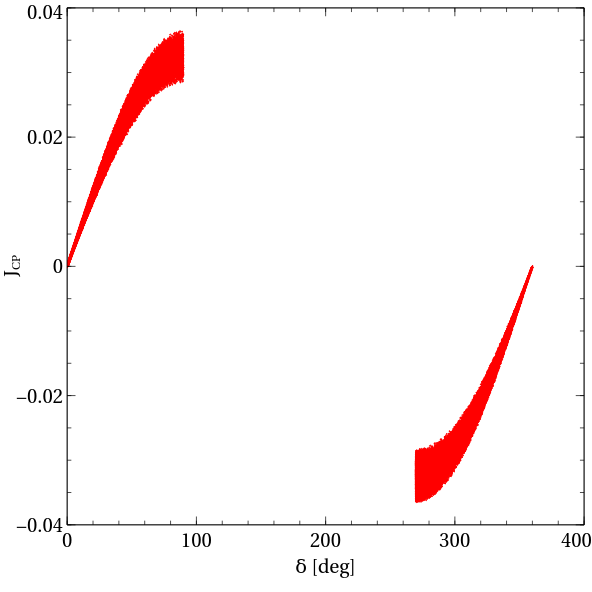}
\includegraphics[angle=0,height=7.50cm,width=8.0cm]{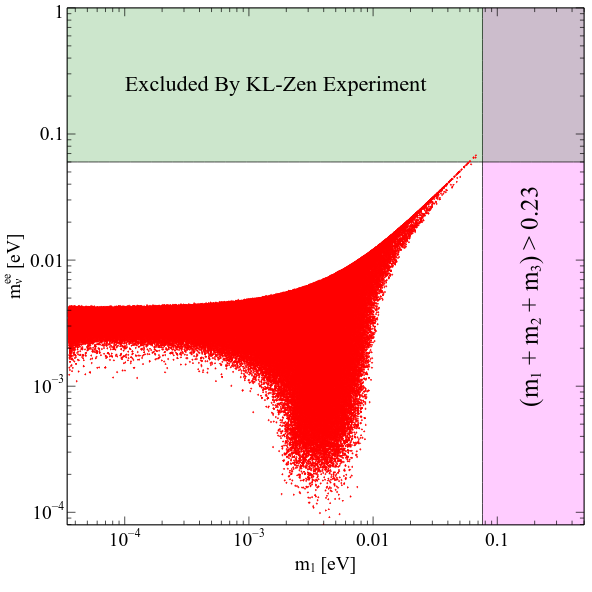}
\caption{{\bf LP:} Variation of $J_{\rm cp}$ with $\delta$.
{\bf RP:} Variation of neutrino less double $\beta$ decay
parameter ${m_{\beta\beta}}$ with $m_1$}      
\label{nu-plot3}
\end{figure}

In the LP of Fig.\,\ref{nu-plot3}, we have shown the variation
of $J_{\rm CP}$ parameter (as defined in Eq.~(\ref{jcp}))
with the Dirac CP phase $\delta$. From this plot one can
easily notice that there are two allowed ranges
of Dirac CP phase $0^{\circ}\leq\delta\leq90^{\circ}$ and
$270^{\circ}\leq\delta\leq360^{\circ}$ respectively
which can reproduce the neutrino oscillation parameters
in $3\sigma$ range. Since the Jarlskog invariant $J_{\rm CP}$
is proportional to $\sin \delta$ (Eq.\,\,\ref{jcp}), hence we get
both positive and negative values of $J_{CP}$ symmetrically placed in the 
first and fourth quadrants. However, the absolute values of $J_{\rm CP}$
always lie below 0.039. Also, here we want to mention that
from the recent results of T2K \cite{Abe:2017vif} experiment,
values of $\delta$ lying in the fourth quadrant are
favourable compared to those in first quadrant. 
In the RP of Fig.\,\,\ref{nu-plot3},
we have shown the variation of neutrino less double
$\beta$ decay parameter $m_{\beta \beta}$ with 
the mass of lightest neutrino $m_1$. $m_{\beta \beta}$
is an important quantity for the study of neutrino less double
$\beta$ decay as the cross section of this process is
proportional to $m_{\beta \beta} = \left|\sum_{i=1}^3
({U_{\rm PMNS}})^2_{\,e\,i}\,\,m_i\right|=(m_{\nu})_{e\,e}$
(see Appendix \ref{app:mbb} for details.), where $(m_{\nu})_{e\,e}$
(Eq.\,\,\ref{mnuelements}) is the (1,1) element of light
neutrino mass matrix $m_{\nu}$.  
The nature of this plot is very to similar to
the usual plot in $m_{\beta \beta}-m_1$ plane for the
normal hierarchical scenario \cite{DellOro:2016tmg}.
In the same plot, we have also shown the current bound
on $m_{\beta\beta}$ from KamLand-Zen experiment
\cite{KamLAND-Zen:2016pfg}. 
\subsection{Baryogenesis via Resonant Leptogenesis}
\label{baryogenesis}

\begin{figure}[h!]
\centering
\includegraphics[angle=0,height=7.5cm,width=8.5cm]{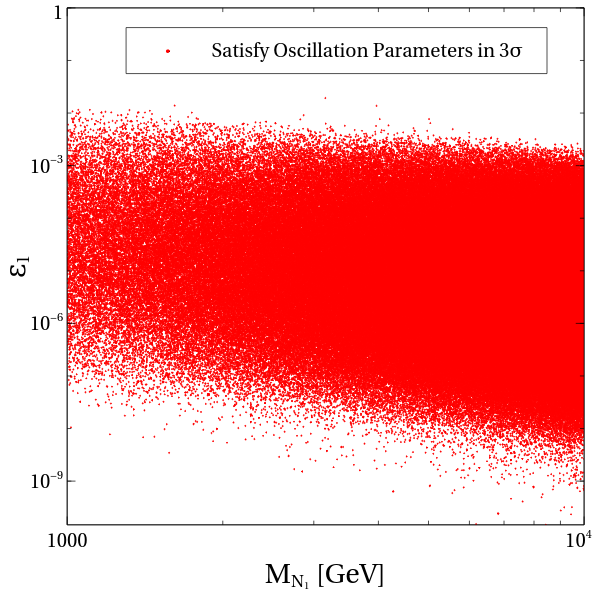}
\includegraphics[angle=0,height=7.50cm,width=8.5cm]{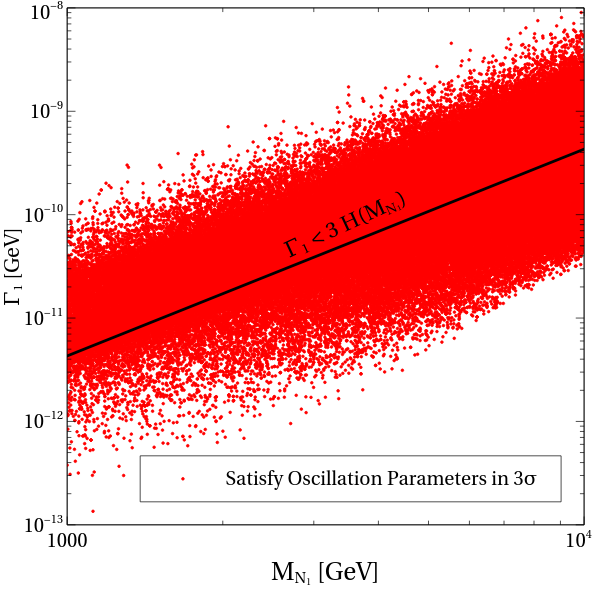}
\caption{{\bf LP:} Variation of CP asymmetry parameter $\varepsilon_1$
with the mass of $N_1$. {\bf RP:} Variation of total decay width of $N_1$
with $M_{N_1}$. Black solid line represents the upper bound of
$\Gamma_1$ coming from out of equilibrium condition of
$N_1$. All the points in both plots satisfy the neutrino
oscillation data in $3\sigma$ range.}      
\label{epsa-gamma-plot}
\end{figure}

As we have three RH neutrinos in the present model,
in this section we have studied the lepton asymmetry 
generated from the CP violating out of equilibrium decays
of these heavy neutrinos at the early stage of the Universe.
The ${\rm B-L}$ asymmetry thus
produced is converted into the baryon asymmetry through sphaleron
transitions which violate ${\rm B+L}$ quantum number while conserving the 
${\rm B-L}$ charge. The sphaleron processes are active between temperatures of 
$\sim 10^{12}$ GeV to $\sim 10^2$ GeV in the early Universe. 
At high temperatures the sphalerons
are in thermal equilibrium and subsequently they freeze-out
at around $T\simeq 100-200$ GeV \cite{Plumacher:1996kc,
Iso:2010mv}, just before electroweak symmetry breaking (EWSB).
To produce sufficient lepton asymmetry, which would eventually be converted
into the observed baryon asymmetry, one requires RH neutrinos with
masses $\ga 10^{8}-10^{9}$ GeV \cite{Plumacher:1996kc, Buchmuller:2002rq}.
This is the well know scenario of the ``normal'' or
``canonical" leptogenesis. However,
detection of these very massive RH neutrinos is beyond the reach of
LHC and other future colliders. 
Here we consider the RH neutrinos to be in TeV mass range to allow 
for their detection at collider experiments.
It has been shown that 
with RH neutrinos in the TeV mass-scale range, it is possible to generate 
adequate lepton asymmetry 
by considering the two lightest RH neutrinos $N_1$ and $N_2$ to be 
almost degenerate. More specifically, we demand that 
$M_{N_2}-M_{N_1}\simeq {\Gamma_1}/{2}$, where $\Gamma_1$
\footnote{The typical value of $\Gamma_1$ is $\sim 10^{-11}$ GeV
(see Fig.\,\ref{epsa-gamma-plot}) while $M_{N_i} \sim \mathcal{O}$(TeV).
Hence we take $M_{N_1}=M_{N_2}$ throughout the work.}
is the total decay width of the
lightest RH neutrino $N_1$. This scenario
is known as Resonant leptogenesis 
\cite{Pilaftsis:1997jf, Pilaftsis:2003gt,
Iso:2010mv, Heeck:2016oda}.
\begin{figure}[]
\centering
\includegraphics[angle=0,height=4cm,width=18cm]{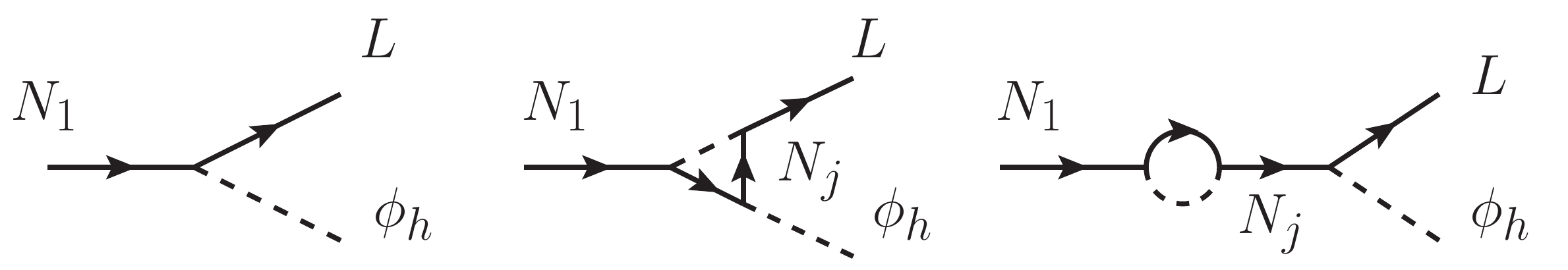}
\caption{Feynmann diagrams for the decay of lightest RH neutrino $N_1$.}
\label{feyn1-lepto}
\end{figure}

Fig.\,\,\ref{feyn1-lepto} shows the tree level as well as one loop  
decay diagrams of the lightest 
RH neutrino $N_1$.
These diagrams are applicable for all the three
RH neutrinos. Here $L$ represents the SM lepton
which can either be a charged
lepton or a left chiral neutrino depending on the nature of the
scalar field (charged \footnote{Since these processes occurred
before EWSB hence we have both charged as well as neutral scalars
in the SM.} or neutral) associated in the vertex while $N_j$
denotes the remaining two RH neutrinos, $N_2$ and $N_3$ for the
case of $N_1$ decay. In order to produce baryon asymmetry in
the Universe we need both C and CP violating interactions,
which is one of the three necessary conditions (see Sakharov
conditions \cite{Sakharov:1967dj} given in Section \ref{intro})
for baryogenesis. Lepton asymmetry generated from
the out of equilibrium decay of RH neutrinos is
determined by the CP asymmetry parameter ($\varepsilon_{i}$), which
is given by (for details see Appendix \ref{App:cp_lepton}),  
\begin{eqnarray}
\varepsilon_2 &\simeq& -\dfrac{1}{2}
\frac{{\rm Im}
\left[(\mathcal{M_{D}} \mathcal{M_{D}}^{\dagger})^{2}_{12}\right]}
{(\mathcal{M_{D}} \mathcal{M_{D}}^{\dagger})_{11}\,(\mathcal{M_{D}}
\mathcal{M_{D}}^{\dagger})_{22}} \,, 
\label{epsa2}\\
\varepsilon_1 &\simeq& -\dfrac{\Gamma_1\,\Gamma_2}{\Gamma^2_1+\Gamma^2_2}\,
\frac{{\rm Im}
\left[(\mathcal{M_{D}} \mathcal{M_{D}}^{\dagger})^{2}_{12}\right]}
{(\mathcal{M_{D}} \mathcal{M_{D}}^{\dagger})_{11}\,(\mathcal{M_{D}}
\mathcal{M_{D}}^{\dagger})_{22}} \,, 
\label{epsa1}\\
&\simeq& \dfrac{2\,\Gamma_1\,\Gamma_2}
{\Gamma^2_1+\Gamma^2_2}\,\varepsilon_2\,.
\label{epsa1-epsa2}
\end{eqnarray}  

In the LP of Fig.\,\,\ref{epsa-gamma-plot}, we show the variation
of CP asymmetry parameter $\varepsilon_1$, generated from the decay
of RH neutrino $N_1$, with the mass of $N_1$. Here we see that
for the considered ranges of $M_{N_1}$ ($1000\,{\rm GeV}\leq M_{N_1}
\leq 10000\,{\rm GeV}$) and other relevant
Yukawa couplings (see Eq.~(\ref{para-ranges})),
the CP asymmetry parameter $\varepsilon_1$ can be as large as $\sim 10^{-2}$,
which is significantly large compared to $\varepsilon_1$ in
the ``normal'' Leptogenesis case ($\varepsilon_1\sim 10^{-8}$ for
$M_{N_1}\sim 10^{10}$ GeV) \cite{Plumacher:1996kc}. In the
RP of Fig.\,\,\ref{epsa-gamma-plot}, we plot
the variation of total decay width of $N_1$ with $M_{N_1}$.
From this plot, one can easily notice that in the present scenario,
$\Gamma_1$ lies between $\sim 10^{-12}$ GeV to $10^{-9}$ GeV
for the entire considered range of $M_{N_1}$. All the points
in both panels satisfy the neutrino oscillations data in the $3\sigma$ range
while the black solid line in the RP provides the upper bound
on $\Gamma_1$, obtained from the out of equilibrium conditions
for $N_1$ i.e. $\Gamma_1<3\,H(M_{N_1})$ \cite{Plumacher:1996kc}
where $H$ is Hubble parameter at $T=M_{N_1}$.

\begin{figure}[]
\centering
\includegraphics[angle=0,height=5cm,width=14cm]{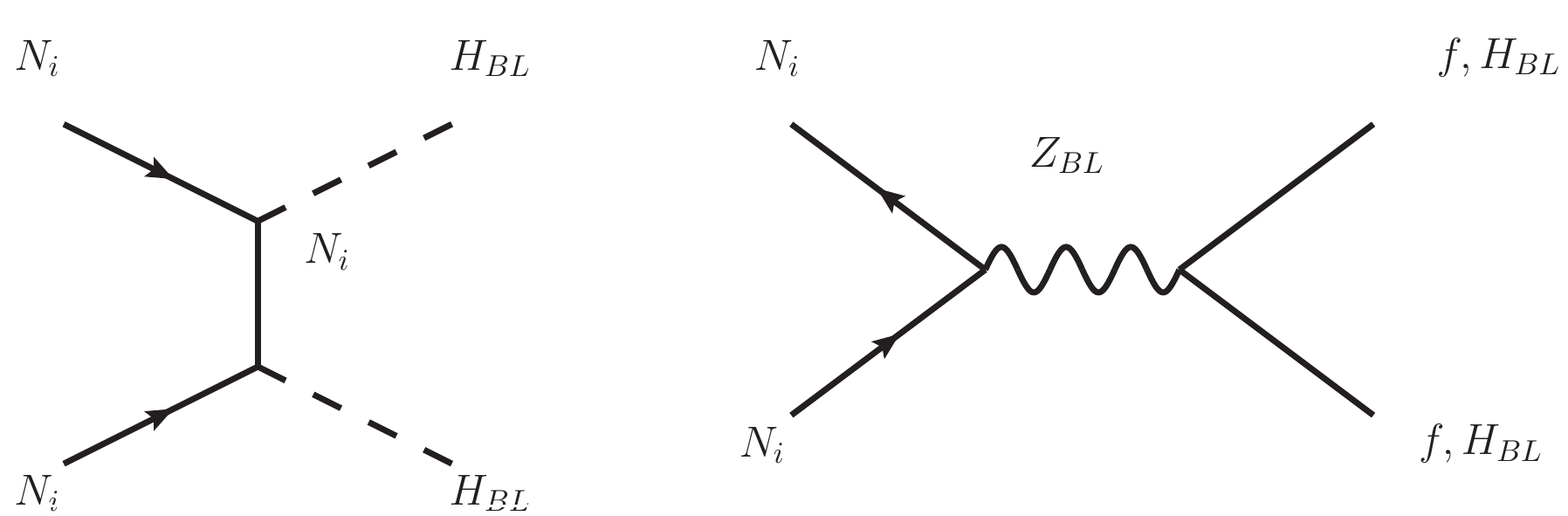}
\caption{Feynman diagrams for the annihilations of RH neutrinos.}
\label{feyn2-lepto}
\end{figure}

Next, we calculate the ${\rm B-L}$ asymmetry generated from
the decays as well as the pair annihilations of the RH neutrinos
$N_1$ and $N_2$. In order to calculate the net ${\rm B-L}$
asymmetry produced from the interactions of $N_1$ and
$N_2$ at temperature of the Universe $T\simeq 150$ GeV (freeze-out
temperature of sphaleron) we have to solve a set of three coupled
Boltzmann equations. The relevant Boltzmann equations
\cite{Plumacher:1996kc,Iso:2010mv} for calculating $Y_{N_i}$
and $Y_{B-L}$ are given below,
\begin{eqnarray}
\frac{d Y_{N_1}}{d z} & = & -\dfrac{M_{pl}}{1.66\,M^2_{N_1}}
\dfrac{z\,\sqrt{g_{\star}(z)}}{g_{\rm s}(z)}
\,\langle \Gamma_{1} \rangle\,\left(Y_{N_1}-Y_{N_{1}}^{\rm eq} \right)
-\dfrac{2\,\pi^2}{45}\,\dfrac{M_{pl}\,M_{N_1}}{1.66}
\dfrac{\sqrt{g_{\star}(z)}}{z^2} \times \nn \\
&&\left({\langle\sigma {\rm v}\rangle}_{N_1,\,\zbl} +
{\langle\sigma {\rm v}\rangle}_{N_1,t,H_{\rm BL}}\right)\,
\left(Y_{N_1}^2-(Y_{N_{1}}^{\rm eq})^2\right),
\label{beN1} \\
\frac{d Y_{N_2}}{d z} & = & -\dfrac{M_{pl}}{1.66\,M^2_{N_1}}
\dfrac{z\,\sqrt{g_{\star}(z)}}{g_{\rm s}(z)}
\,\langle \Gamma_{2} \rangle\,\left(Y_{N_2}-Y_{N_{2}}^{\rm eq} \right)
-\dfrac{2\,\pi^2}{45}\,\dfrac{M_{pl}\,M_{N_1}}{1.66}
\dfrac{\sqrt{g_{\star}(z)}}{z^2} \times \nn \\
&&\left({\langle\sigma {\rm v}\rangle}_{N_2,\,\zbl} +
{\langle\sigma {\rm v}\rangle}_{N_2,t,H_{\rm BL}}\right)\,
\left(Y_{N_2}^2-(Y_{N_{2}}^{\rm eq})^2\right), 
\label{beN2}\\
\frac{d Y_{B-L}}{d z} &=& -\dfrac{M_{pl}}{1.66\,M^2_{N_1}}
\dfrac{z\,\sqrt{g_{\star}(z)}}{g_{\rm s}(z)}
\left[\sum_{j = 1}^{2} \left( \frac{Y_{B-L}}{2}\, 
\frac{Y_{N_{j}^{eq}}}{Y_{L}^{\rm eq}} +
\varepsilon_{j} \left(Y_{N_j}-Y_{N_{j}^{eq}}\right)
\right) \langle\Gamma_{j}\rangle\right]\,,
\label{beYbl}
\end{eqnarray}  
where $Y_{X} = \dfrac{n_X}{\rm s}$ denotes the comoving number density
of $X$, with $n_X$ being the actual number density and
$z=\dfrac{M_{N_1}}{T}$. Planck mass is denoted by $M_{pl}$.
The quantity $g_{\star}(z)$ is a function of $g_\rho$ and $g_{\rm s}$,
the effective degrees of freedom
related to the energy and entropy densities of
the Universe respectively, and it has the following expression
\cite{Gondolo:1990dk},
\begin{eqnarray}
\sqrt{g_{\star}(z)} = \dfrac{g_{\rm s}(z)}
{\sqrt{g_{\rho}(z)}}\,\left(1 -\dfrac{1}{3}
\dfrac{{\rm d}\,{\rm ln}\,g_{\rm s}(z)}{{\rm d} \,{\rm ln} z}\right)\,. 
\label{gstar}
\end{eqnarray}
Before EWSB, the variation of $g_s(z)$ with respect to $z$ is
negligible compared to the first term within the brackets and
hence one can use $\sqrt{g_{\star}(z)} \simeq \dfrac{g_{\rm s}(z)}
{\sqrt{g_{\rho}(z)}}$. The equilibrium comoving number
density of $X$ ($X=N_i$,\,$L$), 
obeying the Maxwell Boltzmann distribution, is given by \cite{Gondolo:1990dk}
\begin{eqnarray}
Y^{\rm eq}_{X}(z) = \dfrac{45\,g_X}{4\,\pi^4}\,
\left(\dfrac{M_X\,z}{M_{N_1}}\right)^2
\,\dfrac{{\rm K}_2\left(\frac{M_{X}}{M_{N_1}}\,z\right)}
{g_s\left(\frac{M_{N_1}}{z}\right)}\,,
\label{yeq}
\end{eqnarray}
where $g_{X}$ and $M_{X}$ are the internal degrees of freedom and
mass of $X$ respectively while $g_s\left(\frac{M_{N_1}}{z}\right)$
is the effective degrees of freedom related to the entropy
density of the Universe at temperature $T=\dfrac{M_{N_1}}{z}$.
${\rm K}_2\left(\frac{M_{X}}{M_{N_1}}\,z\right)$ is the modified
Bessel function of order 2. The relevant Feynman diagrams
including both decay and annihilation
of $N_i$ are shown in Figs.\,\,\ref{feyn1-lepto} and \ref{feyn2-lepto}.
The expression of thermal averaged decay width $\langle {\Gamma_i} \rangle$,
which is related to total decay width $\Gamma_i$ of $N_i$ is given as
\begin{eqnarray}
\langle {\Gamma}_i \rangle = \Gamma_i\
\,\dfrac{{\rm K}_1\left(\frac{M_{N_i}}{M_{N_1}}\,z\right)}
{{\rm K}_2\left(\frac{M_{N_i}}{M_{N_1}}\,z\right)}\,.
\label{thavrdcay}
\end{eqnarray}

The thermally average annihilation cross sections ${\langle \sigma {\rm v}
\rangle}_{N_i,\,\zbl}$ and ${\langle \sigma {\rm v}
\rangle}_{N_i,\,\zbl}$, appearing in Boltzmann equations
(Eqs.~(\ref{beN1}) and (\ref{beN2})) for the processes shown in
Fig.\,\,\ref{feyn2-lepto}, can be defined in a generic form,  

\begin{eqnarray}
{\langle \sigma {\rm v} \rangle}_{N_i,\,x} =
\dfrac{z}{16\,M^4_{N_i}\,M_{N_1}\,g^2_{N_i}\,
{\rm K_2}\left(\frac{M_{N_i}}{M_{N_1}}\,z\right)^2}
\int_{4\,M^2_{N_i}}^{\infty} \hat{\sigma}_{N_i,\,x}
\,{\rm K_1}\left(\dfrac{\sqrt{s}}{M_{N_1}}\,z\right)\,\sqrt{s}\,ds \,,
\end{eqnarray}  
where the $\hat{\sigma}_{N_i,\,x}$ is related to the actual annihilation
cross section ${\sigma}_{N_i,\,x}$ by the following relation
\begin{eqnarray}
\hat{\sigma}_{N_i,\,x} = 2\,g^2_{N_i}\,\left(s-4\,M^2_{N_1}\right)\,
{\sigma}_{N_i,\,x}\,,
\label{sigmahat}
\end{eqnarray}
where $g_{N_i}=2$ is the internal degrees of freedom of RH neutrino
$N_i$. The expression of $\hat{\sigma}_{N_i,\,\zbl}$ and
$\hat{\sigma}_{N_i,\,t,\,H_{BL}}$ for the present model is
given in Ref. \cite{Iso:2010mv}. 
\begin{figure}[]
\centering
\includegraphics[angle=0,height=10cm,width=14cm]{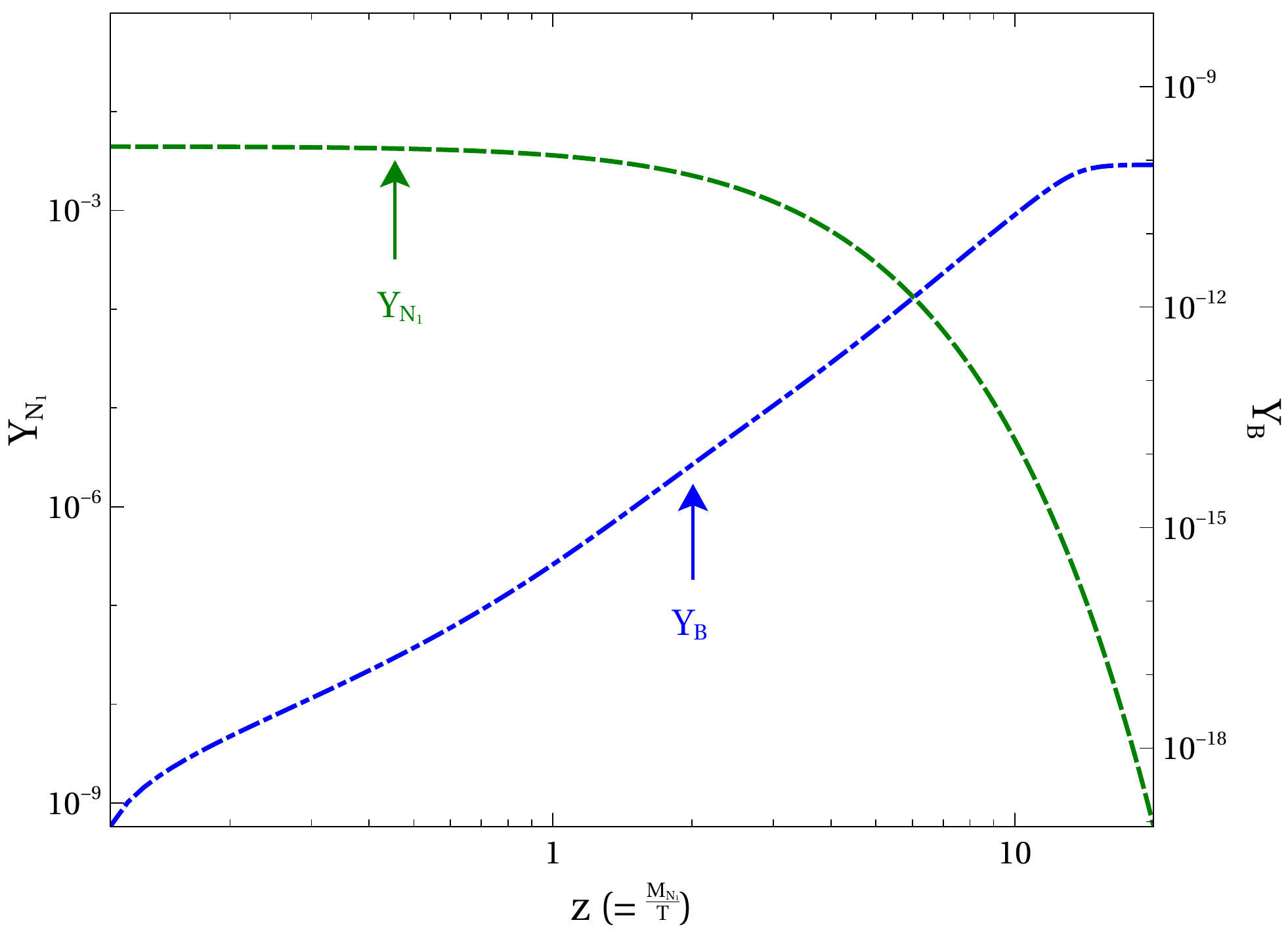}
\caption{Variation of $Y_{N_1}$ (Green dash line) and
$Y_{B-L}$ (blue dash-dot line) with $z$ where other parameters
have kept fixed at $M_{N_1} = 2000$ GeV, $\alpha_{BL} \left(=
\frac{g_{BL}^{2}}{4 \pi}\right) = 3 \times 10^{-4}$, 
$M_{Z_{BL}} = 3000$ GeV.}
\label{b-asym1}
\end{figure} 

To calculate the ${\rm B-L}$ asymmetry at around $T\simeq 150$ GeV,
we have to numerically solve the set of three coupled Boltzmann
equations (Eqs.~(\ref{beN1})-(\ref{beYbl}))
using Eqs.~(\ref{gstar})-(\ref{sigmahat}). However, we can reduce the
two flavour analysis (when both $N_2$ and $N_1$ are separately considered)
into one flavour case by considering the
parameters of $\mD$ matrix in such a way so that the decay widths
of $N_1$ and $N_2$ are of the same order i.e. $\Gamma_{1}\sim\Gamma_{2}$.
Hence, the CP asymmetry generated from the decays of both $N_1$ and $N_2$
are almost identical ($\varepsilon_1\sim\varepsilon_2$,
see Eq.~(\ref{epsa2})-(\ref{epsa1})). In this case, the net ${\rm B-L}$
asymmetry is equal to twice of that is being generated from
the CP violating interactions of the lightest RH neutrino $N_1$
\cite{Iso:2010mv}. Hence instead of solving three coupled differential
equations we now only need to solve Eqs.~(\ref{beN1}) and (\ref{beYbl}).
The results we have found by numerically solving
Eqs.~(\ref{beN1}) and (\ref{beYbl}) are plotted in Fig.\,\,\ref{b-asym1}.
In this plot, we have shown the variation of $Y_{N_1}$ and $Y_{B-L}$
with $z$ for $M_{N_1}=2000$ GeV, $\alpha_{BL}=3\times 10^{-4}$
and $\mzbl=3000$ GeV \footnote{The considered value
of $\mzbl$ and the corresponding gauge coupling $\gbl$
satisfy the upper bounds obtained from LEP
\cite{Carena:2004xs, Cacciapaglia:2006pk} and
more recently from LHC \cite{Guo:2015lxa} as well.}.
While solving the coupled Boltzmann equations
we have considered the following initial conditions: $Y_{N_1}(T_{\rm in}^{B})
=Y^{\rm eq}_{N_1}$ and $Y_{B-L}=0$ with $T_{\rm in}^B$ is the initial
temperature which we have taken as 20 TeV. Thereafter, the evolutions
of $Y_{N_1}$ and $Y_{B-L}$ are governed by their respective
Boltzmann equations. From Fig.\,\,\ref{b-asym1}, one can notice
that initially upto $z\sim 1$ ($T\sim M_{N_1}$), the comoving
number density of $Y_{N_1}$ does not change much as a result of 
the ${\rm B-L}$ asymmetry produced from the decay, and the annihilation of
$N_1$ is also less. However, as the temperature of the Universe drops
below the mass of $M_{N_1}$, there is a rapid change in the
number density of $N_1$, which changes around six orders of
magnitude between $z=1$ and $z=20$. Consequently, the large change in
$Y_{N_1}$ significantly enhances the ${\rm B-L}$ asymmetry $Y_{B-L}$
and finally $Y_{B-L}$ saturates to the desired value
around $\sim 10^{-10}$, when there are practically no
$N_1$ left to produce any further ${\rm B-L}$ asymmetry.  

The produced $B-L$ asymmetry is converted to net baryon
asymmetry of the Universe through the sphaleron transitions  
while they are in equilibrium with the thermal bath.
The quantities $Y_{B-L}$ and $Y_{B}$ are
related by the following equation \cite{Khlebnikov:1988sr}
\begin{eqnarray}
Y_{B} = -\,2\times\dfrac{28}{79}\,Y_{B-L} (T_{\rm f})\,,
\end{eqnarray}
where $T_{\rm f}\simeq 150$ GeV is the temperature of the Universe
upto which the sphaleron process, converting $B-L$
asymmetry to a net $B$ asymmetry, maintains its thermal equilibrium.
The extra factor of two in the above equation
is due to the equal contribution to $Y_{B-L}$ arising
from the CP violating interactions of $N_2$ as well.  
Finally, we calculate the net baryon asymmetry $Y_{B}$ for
three different masses of RH neutrino $N_1$ and CP
asymmetry parameter $\varepsilon_1$. The results are
listed in Table \ref{baryon-table}. In all three
cases, the final baryon asymmetry lies within the
experimentally observed range for $Y_B$
i.e. $(8.239-9.375)\times 10^{-11}$ at 95\% C.L. \cite{Olive:2016xmw}.  

\begin{table}[h!]
\begin{center}
\vskip 0.5cm
\begin{tabular} {||c|c|c||}
\hline
\hline
$M_{N_1}$ [GeV] & $\varepsilon_{1}$ & $Y_{B} = \frac{n_{B}}{S}$\\
\hline
$1600$ & $4.4 \times 10^{-4}$ & $8.7121 \times 10^{-11}$\\
\hline
$1800$ & $2.25 \times 10^{-4}$ & $8.7533 \times 10^{-11}$\\
\hline
$2000$ & $1.8 \times 10^{-4}$ & $8.5969 \times 10^{-11}$\\
\hline
\hline
\end{tabular}
\end{center}
\caption{Baryon asymmetry of the Universe generated for
three different values of $M_{N_1}$ and $\varepsilon_1$.}
\label{baryon-table}
\end{table}
\subsection{FIMP Dark Matter}
\label{fimp-DM}
In the present section we explore the FIMP scenario for dark matter 
in the Universe, by considering the complex scalar field $\dm$ as a
corresponding candidate. As described in the
Section \ref{model}, the residual $\mathbb{Z}_2$ symmetry of $\dm$
makes the scalar field absolutely stable over the cosmological time scale 
and hence can play the role of a dark matter candidate. Since
$\dm$ has a nonzero ${\rm B-L}$ charge $\nbl$, therefore DM talks
to the SM as well as the BSM particles through the exchange of
extra neutral gauge boson $\zbl$ and two Higgs bosons present
in the model, one is the SM-like Higgs $h_{1}$ while
other one is the BSM Higgs $h_{2}$.
The corresponding coupling strengths, in terms of
gauge coupling $\gbl$, ${\rm B-L}$ charge $\nbl$,
mixing angle $\alpha$ and $\lambda$s, are listed in Table\,\,\ref{tab3.1}.
As the FIMP never enters into thermal equilibrium, these couplings
have to be extremely feeble in order to make the corresponding
interactions nonthermal. For the case of $\phi_{DM}\,\phi^\dagger_{DM}
\,{\zbl}_{\mu}$ coupling, we will make the ${\rm B-L}$ charge
of $\dm$ extremely tiny so that this interaction enters into
the nonthermal regime. In principle, one can also choose the gauge
coupling $\gbl$ to be very small, however in the present case
we will keep the values of $\gbl$ and $\mzbl$ fixed at 0.07
and 3 TeV respectively as these values reproduce the observed
baryon asymmetry of the Universe (see Section \ref{baryogenesis}). 
Also, there is another advantage of choosing tiny $\nbl$ as this
will make only $\dm$ out of equilibrium while keeping $\zbl$
in equilibrium with the thermal bath. Moreover, due to the
nonthermal nature, the initial number density of FIMP is
assumed to be negligible and as the temperature
of the Universe begins to fall down, they start to be produced dominantly
from the decays and annihilation of other heavy particles.
\begin{table}[h!]
\begin{center}
\vskip 0.5cm
\begin{tabular} {||c||c||}
\hline
\hline
Vertex & Vertex Factor\\
$a\,b\,c$ & $g_{abc}$\\
\hline
$\phi_{DM}\,\phi^\dagger_{DM}\,{\zbl}_{\mu}$
& $\gbl\,\nbl (p_2-p_1)^\mu$\\
\hline
$\phi_{DM}\,\phi^\dagger_{DM}\,h_1$ & $-\,(\lambda_{Dh} v
\cos \alpha + \lambda_{DH} v_{BL} \sin \alpha)$\\
\hline
$\phi_{DM}\,\phi^\dagger_{DM}\,h_2$ & $\,(\lambda_{Dh} v
\sin \alpha - \lambda_{DH} v_{BL} \cos \alpha)$ \\
\hline
\hline
\end{tabular}
\end{center}
\caption{Couplings of FIMP ($\dm$) with $\zbl$, $h_1$ and $h_2$.}
\label{tab3.1}
\end{table}

\begin{figure}[h!]
\centering
\includegraphics[angle=0,height=8cm,width=14cm]{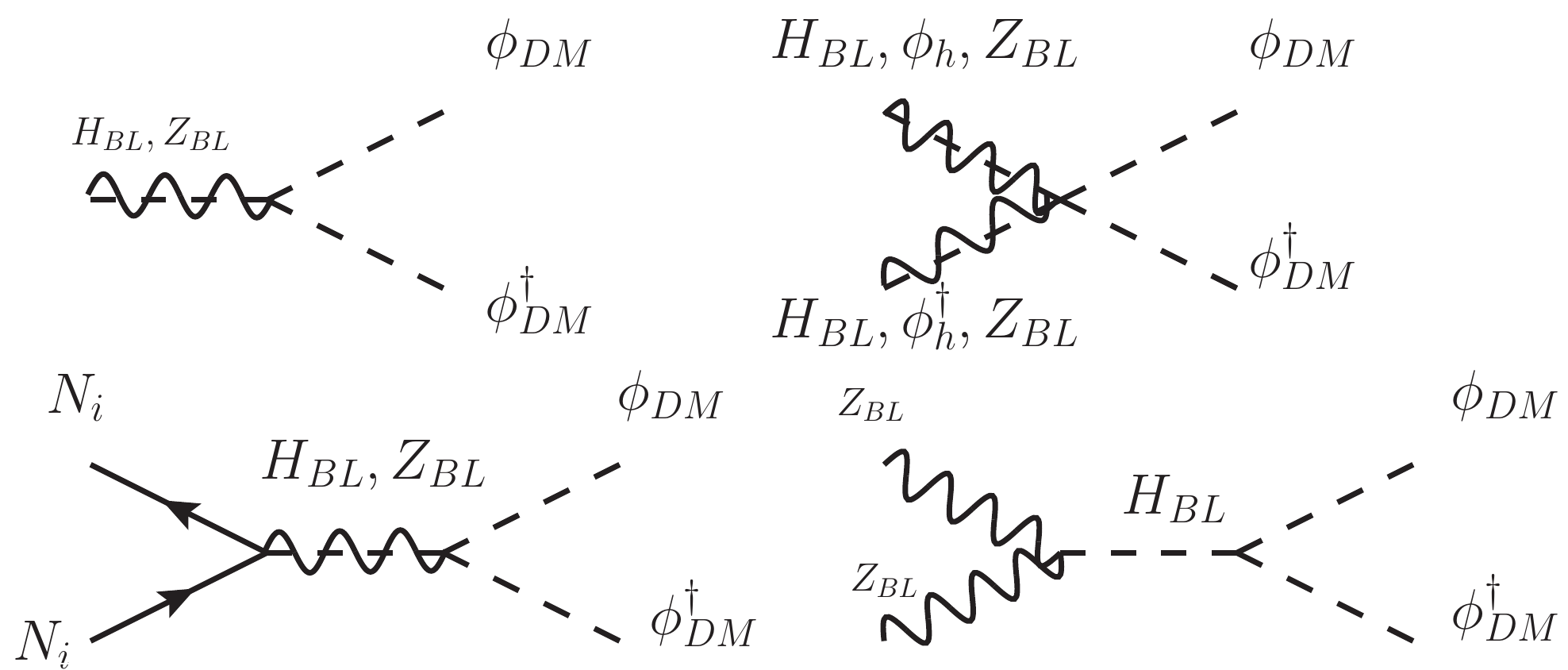}
\caption{Feynman diagrams for the all possible production modes of
$\dm$ before EWSB.}
\label{bfewsb}
\end{figure}

In the present scenario, we have considered all the particles except
$\dm$ to be in thermal equilibrium. Before EWSB, all the SM particles
are massless\footnote{Although the SM particles acquire thermal
masses before EWSB, we have neglected these masses, as in this regime
this approximation will not affect the DM production
processes significantly.}. In this regime,
production of $\dm$ occurs mainly from the
decay and/or annihilation of BSM particles namely $\zbl$, $H_{BL}$,
and $N_i$. Also, before EWSB the annihilation of all four
degrees of freedom of SM Higgs doublet $\phi_h$ can produce $\dm$.
Feynman diagrams for all the production processes of $\dm$
before EWSB are shown in Fig.\,\ref{bfewsb}.

\begin{figure}[h!]
\centering
\includegraphics[angle=0,height=12cm,width=16cm]{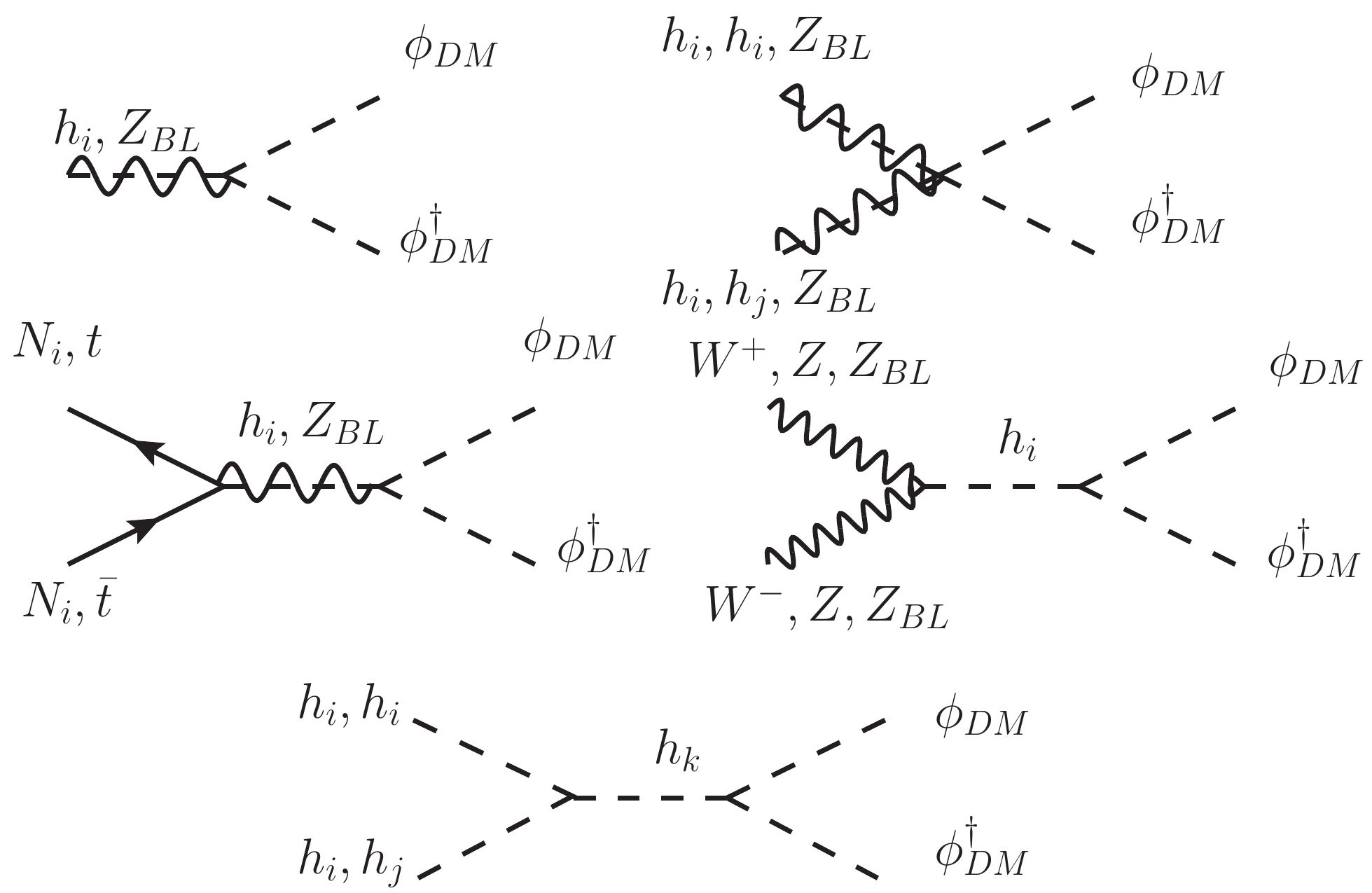}
\caption{Production processes of $\dm$ from both SM as well as
BSM particles after EWSB.}
\label{afewsb}
\end{figure}

After EWSB, all the SM particles become massive and consequently
besides the BSM particles, $\dm$ can now also be produced from
the decay and/or annihilation of the SM particles as well. The
corresponding Feynman diagrams are shown in Fig.\,\ref{afewsb}.
In generating the vertex factors for different vertices to compute
the Feynman diagrams as listed in Fig.\,\ref{bfewsb} and Fig.\,\ref{afewsb}
we have used the LanHEP \cite{Semenov:2010qt} package.  

In order to compute the relic density of a species at the present epoch,
one needs to study the evolution of the number density of the corresponding
species with respect to the temperature of the Universe. The evolution
of the number density of $\dm$ is governed by the Boltzmann equation containing
all possible number changing interactions of $\dm$. The Boltzmann
equation of $\dm$ in terms of its comoving number density
$Y_{\dm}=\dfrac{n_{\dm}}{\rm s}$, where $n$ and ${\rm s}$
are actual number density and entropy density of the Universe 
is given by
\begin{eqnarray}
&&\dfrac{dY_{\phi_{DM}}}{dz} = \dfrac{2 M_{pl}}{1.66 M_{h_1}^{2}}
\dfrac{z \sqrt{g_{\star}(z)}}{g_{\rm s}(z)}\,\,\Bigg[\sum_{X = \zbl,\,h_1,\,h_2}
\langle\Gamma_{X\rightarrow  \phi_{DM} \phi^\dagger_{DM}} \rangle
(Y_{X}^{\rm eq} - Y_{\phi_{DM}})\Bigg]\nn \\
&&~~~~~~~~~
+\dfrac{4 \pi^{2}}{45} \dfrac{M_{pl} M_{h_1}}{1.66}
\dfrac{\sqrt{g_{\star}(z)}}
{z^{2}}\,\,\Bigg[\sum_{p}
\langle\sigma {\rm v}_{p\bar{p}\rightarrow \phi_{DM}
\phi^\dagger_{DM}}\rangle (Y_{p}^{\rm eq\,\,2} - Y_{\phi_{DM}}^{2})\nn \\
&&~~~~~~~~~+
\langle\sigma {\rm v}_{h_{1} h_{2}\rightarrow \phi_{DM}
\phi^\dagger_{DM}}\rangle (Y_{h_1}^{\rm eq} Y_{h_2}^{\rm eq} -
Y_{\phi_{DM}}^{2})\Bigg]\,, 
\label{BE}
\end{eqnarray}
where $z=\dfrac{M_{h_1}}{T}$, while $\sqrt{g_{\star}(z)}$, $g_{\rm s}(z)$
and $M_{pl}$ are same as those in Eqs.~(\ref{beN1})--(\ref{beYbl})
of Section \ref{baryogenesis}. 
In the above equation (Eq.~(\ref{BE})), first term represents
the contribution coming from the decays of $\zbl$, $h_1$ and
$h_2$. The expressions of equilibrium number density $Y^{\rm eq}_{X}(z)$
($X$ is any SM or BSM particle expect $\dm$) and the thermal
averaged decay width $\langle \Gamma_{X \rightarrow \dm \dm}\rangle$
can be obtained from Eqs.~(\ref{yeq}) and (\ref{thavrdcay}), respectively
by only replacing $M_{N_1}$ with $M_{X}$, the mass of decaying
mother particle. As mentioned above, before EWSB, the
summation in the first terms is over $h_2$ and $\zbl$ only,
as there will be no contribution from the SM Higgs decay as
such trilinear vertex ($h_1\dm\dm^{\dagger}$) is absent
before EWSB and after EWSB there will be contributions
to the relic density of $\dm$ from all there decaying particles.
The dark matter production from the pair annihilations of
SM and BSM particles are described by the second term
of the Boltzmann equation. Here, summation over $p$
includes all possible pair annihilation channels namely
$W^+W^-,\,ZZ,\,\zbl\zbl,\,N_iN_i,\,h_ih_i,\,t\bar{t}$.
However before EWSB, pair annihilations of the BSM particles
and SM Higgs doublet $\phi_h$ contribute to the production processes
(i.e. $p=\zbl,\,N_i,\,H_{BL},\,\phi_h$, see Fig.\,\,\ref{bfewsb}).
The third term,
which is present only after EWSB, is another the production
mode of $\dm$ from the annihilation of $h_1$ and $h_2$.
The expressions of all the relevant cross sections and decay widths
for computing the DM number density are given
in Appendix \ref{App:AppendixA}.
The most general form of thermally averaged annihilation cross
section for two different annihilating particles of
mass $M_A$ and $M_B$ is given by \cite{Biswas:2016yjr}, 
\begin{eqnarray}
f_{1} &=& \sqrt{s^{2} + (M_{A}^{2} - M_{B}^{2})^{2}
- 2\,s\,(M_{A}^{2} + M_{B}^{2})}\,,\nn \\
f_{2} &=& \sqrt{s - (M_{A} - M_{B})^{2}} \,\,\sqrt{s
- (M_{A} + M_{B})^{2}}\,\,, \nn \\
\langle\sigma v_{A\,B \rightarrow \phi_{DM}
\phi_{DM}}\rangle &=& \dfrac{1}{8 \,M_{A}^{2}\, M_{B}^{2}\,T\,
{\rm K}_{2}\left(\dfrac{M_{A}}{T}\right)
\,{\rm K}_{2}\left(\dfrac{M_{B}}{T}\right)}
\times \nonumber \\
&& \int_{(M_{A} + M_{B})^{2}}^{\infty}
\dfrac{\sigma_{A\,B\rightarrow} \phi_{DM}
\phi_{DM}}{\sqrt{s}}\,f_{1}\,f_{2}\,{\rm K}_{1}
\left(\dfrac{\sqrt{s}}{T}\right)\,{\rm d}s \,. 
\label{avg_th}
\end{eqnarray} 

Finally, the relic density of $\dm$ is obtained using
the following relation between $\Omega h^2$ and $Y_{\dm}(0)$
\cite{Edsjo:1997bg, Biswas:2011td},
\begin{eqnarray}
\Omega h^2 = 2.755 \times 10^8 \left(\dfrac{\mdm}
{\rm GeV}\right)\,Y_{\dm}(0)\,,
\label{omegadm}
\end{eqnarray}
where $Y_{\dm}(0)$ is the value of comoving number density at
the present epoch, which can be obtained by solving the Boltzmann
equation. 

The contribution to dark matter production processes from decays
as well as annihilations of various SM and BSM particles
depend on the mass of $\dm$. Accordingly, We have divided
our rest of the dark matter analysis into four different
regions depending on $M_{DM}$ and the dominant production
modes of $\dm$.
\subsubsection{$M_{DM} < \dfrac{M_{h_1}}{2}$,
\,$\dfrac{M_{h_2}}{2}$,\,$\dfrac{M_{Z_{BL}}}{2}$,
{\bf SM and BSM particles decay dominated region.}}
\label{region1}
\begin{figure}[h!]
\centering
\includegraphics[angle=0,height=7cm,width=8cm]{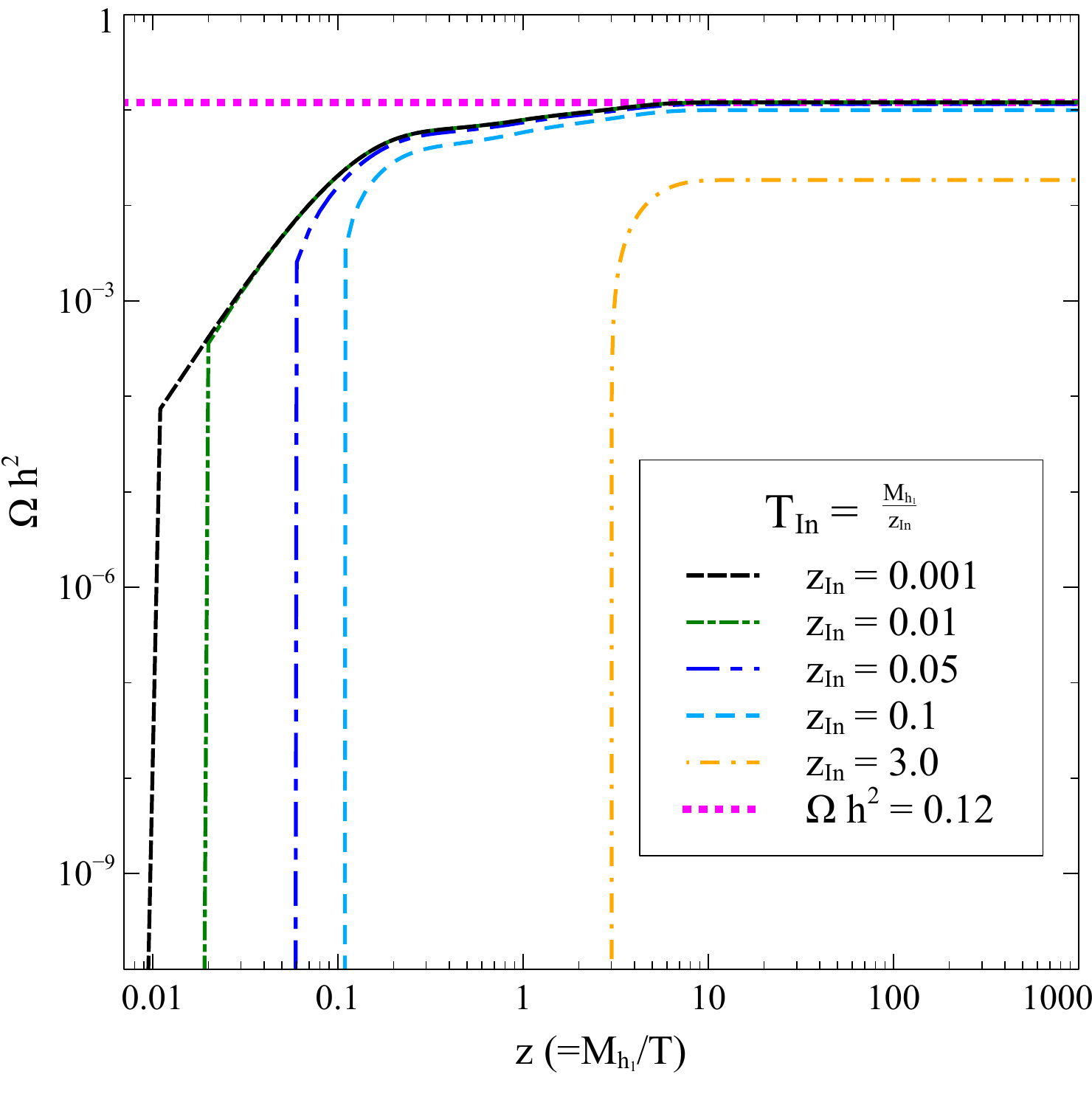}
\includegraphics[angle=0,height=7cm,width=8cm]{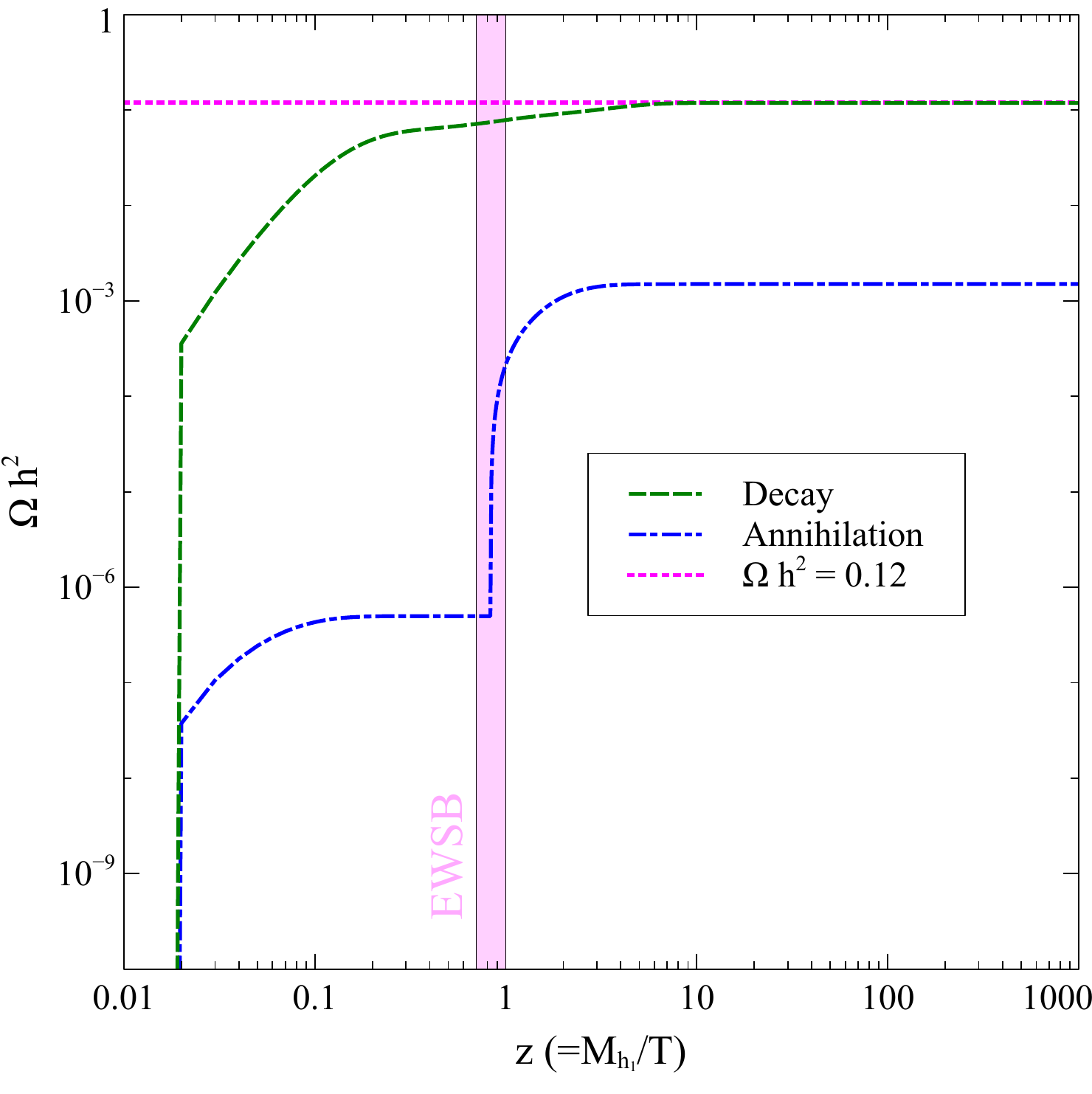}
\caption{Left (Right) panel: Variation of relic density $\Omega h^{2}$
with $z$ for different initial temperature
(Contributions to $\Omega h^2$ coming from decay and annihilation),
where other parameters are fixed at $\lambda_{Dh} = 8.75\times10^{-13}$,
$\lambda_{DH} = 5.88\times10^{-14}$, $n_{BL} = 1.33 \times 10^{-10}$,
$M_{DM}$ = 50 GeV, $\mzbl$ = 3000 GeV, $\gbl$ = 0.07,
$M_{h_{1}} = 125.5$ GeV
and $M_{h_{2}}$ = 500 GeV, $\alpha = 10^{-4}$.}
\label{s1a}
\end{figure}

In this case DM is dominantly produced from the decays of all three 
particles namely $h_1$, $h_2$ and $Z_{BL}$. Therefore, in this
case $\ubl$ part of the present model directly enters
into the dark matter production. Moreover in this mass range,
$\dm$ can also be produced from the annihilations of SM and
BSM particles, however, we find that their contributions
are not as significant as those from the decays of $h_1,\,h_2$ and $\zbl$.
In the left panel (LP) and right panel (RP) of Fig.\,\ref{s1a},
we have shown the variation of DM relic density with $z$.
In LP, we have shown the dependence of DM relic density with
the initial temperature $T_{in}$. Initial temperature ($T_{in}$) is
the temperature upto which we have assumed that the number
density of DM is zero and its production processes
start thereafter. We can clearly see from the
figure that as long as the initial temperature
is above the mass of BSM Higgs ($M_{h_2} \sim 500$ GeV),
the final relic density does not depend on the choice of the initial
temperature and reproduces the observed DM relic density of
the Universe for the chosen values of model parameters as written
in the caption of Fig.\,\ref{s1a}.
If we reduce the initial temperature
from $500$ GeV, i.e. for $T_{in} = 251$ GeV, the decay contribution
of BSM Higgs $h_{2}$ becomes less since corresponding the number
density of $h_2$ for $T_{in}<M_{h_2}$ is Boltzmann suppressed
(exponentially suppressed), which is clearly shown by the blue dashed-dotted line.
Hence, if we reduce the initial temperature ($T_{in}$)
further i.e. $T_{in}<M_{h_2}, M_{h_1}$ $\sim 42$ GeV then
the number densities of both SM-like Higgs $h_1$ as well
as BSM Higgs $h_{2}$ become Boltzmann suppressed and
hence, less amount of DM production will take place
which is evident from the LP of Fig.\,\ref{s1a}
(represented by the yellow dashed-dot line). On the other hand
in the RP of Fig.\,\ref{s1a}, we have shown the contributions
to DM relic density coming from decay and annihilation. Magenta
dotted horizontal line represents the present day observed
DM relic density of the Universe. Green dashed line represents
the total decay contribution arising from the decays of both
$h_1$, $h_2$ and $\zbl$ whereas the net annihilation contribution coming
from the annihilation of all the SM as well as BSM particles
has been shown by the blue dashed-dotted line.
There is a sudden rise in the annihilation contribution which occurs
around the Universe temperature $T \sim 154$ GeV (i.e. EWSB temperature).
After the EWSB temperature, all the SM particles become massive
and hence the sudden rise in the annihilation part because of
the appearance of the following annihilation channels
$W^{+}\,W^{-}$, $Z\,Z$, $h_{1}\,h_{1}$, $h_{1}\,h_{2}$.
The plot clearly implies that the lion share of the contribution
comes from the decay of both Higgses $h_1$, $h_2$ and $Z_{BL}$,
while for the considered values of model parameters
the annihilation contribution is subdominant. Moreover,
in this case we cannot enhance the annihilation contribution
by increasing parameters $\ldh$, $\ldH$ and $\nbl$ as
these changes will result in the over production of dark matter
from the decays of $h_1$, $h_2$ and $\zbl$.            
\begin{figure}[]
\centering
\includegraphics[angle=0,height=7cm,width=8cm]{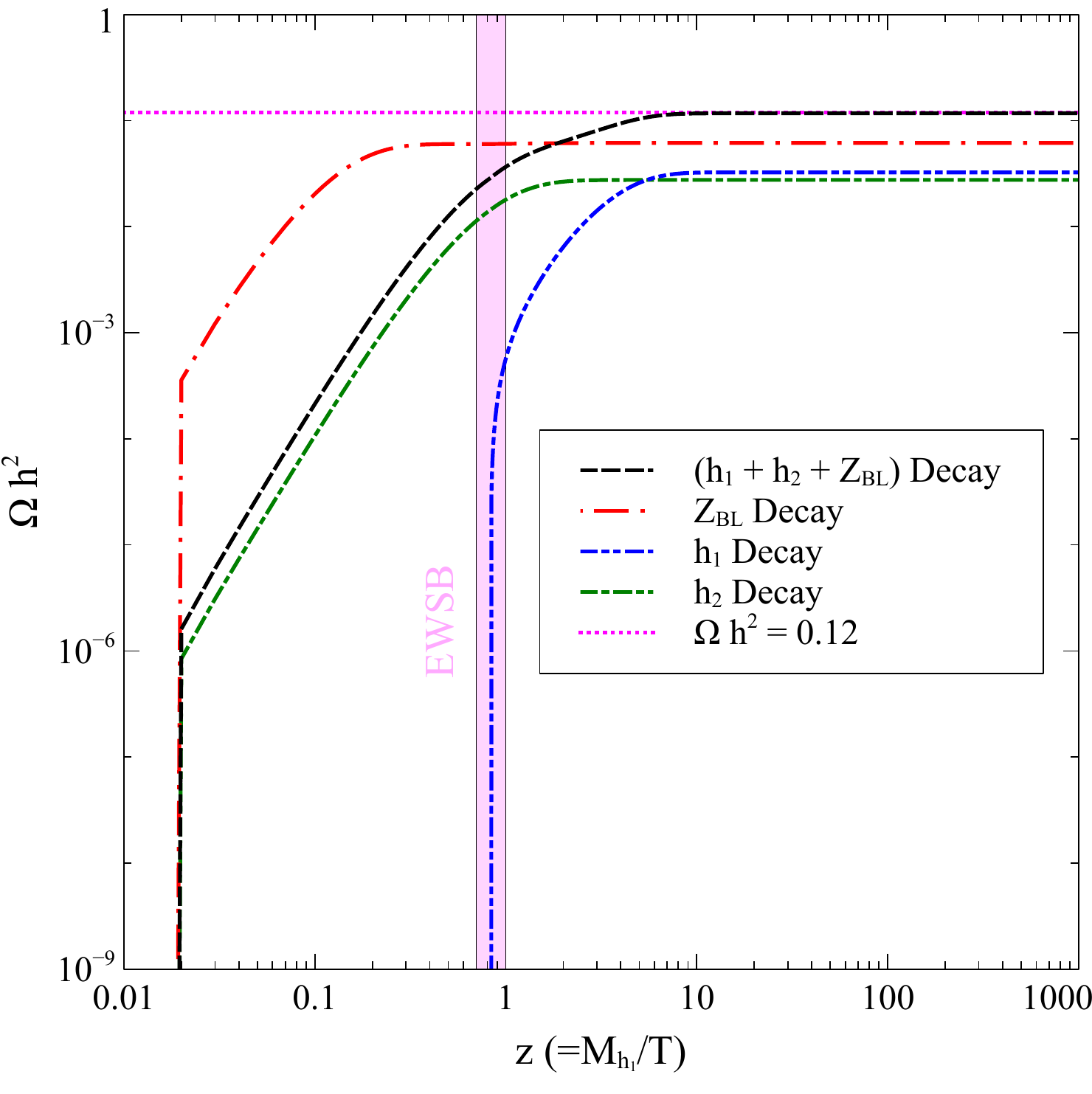}
\includegraphics[angle=0,height=7cm,width=8cm]{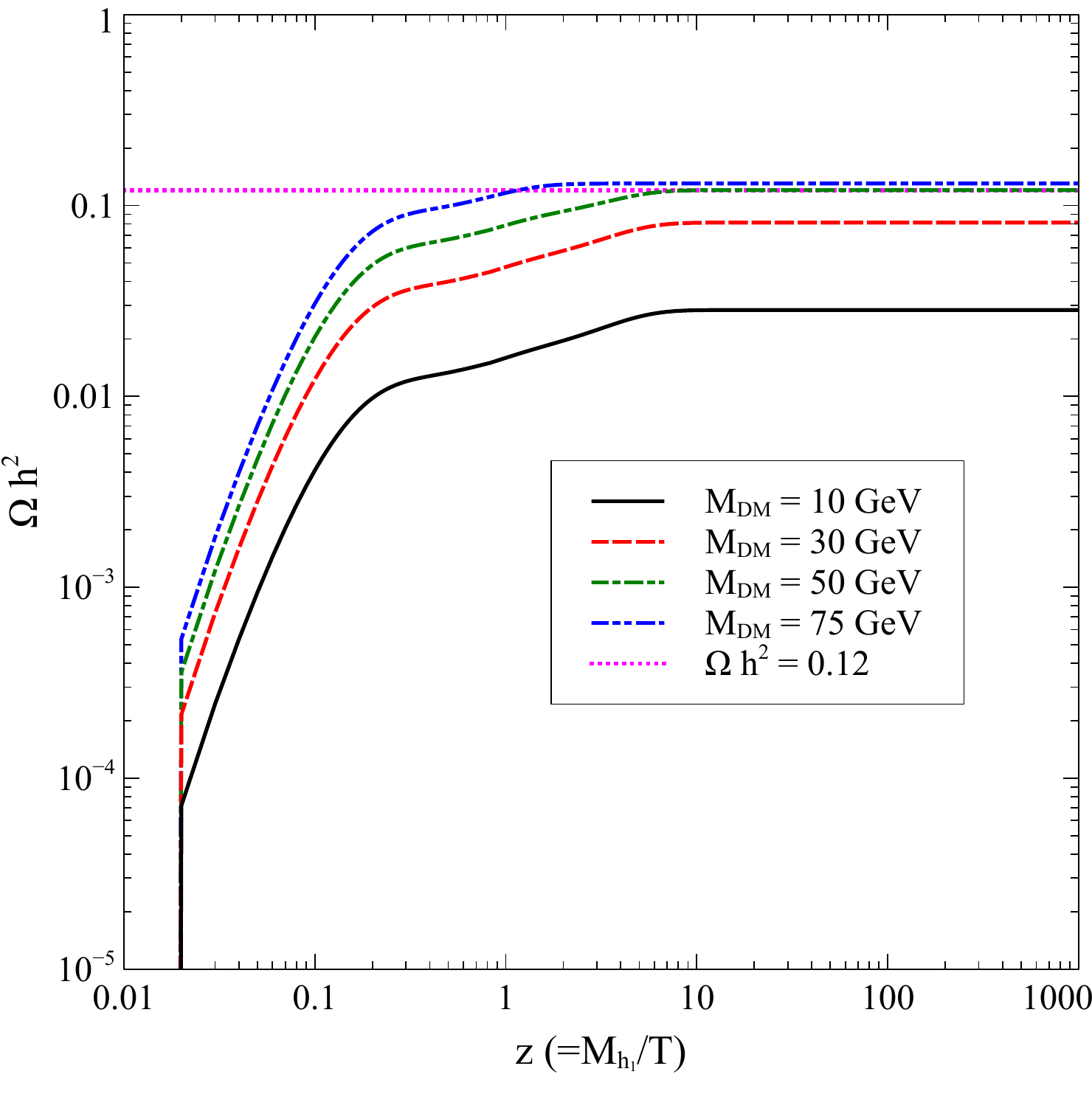}
\caption{Left panel: Showing variation of decay contributions
of both the Higgs bosons to $\Omega h^2$ separately
with $z$. Right panel: Variation of relic density
$\Omega h^{2}$ with $z$ for different values of DM mass
$M_{DM}$. Other parameters value have been kept fixed
at $\lambda_{Dh} = 8.75\times10^{-13}$, $\lambda_{DH} =
5.88\times10^{-14}$, $n_{BL} = 1.33 \times 10^{-10}$, $M_{DM}$ = 50 GeV (for LP),
$\mzbl$ = 3000 GeV, $\gbl$ = 0.07, $M_{h_{1}} = 125.5$
GeV and $M_{h_{2}}$ = 500 GeV, $\alpha = 10^{-4}$.}
\label{s2a}
\end{figure}

In the LP of Fig.\,\ref{s2a}, we have shown how 
the individual decay contribution from each scalar
varies with $z$. Here we consider the following values
of the scalar quartic couplings $\lambda_{Dh} = 8.75
\times 10^{-13}$ and $\lambda_{DH} = 5.88 \times 10^{-14}$
and the ($B-L$) charge of $\phi_{DM}$ $n_{BL} = 1.33 \times 10^{-10}$. 
From this plot we can see that before EWSB 
SM-like Higgs $h_1$ cannot decay to a pair of $\dm$
as in this epoch it has no coupling with the latter.
In this regime the decay of BSM Higgs $h_2$ and $Z_{BL}$ contribute, 
while after EWSB even the SM-like Higgs starts contributing
to the DM production and hence we get an increased
relic density (right side of EWSB). Its worth 
mentioning here that while generating the plot in the
LP of Fig.\,\,\ref{s2a}, we have taken the scalar quartic
couplings $\lambda_{Dh}$, $\lambda_{DH}$ and ${\rm B-L}$
charge of $\phi_{DM}$ $n_{BL}$ of different
strengths such that the contributions of both
the scalars ($h_1$ and $h_2$) and the extra gauge
boson to the DM relic density are of equal order.
This is because for the case of BSM Higgs $h_{2}$ decay 
the coupling $\lambda_{DH}$ multiplied by the ${\rm B-L}$
symmetry breaking VEV $v_{BL}$ is relevant,
while for the decay of the SM-like
Higgs $h_{1}$, the product of the parameter $\lambda_{Dh}$ and 
the EWSB VEV $v$ is relevant and the contribution from the 
decay of $Z_{BL}$, DM charge
$n_{BL}$ is relevant. 
Since in the present case $v_{BL}>v$, 
the magnitudes of the two quartic couplings $\lambda_{Dh}$
and $\lambda_{DH}$ are of different  order (see Table\,\ref{tab3.1}).
On the other hand, in the RP of Fig.\,\,\ref{s2a}, we have shown
the variation of the relic density with $z$ for four different
values of the DM mass $M_{DM}$. From Eq.~(\ref{omegadm}), one can
see that the DM relic density is directly proportional to
the mass $M_{DM}$ and as a result when other relevant couplings
remain unchanged $\Omega h^2$ increases with $\mdm$.
This feature is clearly visible in the RP for
the cases with $M_{DM}=10$ GeV (black solid line),
$M_{DM}=30$ GeV (red dashed line) and
50 GeV (green dashed line) respectively. However for
$M_{DM} = 75$ GeV (blue dashed dot line) $\Omega h^2$ does not rise
equally because for this value of DM mass the decay of $h_1$
to a pair of $\dm$ and $\dm^{\dagger}$ becomes kinematically
forbidden and hence, there is no equal increment in this case. 

\begin{figure}[]
\centering
\includegraphics[angle=0,height=7cm,width=8cm]{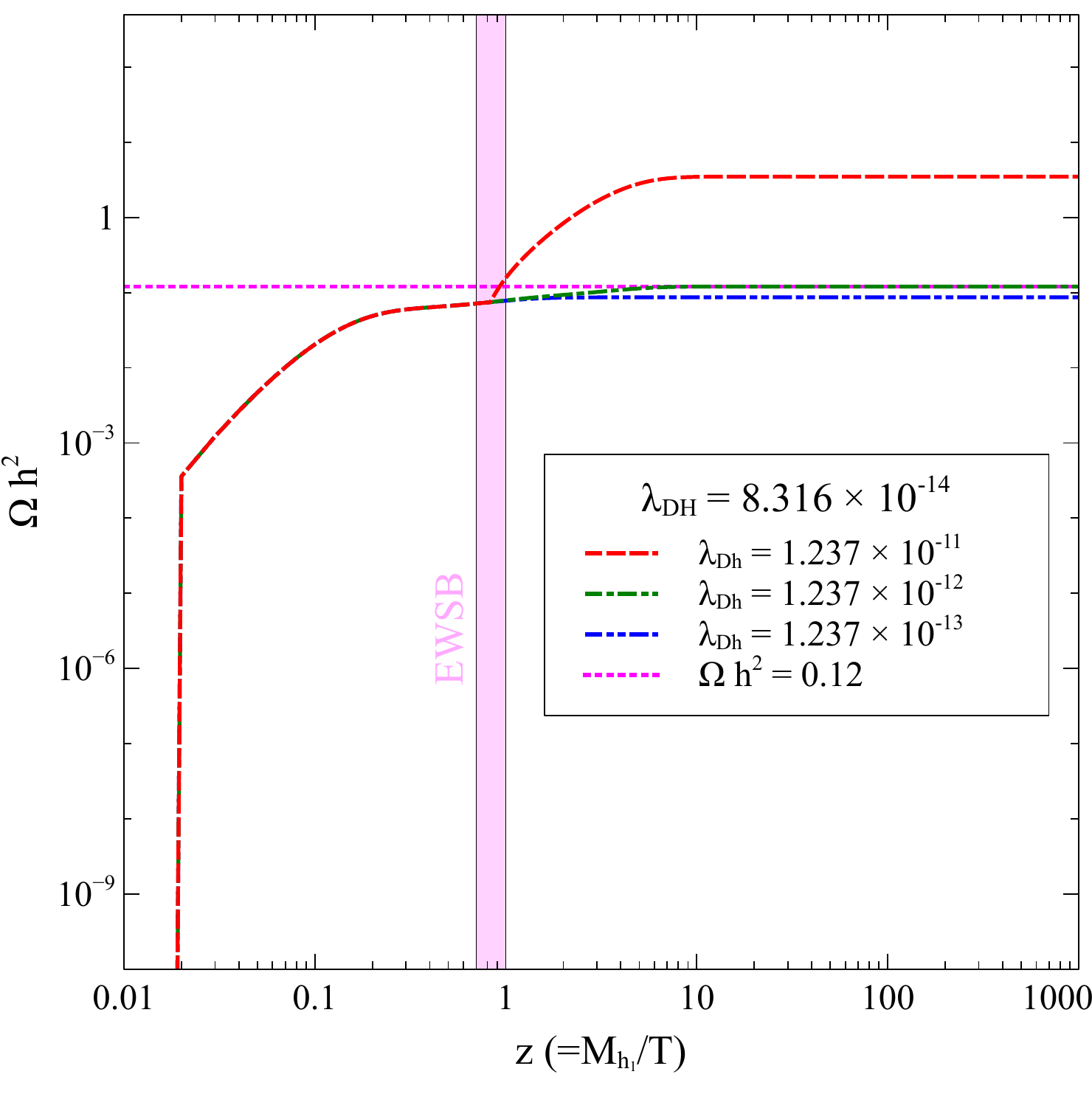}
\includegraphics[angle=0,height=7cm,width=8cm]{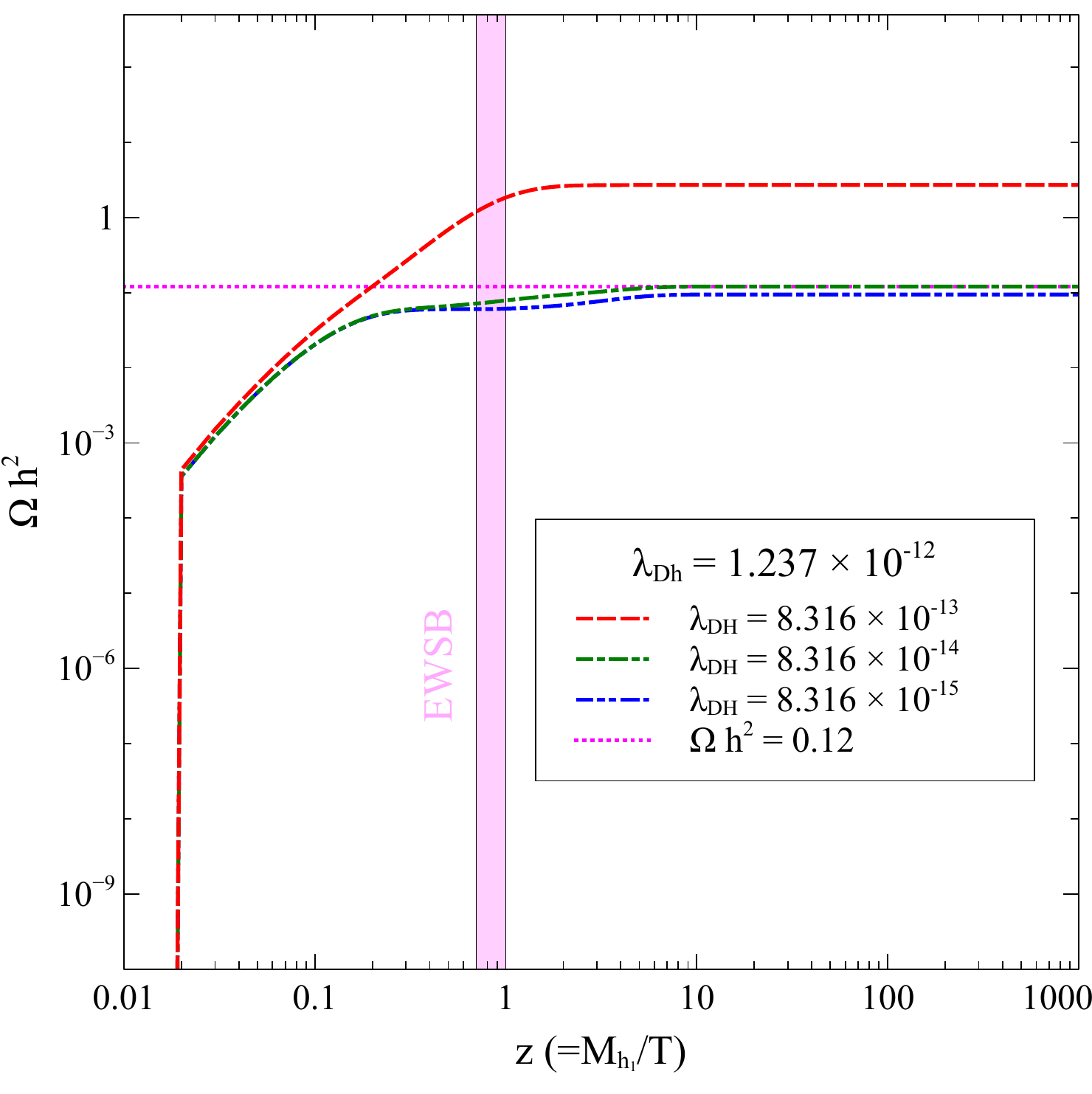}
\caption{Left (Right) panel: Variation of relic density $\Omega h^{2}$
with $z$ for three different values of $\lambda_{Dh}$ ($\lambda_{DH}$),
where other parameters are fixed at $\lambda_{DH} = 5.88\times10^{-14}$
($\lambda_{Dh} = 8.75\times10^{-13}$), $n_{BL} = 1.33 \times 10^{-10}$, 
$M_{DM}$ = 50 GeV, $\mzbl$ = 3000 GeV, $\gbl$ = 0.07,
$M_{h_{1}} = 125.5$ GeV and $M_{h_{2}}$ = 500 GeV, $\alpha = 10^{-4}$.}
\label{s3a}
\end{figure}

In LP and RP of Fig.\,\ref{s3a}, we have shown how the relic density
varies with $z$ for different values of scalar quartic couplings
$\lambda_{Dh}$ and $\lambda_{DH}$, respectively. In each panel, one
can easily notice that there exists a kink around the EWSB region.
However in the LP, the kink occurs for a higher value of $\lambda_{Dh}$
while in the RP, the situation is just opposite. We have already seen
in the LP of Fig.\,\ref{s2a} that before EWSB only $h_{2}$ decay is
contributing to DM relic density and at the EWSB region SM-like Higgs
$h_1$ also starts contributing. A kink will always appear in the
relic density curve when contribution of the SM-like Higgs
boson $h_1$ to $\Omega h^2$ is larger compared to that of
the BSM Higgs $h_2$ and extra gauge boson $Z_{BL}$ i.e.
$\Gamma_{h_1\rightarrow \dm \dm^{\dagger}}>
\Gamma_{h_2\rightarrow \dm \dm^{\dagger}}$,
$\Gamma_{Z_{BL}\rightarrow \dm \dm^{\dagger}}$. 
The values of scalar quartic couplings $\ldh$ and
$\ldH$ in the LP of Fig.\,\,\ref{s2a}
are such that $\Gamma_{h_2\rightarrow \dm \dm^{\dagger}}$
and $\Gamma_{Z_{BL}\rightarrow \dm \dm^{\dagger}}$  always remain
large compared to $\Gamma_{h_1\rightarrow \dm \dm}$ and hence
no kink is observed in the total relic density curve.
However, in the present figure (in the left panel of
Fig.\,\,\ref{s3a}) we do have kinks around the
EWSB region, because in the LP with $\ldH=8.316\times10^{-14}$
and $n_{BL} = 1.33 \times 10^{-10}$,
$\Gamma_{h_1\rightarrow \dm \dm^{\dagger}}>
\Gamma_{h_2\rightarrow \dm \dm^{\dagger}}$, 
$\Gamma_{Z_{BL}\rightarrow \dm \dm^{\dagger}}$
condition is satisfied only for the case with larger
value of $\ldh=1.237\times10^{-11}$ ($\ldh>>\ldH$)
while in the RP with a fixed value of $\ldh=1.237\times10^{-12}$,
the above condition is not maintained because $Z_{BL}$ decay channel
dominates.
       
\begin{figure}[]
\centering
\includegraphics[angle=0,height=7cm,width=8cm]{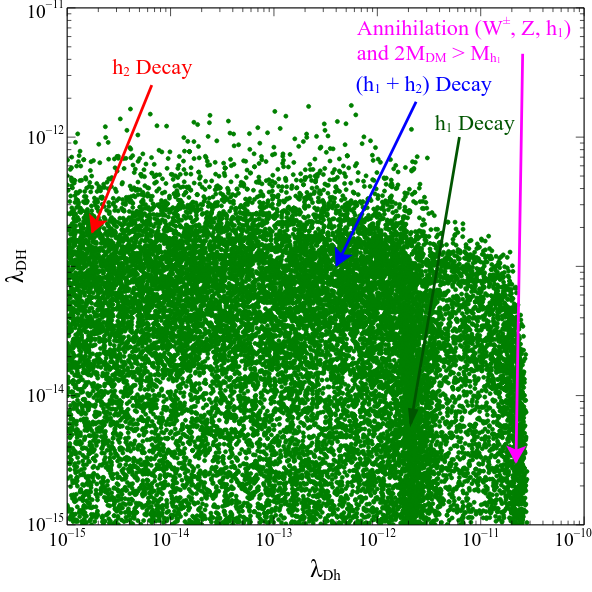}
\includegraphics[angle=0,height=7cm,width=8cm]{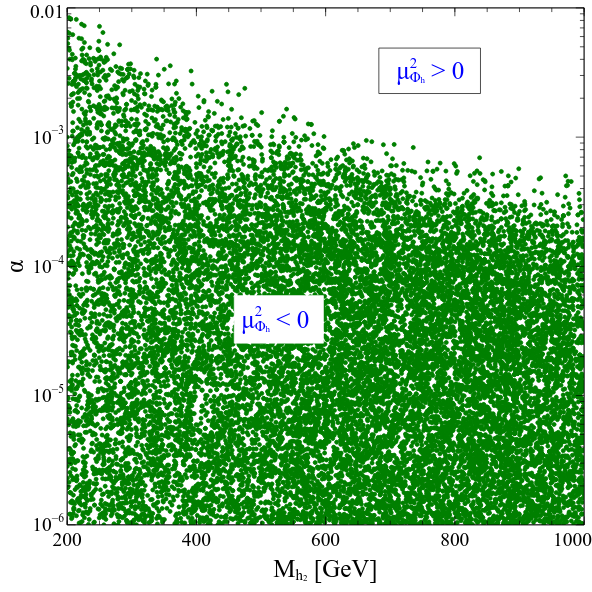}
\caption{Left (Right) panel: Allowed region in the $\lambda_{Dh}-\lambda_{DH}$
($M_{h_2}-\alpha$) plane where other parameters are fixed at
$\mzbl$ = 3000 GeV, $\gbl$ = 0.07, $n_{BL} = 1.33 \times 10^{-10}$, 
$M_{h_{1}} = 125.5$.}
\label{s5a}
\end{figure}

In the LP of Fig.\,\ref{s5a}, we have shown the allowed region
in the coupling plane ($\lambda_{Dh}-\lambda_{DH}$) which
reproduces the observed DM relic density ($0.1172\leq\Omega h^2\leq0.1226$).
In this figure, we have clearly indicated the
dominant DM production processes when $M_{DM}$ varies between
10 GeV to 100 GeV i.e.\,\,DM production from the decays
of $h_1$, $h_2$ or both or entirely from the annihilations
of SM particles like $W^\pm,\,\,Z,\,\,h_1$ etc.
The parameters which are related to
the $Z_{BL}$ decay ($g_{BL}$, $n_{BL}$) have been kept fixed at
0.07 and $1.33\times10^{-10}$ respectively, so at every time
an equal amount of $Z_{BL}$ decay contribution remains present.
As illustrated in the figure, when the parameter $\ldh$ is
small compared to the other parameter $\ldH$ then among
the two scalars it is the BSM Higgs $h_2$ which is mainly
contributing to the DM production while for the opposite
case, the production of $\dm$ becomes $h_1$ dominated
and in between both the scalars contribute equally.
Apart from that, if the mass of $\dm$ is greater than the half of the
SM-like Higgs mass (i.e. $M_{DM} > \dfrac{M_{h_1}}{2}$)
then DM production from $h_1$ decay becomes kinematically
forbidden. In this case, however, the production from
the decays of $h_2$ and $\zbl$ are still possible.
Now, the deficit in DM production can be compensated
by the production from self annihilation of SM particles
like $h_1$, $W^\pm$ and $Z$ and for this we need to increase
the parameter $\ldh$. Moreover, by increasing
$\ldh$ (decreasing $\ldH$ simultaneously)
we can arrive a situation where DM production is entirely
dominated by the annihilations of SM particles and this situation
has been indicated by a pink coloured arrow in the
LP of Fig.\,\ref{s5a}.
On the other hand, in the RP of Fig.\,\ref{s5a}
we have presented the allowed region in $M_{h_2}-\alpha$
plane which satisfies the relic density
bound. From this figure one can see that with the increase of
$M_{h_2}$, the allowed values of mixing angle $\alpha$ decrease.
The reason behind this decrement is related to the vacuum
stability conditions as given in the Eq.~(\ref{vsc}).
The region satisfying both the relic density bound as well as
the vacuum stability conditions is shown by the green dots
while in the other part of $M_{h_2}-\alpha$ plane the quantity
$\mu_{\phi_{h}}^{2}$ becomes positive which is undesirable
in the context of the present model (see Eq.~(\ref{vsc})). 
\subsubsection{$\dfrac{M_{h_1}}{2}<M_{\dm}
<\dfrac{M_{h_2}}{2},\,\dfrac{M_{\zbl}}{2}$,
{\bf BSM particles decay and SM particles annihilation
dominated region.}}
\label{region2}
\begin{figure}[h!]
\centering
\includegraphics[angle=0,height=8cm,width=10cm]{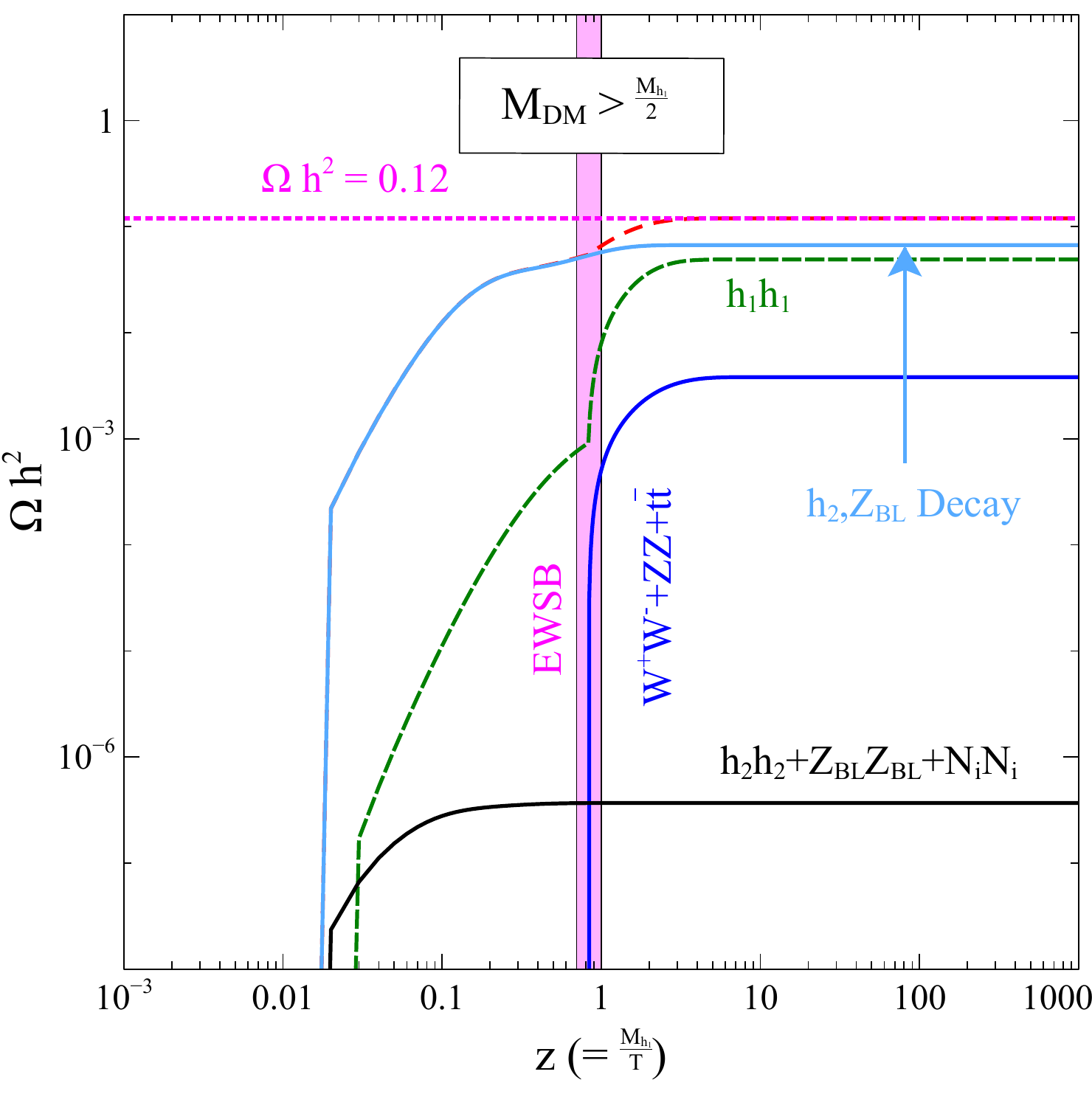}
\caption{Variation of DM relic density $\Omega h^2$
with $z$. Other parameters value have been kept fixed
at $\lambda_{Dh} = 6.364\times10^{-12}$, $\lambda_{DH} =
7.637\times10^{-14}$, $n_{BL} = 8.80 \times 10^{-11}$, $M_{DM}$ = 70 GeV,
$\mzbl$ = 3000 GeV, $\gbl$ = 0.07, $M_{h_{1}} = 125.5$
GeV, $M_{h_{2}}$ = 500 GeV, $\alpha = 10^{-5}$, $M_{N_2}$
$\approx$ $M_{N_1}$ = 2000 GeV and $M_{N_3} = 2500$ GeV.}
\label{lp1}
\end{figure}

Clearly in this mass region, DM production from the
decay of SM-like Higgs $h_1$ is kinematically forbidden
and hence DM has been produced from the
decays of $h_2$, $Z_{BL}$ only. However,
unlike the previous case, here we find significant
contribution to DM relic density arising from
the self annihilation of the SM particles namely,
$h_1$, $W^\pm$, $Z$ and $t$. On the other hand,
the annihilations of BSM particles like $\zbl$,
$h_2$ and $N_i$ have negligible effect on DM
production processes. In Fig.\,\ref{lp1}, we have
shown the variation of DM relic density
with $z$ for $\dfrac{M_{h_1}}{2}<M_{\dm}
<\dfrac{M_{h_2}}{2},\,\dfrac{M_{\zbl}}{2}$. 
Since now the decay of the $h_1$ to $\dm{\dm}^\dagger$ is kinematically
forbidden, hence we can increase the parameter $\lambda_{Dh}$ safely and this
will not overproduce DM in the Universe. Due to this
moderately large value of $\lambda_{Dh}$,
the annihilation channels become important.
From Fig.\,\ref{lp1} it is clearly seen that in this case the
annihilation channel $h_1 h_1 \rightarrow \dm \dm^{\dagger}$
(Green dashed line) contributes significantly to the DM production.
Therefore in the present case, production of DM has been controlled
by the decays of $h_2$, $\zbl$ and the self annihilations of
the SM particles and thus directly relates to the $\ubl$ sector
of this model.
\subsubsection{$\dfrac{M_{h_1}}{2},\,
\dfrac{M_{h_2}}{2}<M_{\dm}<\dfrac{M_{\zbl}}{2}$,
{\bf BSM particles decay and annihilation dominated
region.}}
\label{region3}
\begin{figure}[h!]
\centering
\includegraphics[angle=0,height=7cm,width=8cm]{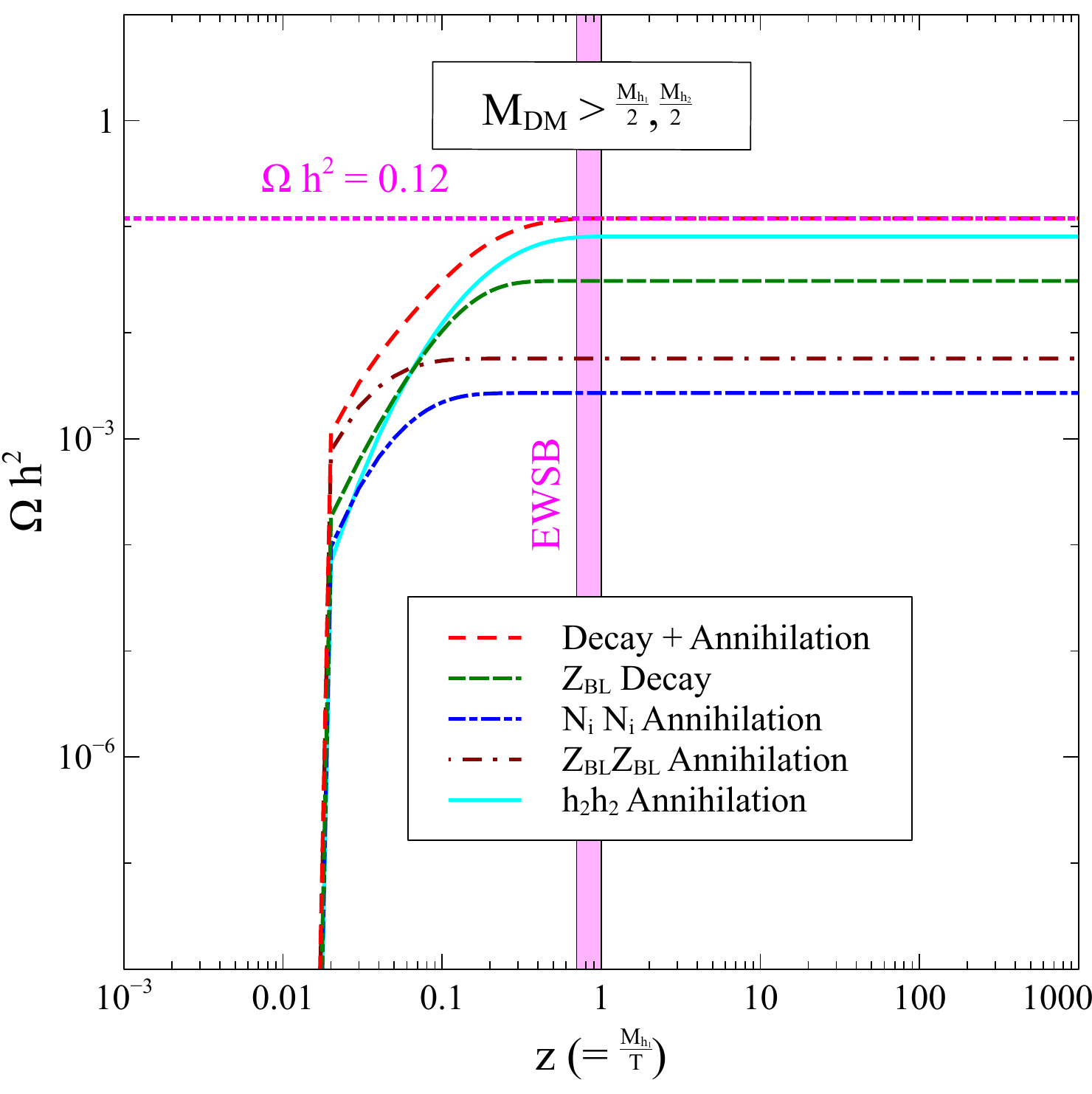}
\includegraphics[angle=0,height=7cm,width=8cm]{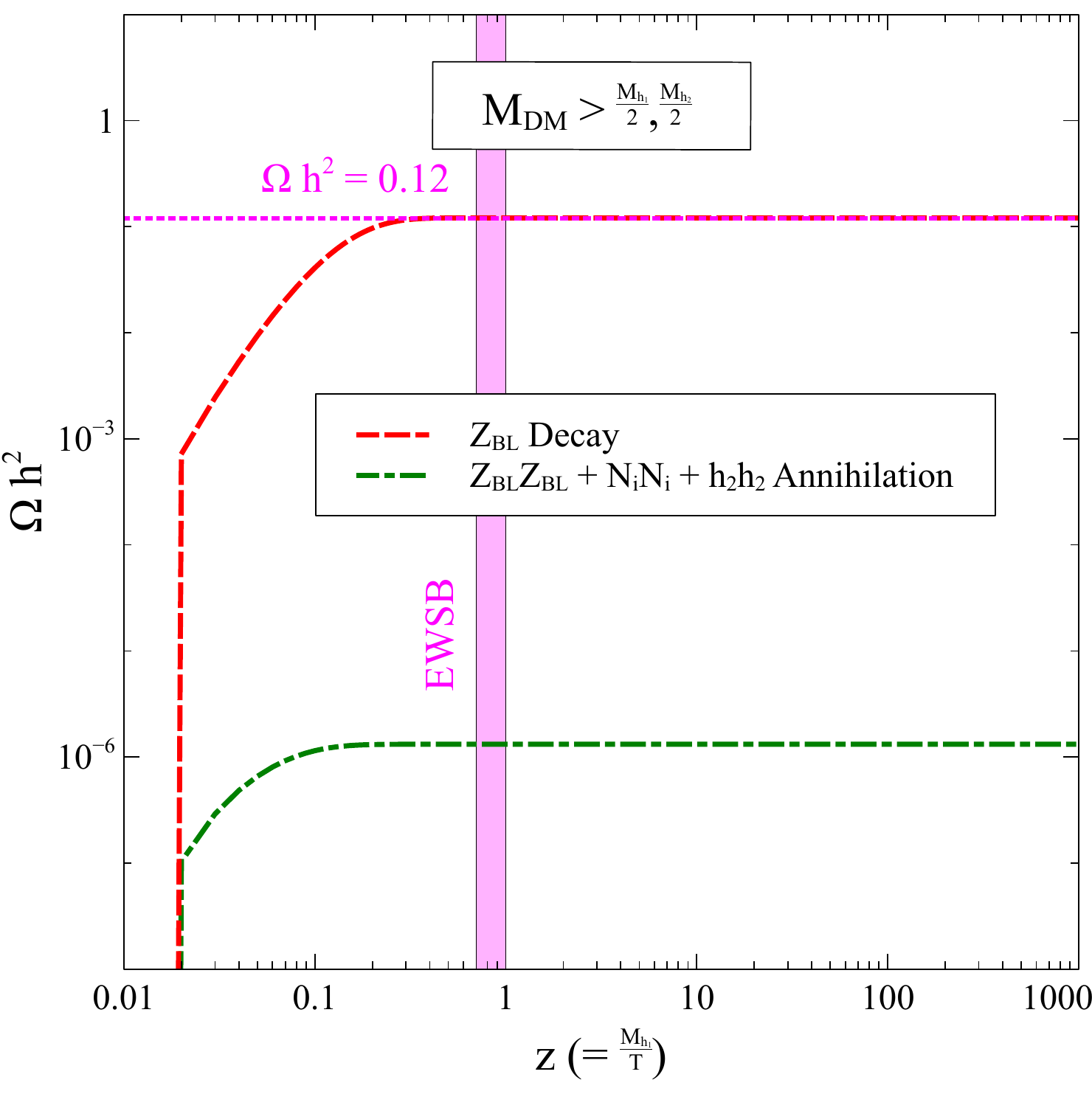}
\caption{Left (Right) panel: Variation of DM relic density $\Omega h^2$
with $z$. Other parameters value have been kept fixed
at $\lambda_{Dh} = 2.574\times10^{-12}$ ($7.212\times10^{-14}$),
$\lambda_{DH} = 3.035\times10^{-11}$ ($8.316\times10^{-14}$),
$n_{BL} = 3.4 \times 10^{-11}$ $(6.2 \times 10^{-11})$,
$M_{DM}$ = 450 GeV (600 GeV),
$\mzbl$ = 3000 GeV, $\gbl$ = 0.07, $M_{h_{1}} = 125.5$
GeV, $M_{h_{2}}$ = 500 GeV, $\alpha = 10^{-5}$,
$M_{N_2}$ $\approx$ $M_{N_1}$ = 2000 GeV
and $M_{N_3} = 2500$ GeV.}
\label{lp2}
\end{figure}
 
In this regime of the DM mass, the only surviving decay
mode is the decay of ${\rm B-L}$ gauge boson $\zbl$ to a pair
of $\dm$. Apart from that, depending on the choice of mass
of $\dm$ a significant fraction of DM has been produced
from the self annihilation of either BSM Higgs
$h_2$. In other word, we can say that in this
region the production of DM is BSM particles dominated. 
In LP of Fig.\,\ref{lp2} we show the relative
contribution of dominant production modes of DM to
$\Omega h^2$ for a chosen value of $M_{DM}=450$ GeV.
From this plot one can easily notice that in the
case when $M_{DM}<M_{h_2}$, the almost entire fraction of DM
is produced from the decay of $\zbl$
(green dashed line) and self annihilation
of BSM Higgs $h_2$ (solid turquoise line).
This is because, as in this case the production
of $\dm$ from $h_2$ decay is kinematically forbidden
hence one can increase the parameter $\ldH$ so that
the annihilation channel $h_2h_2 \rightarrow \dm {\dm}^\dagger$,
which is mainly proportional to $\ldH^2$
(due to four point interaction) becomes significant.   

\begin{figure}[]
\centering
\includegraphics[angle=0,height=8cm,width=11cm]{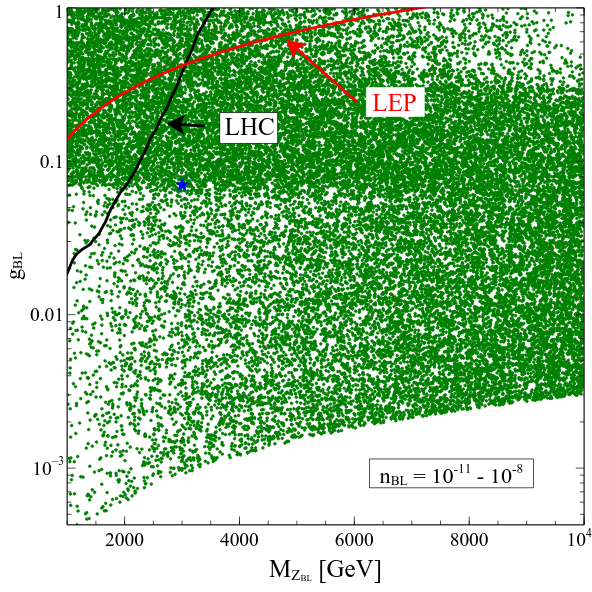}
\caption{Allowed region in $M_{\zbl}-\gbl$ plane which produces
observed DM relic density. Solid lines (black and red) are the upper limits
on the gauge coupling $\gbl$ for a particular mass of $\zbl$
obtained from LHC and LEP respectively. Other relevant parameters
used in this plot are $250\,{\rm GeV}\leq\mdm\leq
5000\,{\rm GeV}$, $\lambda_{Dh} = 7.212\times10^{-14}$,
$\lambda_{DH} = 8.316\times10^{-14}$, $M_{h_{2}}$ = 500 GeV,
$\alpha = 10^{-5}$, $M_{N_2}$ $\approx$ $M_{N_1}$ = 2000 GeV
and $M_{N_3} = 2500$ GeV.}
\label{scatt2}
\end{figure} 

On the other hand, in the RP we have considered a situation
where almost the entire DM has been produced from the decay of
${\rm B-L}$ gauge boson. For this, we have chosen $M_{DM}>M_{h_2}$
and a larger value of $\nbl = 6.2\times 10^{-11}$. Similar to the previous
case (i.e. $M_{DM}< M_{h_2}$) here also,
the production of $\dm$ from $h_2$ decay still remains
forbidden. However, as the sum of final state particles masses
are larger than that of initial state hence, in this case
$h_2 h_2$ annihilation mode becomes suppressed. Moreover,
to make the contribution of $h_2$ annihilation even more
suppressed we have reduced the quartic couplings $\ldh$
and $\ldH$ accordingly. As a result other annihilation
channels e.g. $\zbl \zbl$, $N_iN_i$ also become inadequate
as these channels are mediated by the exchange of $h_1$
and $h_2$. Although, RH neutrinos can annihilate to $\dm {\dm}^{\dagger}$
via $\zbl$, we cannot increase the contribution of $Z_{BL}$
mediated diagrams because for that one has to further increase 
the ${\rm B-L}$ charge of $\dm$ ($n_{BL}$), which
results in an over production of DM in the Universe
from $\zbl$ decay. From the RP of Fig.\,\ref{lp2},
one can easily notice that in this situation $\zbl$ decay
is the most dominant DM production channel (red dashed line)
while the total contributions from the annihilations of
$h_2$, $\zbl$ and $N_i$ are negligible. 
Therefore, for the entire mass range of $\dm$
i.e. $\dfrac{M_{h_1}}{2},\,\dfrac{M_{h_2}}{2}<M_{\dm}<\dfrac{M_{\zbl}}{2}$,
the DM production processes are always related
to the $\ubl$ sector of the present model by
receiving a sizeable contribution from $\zbl$ decay.

In Fig.\,\,\ref{scatt2}, we have shown the allowed
region (green coloured points) in $M_{\zbl}-g_{BL}$ plane
which reproduces the observed DM relic density. While
generating this plot we have varied $250\,{\rm GeV}\leq\mdm\leq$
5000 GeV and $10^{-11}\leq\nbl\leq10^{-8}$. In this region
as mentioned above dominant contributions to DM relic
density arise from $\zbl$ decay and annihilation of
BSM Higgs $h_2$. In this figure, the black solid line
represents the current upper bound
\cite{Chatrchyan:2012oaa, Aad:2014cka, Guo:2015lxa}
on $\gbl$ for a particular mass of $\zbl$ from LHC
\footnote{To get the bound in $M_{\zbl}-\gbl$ plane
from LHC, ATLAS and CMS collaborations consider
the Drell-Yan processes ($p\,p \rightarrow \zbl
\rightarrow \bar{l}\,l$, with $l$ = e or $\mu$)
and by searching the dilepton resonance they put
lower bound on $M_{\zbl}$ for a particular value of
extra gauge coupling $\gbl$.} 
while the limit \cite{Carena:2004xs, Cacciapaglia:2006pk,
Schael:2013ita} from LEP
\footnote{LEP consider the processes $e^{+}\,e^{-}
\rightarrow \bar{f}\,f$ ($f \neq e$) above the
Z-pole mass and by measuring its cross section
they put lower limit on the ratio between the
gauge boson mass and guage coupling, which is
$\frac{M_{\zbl}}{g_{BL}} \geq 6-7$ TeV.}
has been indicated by the red solid line respectively.
Therefore, the region below the red and black solid line
is allowed by the collider experiments like LHC and LEP.
The benchmark value of $\gbl$, $\mzbl$ (= 0.07, 3000 GeV)
for which we have computed the baryon asymmetry in the
previous section (Section \ref{baryogenesis}) is highlighted 
by a blue coloured star. Hence, in this regime
the extra gauge boson $Z_{BL}$ immensely takes part in
achieving the correct ballpark value of the DM relic density
and also at the same time $Z_{BL}$ plays a significant
role to obtain the observed value of the matter-antimatter
asymmetry of the Universe.  
\subsubsection{$M_{\dm}>\dfrac{M_{h_1}}{2},\,
\dfrac{M_{h_2}}{2},\,\dfrac{M_{\zbl}}{2}$,
{\bf BSM particles annihilation dominated region.}}
\label{region4}

Finally, in this range of DM mass the entire
production of $\dm$ from the decays of 
$h_1$, $h_2$ and $Z_{Bl}$ become kinematically
inaccessible. Therefore, in this case all three
parameters namely $\ldh$, $\ldH$ and $\nbl$
become free and we can make sufficient
increment to these parameters so that
either scalar medicated ($h_1$, $h_2$) or gauge boson
mediated ($\zbl$) annihilation processes of $N_i$, $\zbl$
or both can be the dominant contributors in DM production.

\begin{figure}[]
\centering
\includegraphics[angle=0,height=7cm,width=8cm]{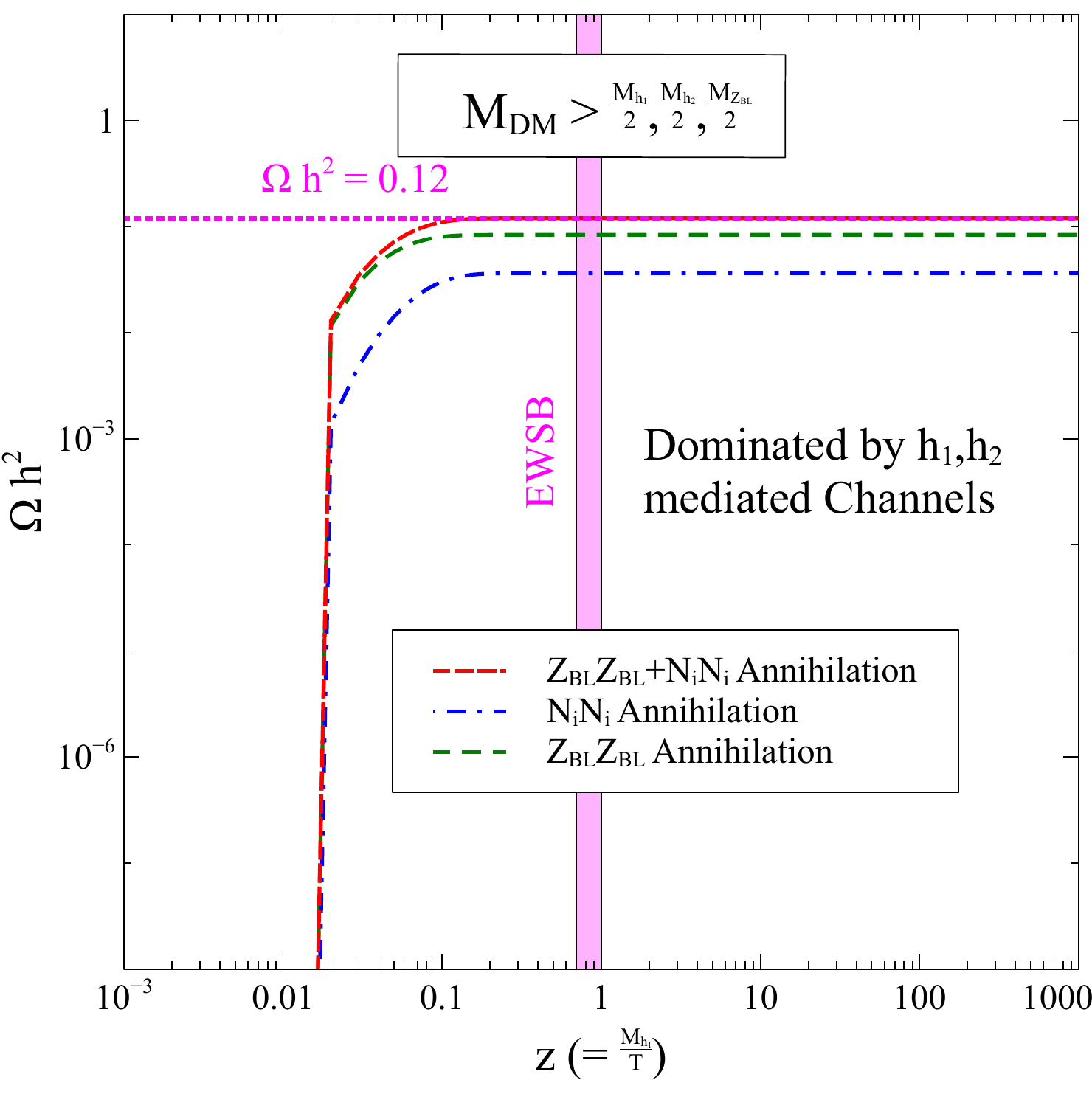}
\includegraphics[angle=0,height=7cm,width=8cm]{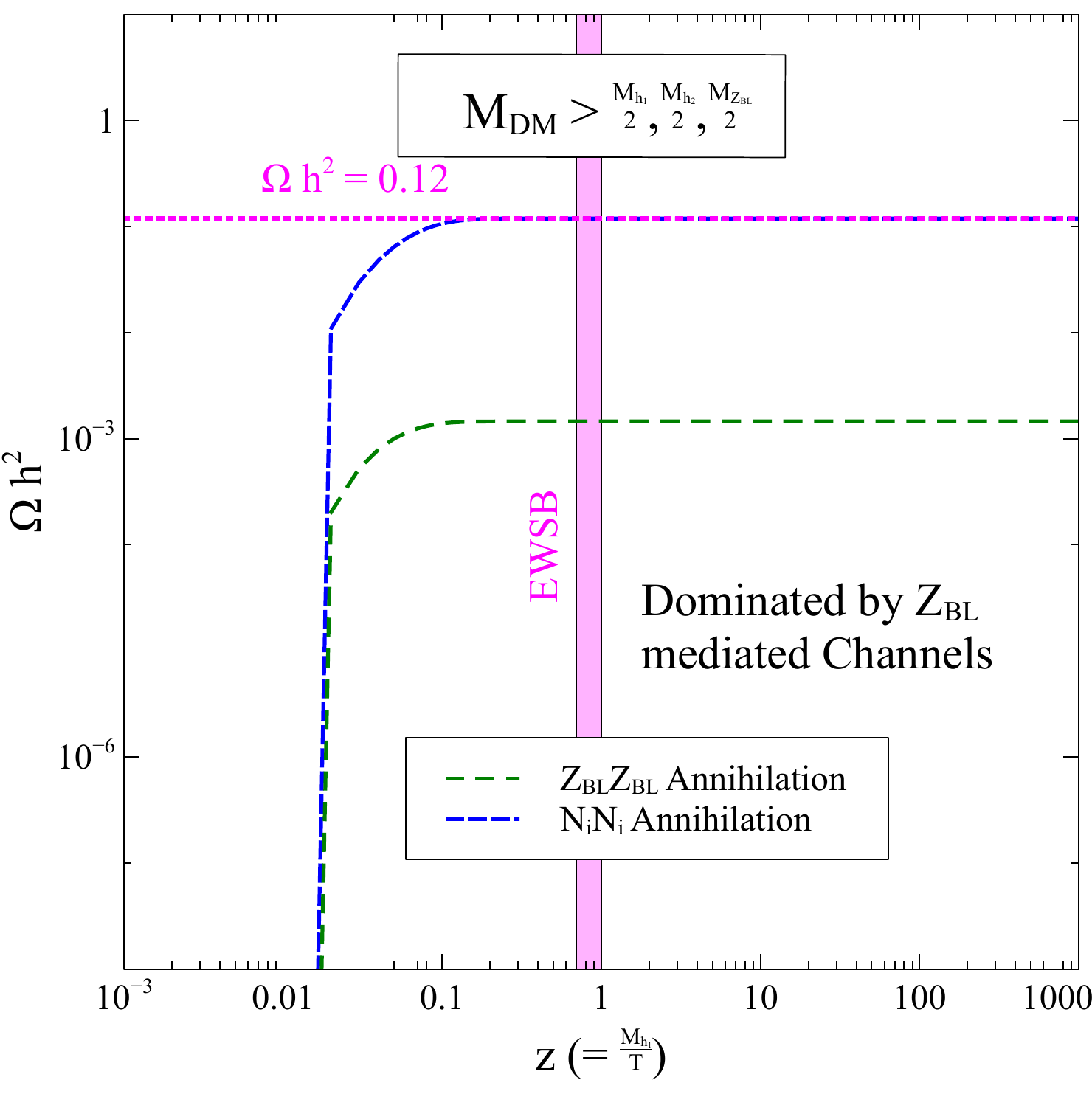}
\caption{Left (Right) Panel: Variation of DM relic density
$\Omega h^2$ with $z$ when dominant contributions are coming from
scalar $h_i$, (gauge boson $Z_{BL}$) mediated annihilation channels.
Other relevant parameters value have been kept fixed
at $\lambda_{Dh} = 7.017\times10^{-12}$
($7.212\times10^{-13}$), $\lambda_{DH} =
6.307\times10^{-11}$ ($8.316\times10^{-12}$),
$n_{BL} = 1.0 \times 10^{-10}$
($1.34 \times 10^{-8}$), $M_{DM}$ = 1600 GeV,
$\mzbl$ = 3000 GeV, $\gbl$ = 0.07, $M_{h_{1}} = 125.5$
GeV, $M_{h_{2}}$ = 500 GeV, $\alpha = 10^{-5}$,
$M_{N_2}$ $\approx$ $M_{N_1}$ = 2000 GeV
and $M_{N_3} = 2500$ GeV.}
\label{lp3}
\end{figure}

Similarly, in the LP and RP of Fig.\,\ref{lp3}, we have shown
two different situations where the DM production are
dominated by scalar ($h_1$, $h_2$) mediated
diagrams and gauge boson $Z_{BL}$ mediated diagrams respectively.
In the LP, by keeping the $n_{BL}$ value low and adjusting
the parameters $\lambda_{Dh}$ and $\lambda_{DH}$ one can achieve
the correct value DM relic density and on the other hand, in
the RP we have kept the values of $\lambda_{Dh}$ and $\lambda_{DH}$
sufficiently low and by suitably adjusting the DM charge $n_{BL}$ we have
achieved the correct value of the DM relic density. Therefore,
in this region, a strong correlation exists among the neutrino sector,
$\ubl$ sector and DM sector as the entire DM is now being produced from
$N_iN_i$ and $\zbl\zbl$ annihilations.

\begin{figure}[]
\centering
\includegraphics[angle=0,height=8cm,width=11cm]{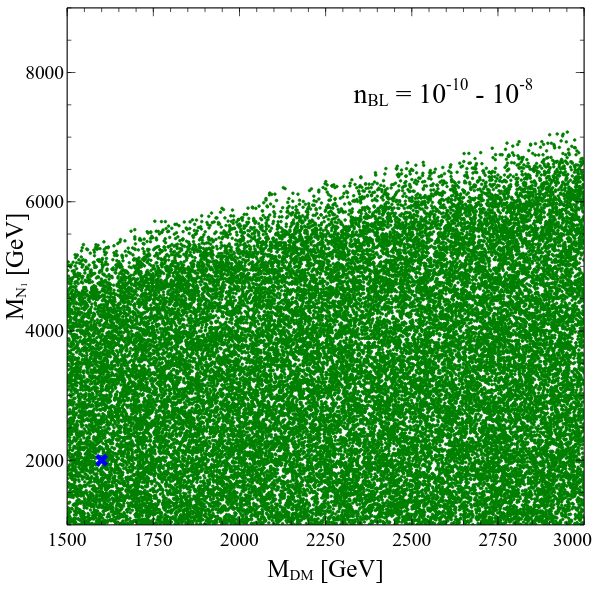}
\caption{Allowed region in $M_{DM}-M_{N_1}$ plane which
mimics the observed DM relic density. The blue
coloured star represent our benchmark point ($\mdm=1600$
GeV, $M_{N_1}=2000$ GeV).}
\label{scatt3}
\end{figure} 
In Fig.\,\ref{scatt3}, we have shown the allowed parameter space
in $M_{DM}-M_{N_1}$ plane by DM relic density. In order to generate
this plot we have varied DM mass in the range
$1500\,{\rm GeV}\leq \mdm \leq 3000\,{\rm GeV}$,
RH neutrino masses $1500\,{\rm GeV}\leq M_{N_i} \leq
10000\,{\rm GeV}$ ($i=1,\,2$), $M_{N_1}<M_{N_3}\leq M_{N_1} + 5000$ GeV
and $10^{-10}\leq\nbl\leq 10^{-8}$. Other relevant
parameters have been kept fixed at
$\lambda_{Dh} = 7.212\times10^{-13}$,
$\lambda_{DH} = 8.316\times10^{-12}$,
$\mzbl$ = 3000 GeV, $\gbl$ = 0.07,
$M_{h_{2}}$ = 500 GeV, $\alpha = 10^{-5}$
As discussed above,
in this regime ($\mdm>\dfrac{M_{h_1}}{2},\,
\dfrac{M_{h_2}}{2},\,\dfrac{M_{\zbl}}{2}$) $\dm$
is dominantly produced from the annihilations of
$\zbl$ and RH neutrinos. From this plot one can
observe that in this high DM mass range to obtain
the observed DM relic density, the mass of the lightest
RH neutrino cannot be larger than $\sim 6000$ GeV.
Analogous to the Fig.\,\ref{scatt2}, here also
we have indicated the benchmark point for which
we have computed baryon asymmetry in the previous
section (Section \ref{baryogenesis}) by a blue coloured
star. Therefore, in this case RH neutrinos are very actively
taking part in all three processes we have considered
in this work namely DM production processes, tiny neutrino mass
generation and also the generation of required lepton asymmetry
to reproduce the observed baryon asymmetry of the Universe.

From the above four regions, which are based on the
mass of our FIMP DM, it is evident that in the first region DM production
mainly happens from the decay of $h_1$, $h_2$ and $Z_{BL}$
and all annihilations are subdominant. Therefore, in this region only
the extra neutral gauge boson ($\zbl$), BSM Higgs ($h_2$)
and SM-like Higgs ($h_1$) are taking part in the
DM relic density estimates and there is no significant
role of the RH neutrinos.
In the second region, SM-like Higgs decay does not contribute
to DM production processes, hence one can safely increase
the quartic coupling $\lambda_{Dh}$ and consequently $h_1 h_1$
annihilation contribution increases. Similar to the previous regime,
here also RH neutrinos have less importance in determining the DM
relic density. In the third region, the only decay mode that
involves in DM production is $\zbl\rightarrow\dm{\dm}^\dagger$.
Since all other decay modes correspond to $h_1$ and $h_2$
are kinematically forbidden, hence we can increase
both the quartic couplings $\lambda_{Dh}$ and $\lambda_{DH}$
appropriately which eventually enhance the annihilation contribution
from the BSM Higgs significantly. Moreover, due to the
increment of quartic couplings in this region $Z_{BL} Z_{BL}$
and $N_{i} N_{i}$ annihilation channels start contributing
in the DM production processes.
Lastly in region four, due to the high value of the DM mass no
decay process contributes to DM relic density and only
the BSM particles annihilation contributes. Therefore,
in this region by properly adjusting the extra gauge coupling
$\gbl$, one can get a sizeable fraction of DM
production from the annihilation of RH neutrinos. Since apart from
the masses of the involving particles, the annihilation of RH neutrinos
mediated by $\zbl$ depends on the extra (B-L) gauge coupling
$\gbl$ solely. Thus, depending on the mass range of our FIMP DM,
we can say that the different model parameters and the
additional BSM particles (e.g. $\zbl$, $N_i$, $h_2$)
are fully associated to the DM production processes
in the early Universe.
         
\subsubsection{\bf Analytical Estimates}
So far, we have solved the full Boltzmann equation (Eq.\,\ref{BE})
for a FIMP $\dm$ numerically. Apart from this, one can estimate the FIMP
relic density (or comoving number density) by using the approximate
analytical formula. Let us consider a FIMP ($\dm$) which is produced
from the decay of a particle $A$ {\it i.e.}, $A \rightarrow
\dm\,{\dm}^{\dagger}$, where $A$ in the present model can be $h_1$, $h_2$ or $Z_{BL}$. The contribution of $A$
to the FIMP relic density at the present epoch, considering the
effect of both $\dm$ and ${\dm}^{\dagger}$, is given by \cite{Hall:2009bx},

\begin{eqnarray}
\Omega^{FIMP} h^{2} \simeq \frac{2.18 \times 10^{27} g_{A}}
{g_{\rm s}\sqrt{g_{\rho}}} \frac{M_{DM} \Gamma_{A}}{M_{A}^{2}}\,,
\label{fimp_analytical}
\end{eqnarray} 
where $M_A$ and $g_A$ are mass and internal degrees of freedom
of the mother particle $A$, respectively, while $\Gamma_A$ is the decay
width of the process $A \rightarrow \dm\,{\dm}^{\dagger}$. 
The analytic expressions for $\Gamma_A$  corresponding to $h_2$, $h_1$ and $Z_{BL}$ are 
given in Eqs.~(\ref{h2dmdm}), (\ref{h1dmdm}) and (\ref{zbldmdm}) in the Appendices. 
Moreover,
$g_{\rho}$ and $g_{\rm s}$, as define earlier, are the degrees of freedom
related to the energy and entropy densities of the Universe, respectively. 
Let us now compare the analytical result with the numerical
value which we obtain by solving the Boltzmann equation Eq.\,(\ref{BE}).
For this, let us consider a situation when a significant
fraction of our FIMP candidate ($\dm$) is produced from the
decay mode of BSM Higgs {\it i.e.}, $h_{2} \rightarrow \phi_{DM}
\phi_{DM}^{\dagger}$. Substituting the values of model parameters
given in the caption of Fig.\,\ref{s2a} to Eq.\,(\ref{fimp_analytical}),
we get the contribution of $h_2$ to DM 
relic density, which is
\begin{eqnarray}
\Omega^{FIMP} h^{2} \simeq 0.027,
\end{eqnarray}
where we consider $g_{\rho}=g_{\rm s} \approx 100$ and $g_{A} = 1$.
This can be compared to the contribution of $h_2$ obtained from exact numerical estimate shown in  
the LP of Fig.\,\ref{s2a} which is, 
\begin{eqnarray}
\Omega_{h_{2}\rightarrow \phi_{DM} \phi_{DM}^{\dagger}} h^{2}
= 0.0276.
\end{eqnarray}
Therefore, from the above two estimates it is clearly evident that
the analytical result agrees well with the full numerical result.
Similarly, for the other decay modes also ({\it i.e.} $h_1$, $\zbl$)
one can match the analytical and numerical results.



\section{Conclusion}
\label{conclusion}
In this work we considered a local $\ubl$ extension of
the SM and to cancel the additional anomalies associated with 
this gauge symmetry we introduced
three RH neutrinos ($N_{i}$, $i$ = 1, 2, 3). Besides the
three RH neutrinos, we also introduced two SM gauge singlet
scalars $\phi_{H}$ and $\dm$. The scalar field $\phi_{H}$,
being charged under $\ubl$, takes a nonzero VEV and
breaks the proposed ${\rm B-L}$ symmetry spontaneously.
Moreover, as the scalar field $\dm$ has also a nonzero ${\rm B-L}$
charge, one can adjust this charge suitably so that
after symmetry breaking the model has left with
a residual $\mathbb{Z}_2$ symmetry and only $\dm$
behaves as a odd particle under this leftover symmetry.
This makes $\dm$ absolutely stable over the cosmological time scale
and hence acts as a dark matter candidate.
After spontaneous breaking of $\ubl$ gauge
symmetry, all RH neutrinos and extra neutral
gauge boson $\zbl$, acquired mass. Due to the
presence of the three RH neutrinos in the model,
we easily generated Majorana masses for the
three light neutrinos by the Type I seesaw mechanism.
This model is also able to explain baryogenesis via leptogenesis, where we
generated the lepton asymmetry in the Universe
from out of equilibrium, CP violating decays of
two degenerate RH neutrinos and converted this
lepton asymmetry to the observed baryon asymmetry
through the sphaleron transitions. 

In explaining the neutrino masses by Type I seesaw mechanism,
we considered a complex Dirac mass matrix $\mD$ and
a diagonal Majorana mass matrix $\mR$ for the RH neutrinos.
In determining the allowed model parameter space,
we used the measured values of
neutrino oscillation parameters namely three mixing angles
($\theta_{12}$, $\theta_{13}$ and $\theta_{23}$) and
two mass square differences ($\Delta m_{21}^{2}$, $\Delta m_{atm}^{2}$)
in their current $3\sigma$ range. In particular, in the current model
we could reproduce the whole allowed $3 \sigma$ range of 
the neutrino oscillation parameters by different combinations of the 
relevant model parameters.
The Dirac CP phase was constrained to lie
within two distinct regions. One is the entire first
quadrant ($0^\circ-90^\circ$) while the other one
spans the entire fourth quadrant ($270^\circ-360^\circ$).
However, if we considered the T2K result on Dirac CP phase
then the values of $\delta$ lying in the fourth quadrant
are more favourable compared to those in the first quadrant.  
We also computed the magnitudes of the Jarlskog invariant
$J_{\rm CP}$ and found that the values of $J_{\rm CP}$,
for the model parameters which satisfy neutrino oscillation
data, always lie below $0.039$. Finally, 
we calculated
the values of $m_{\beta\beta}$, the quantity
relevant to neutrino less double $\beta$ decay,
for the allowed model parameter space.  

Since we allowed complex Yukawa
couplings in the Dirac mass matrix $\mD$, the decays of
RH neutrinos were CP violating. We took the 
masses of the RH neutrinos in the TeV range and 
worked in the parameter space where the lightest two 
RH neutrino states were nearly degenerate, with their masses 
separated by their tree level decay width. This scenario led 
to resonant leptogenesis  (or TeV scale
leptogenesis) for the production of observed
baryon asymmetry in the Universe from the
out of equilibrium decays of RH neutrinos.
We generated the
observed baryon asymmetry for three different values
of RH neutrino masses namely $M_{N_1}=1600$ GeV,
1800 GeV and 2000 GeV, respectively, where
required values of CP asymmetry parameter
parameter ($\varepsilon_1$) were $4.4\times10^{-4}$,
$2.25\times10^{-4}$ and $1.8\times10^{-4}$, respectively.
These values of $M_{N_1}$ and $\epsilon_1$ were 
also seen to be allowed by the neutrino oscillation data.  

Lastly, we studied the DM phenomenology
by considering a FIMP type DM candidate $\dm$.
We took into account all the production
modes of $\dm$ (both before and after EWSB)
arising from the annihilations and decays of
SM as well as BSM particles. We found
that depending on the mass of $\dm$, the  
production processes of $\dm$ can be
classified into four distinct categories.
These are (1) SM and BSM particles decay dominated region,
(2) BSM particles decay and SM particles annihilation dominated region,
(3) BSM particles annihilation and $Z_{BL}$ decay dominated region
and finally (4) BSM particles annihilation dominated region. 
The first region is characterised by $M_{DM} < \dfrac{M_{h_1}}{2},
\,\dfrac{M_{h_2}}{2},\,\dfrac{M_{Z_{BL}}}{2}$ and
here DM is mainly produced from the decays of
$h_1$, $h_2$ and $\zbl$. In the second region,
DM mass is concentrated between $\dfrac{M_{h_1}}{2}$
and min$\left[\dfrac{M_{h_2}}{2}, \dfrac{M_{\zbl}}{2}\right]$ i.e.
$\dfrac{M_{h_1}}{2} < M_{DM} < 
\dfrac{M_{h_2}}{2},\,\dfrac{M_{Z_{BL}}}{2}$. In this case,
$h_2$, $\zbl$ decays and $h_1h_1$, $W^+W^-$, $ZZ$
annihilations act as the dominant production
modes of FIMP DM. In the third region where
$\dfrac{M_{h_1}}{2},\,\dfrac{M_{h_2}}{2}
< M_{DM} < \dfrac{M_{Z_{BL}}}{2}$, DM
has mainly been produced from $\zbl$ decay 
and also from the annihilation of BSM Higgs $h_2$ (for $M_{DM}<M_{h_2}$).
Finally, in the last region all three decay modes
become kinematically forbidden as $M_{DM} >
\dfrac{M_{h_1}}{2},\,\dfrac{M_{h_2}}{2},\,\dfrac{M_{Z_{BL}}}{2}$
and hence entire DM is produced from the self annihilations of
$\zbl$ and right handed neutrinos ($N_i$).  
Therefore in all four regions the $\ubl$ gauge
boson has played a significant role in DM production
while the effects of right handed neutrinos are
important in the last two regions only.
We also found that, since for a FIMP candidate $\dm$ the observed
DM relic density ($0.1172\leq\Omega_{h^2}\leq0.1226$)
is generated via the freeze-in mechanism, this puts
upper bounds on the scalar and gauge portal couplings of $\dm$
to restrict its over production i.e.
$\ldh\la10^{-11}$, $\ldH\la10^{-10}$ and $\nbl \la 10^{-8}$.
Hence, due to such extremely feeble couplings $\dm$ can
easily evade all the constrains coming from any terrestrial
DM direct detection experiment.

In conclusion, our spontaneously broken local $\ubl$ extension of
the SM with three additional RH and two additional scalars can explain the 
three main evidences for physics beyond the SM, {\it viz.}, small 
neutrino masses, matter-antimatter asymmetry of the Universe and 
dark matter. Tiny neutrino masses and all mixing angles can be 
obtained via Type I seesaw mechanism where we chose a certain 
pattern for the real and complex Yukawa couplings. The model gave a 
definite prediction for the CP violating phase to be measured in
the next generation long baseline experiments.
The dark matter candidate is a scalar which is 
neutral under the SM gauge group and has a nonzero ${\rm B-L}$ charge.
DM is made stable by virtue of a remnant $\mathbb{Z}_2$ symmetry
arises after the spontaneous breaking of $\ubl$ gauge symmetry.
This can be achieved by imposing a suitable ${\rm B-L}$ charge
on $\dm$ so that the Lagrangian does not contain any odd
term of $\dm$. This scalar DM can easily be taken as a
FIMP candidate which is produced from the decays and
annihilations of SM and BSM particles. Therefore, even if the
WIMP type DM is ruled out in near future from direct detection
experiments this present variant of $\ubl$ scenario with FIMP
DM will still survive. Further, since $\gbl$ is of the order
of SM gauge couplings, this model has the potential to be tested 
in the LHC or in other future collider experiments by detecting
${\rm B-L}$ gauge boson $\zbl$ from its SM decay products.
Moreover, considering the masses of RH neutrinos in TeV scale 
allow us to simultaneously explain the baryon asymmetry of the Universe from
resonant leptogenesis, FIMP DM production via Freeze-in mechanism
and also neutrino masses and mixing from TeV scale Type-I seesaw. 
Thus, all three phenomena addressing in this article are interconnected
to each other.
\section{Acknowledgement}
\label{acknowledge}
SK and AB also acknowledge the HRI cluster
computing facility (http://cluster.hri.res.in).
The authors would also like to thank the Department of Atomic Energy
(DAE) Neutrino Project under the XII plan of Harish-Chandra
Research Institute.
This project has received funding from the European Union's Horizon
2020 research and innovation programme InvisiblesPlus RISE
under the Marie Sklodowska-Curie
grant  agreement  No  690575. This  project  has
received  funding  from  the  European
Union's Horizon  2020  research  and  innovation
programme  Elusives  ITN  under  the Marie
Sklodowska-Curie grant agreement No 674896.

\appendix
\section{Expression for the Majorana mass matrix of light neutrinos}
\label{app:mnu}
Here we have given the expression of all the elements of the light neutrino
mass matrix $m_{\nu}$ in terms of the Yukawa couplings and the RH neutrino masses.
\begin{eqnarray}
&& (m_{\nu})_{11} = -\frac{y_{ee}^{2}}{M_{N_1}}
-\frac{y_{e\mu}^{2}}{M_{N_2}} 
-\frac{y_{e\tau}^{2}}{M_{N_3}},
\nn \\
&& (m_{\nu})_{12} = -\frac{y_{e\mu}\,y_{\mu\mu}}{M_{N_2}}
-\frac{y_{e\tau}\,y_{\mu\tau}}{M_{N_3}}
-\frac{y_{ee}\,y_{\mu e}}{M_{N_1}}-
i\,\frac{y_{ee}\,\tilde{y}_{\mu e}}{M_{N_1}}, \nn \\
&& (m_{\nu})_{13} = -\frac{y_{e\tau}\,y_{\tau \tau}}{M_{N_3}}
-\frac{y_{e e}\,y_{\tau e}}{M_{N_1}}
-\frac{y_{e\mu}\,y_{\tau \mu}}{M_{N_2}}
-i\,\left(\frac{y_{e e}\,\tilde{y}_{\tau e}}{M_{N_1}}
+\frac{y_{e\mu}\,\tilde{y}_{\tau \mu}}{M_{N_2}}\right), \nn \\
&& (m_{\nu})_{21} = (m_{\nu})_{12}, \nn \\
&& (m_{\nu})_{22} = -\frac{y_{\mu\mu}^{2}}{M_{N_2}}
-\frac{y_{\mu \tau}^{2}}{M_{N_3}}
-\frac{y^2_{\mu e}}{M_{N_1}}
+\frac{\tilde{y}^2_{\mu e}}{M_{N_1}}
-i\,\frac{2\,y_{\mu e}\,\tilde{y}_{\mu e}}{M_{N_1}}, \nn \\
&& (m_{\nu})_{23} = -\frac{y_{\mu \tau}\,y_{\tau \tau}}{M_{N_3}} -
\frac{y_{\mu e}\,y_{\tau e}}{M_{N_1}}
-\frac{y_{\mu \mu}\,y_{\tau \mu}}{M_{N_2}}
+\frac{\tilde{y}_{\mu e}\,\tilde{y}_{\tau e}}{M_{N_1}}
-i\,\left(\frac{y_{\tau e}\,\tilde{y}_{\mu e}}{M_{N_1}}
+\frac{y_{\mu e}\,\tilde{y}_{\tau e}}{M_{N_1}}
+\frac{y_{\mu \mu}\,\tilde{y}_{\tau \mu}}{M_{N_2}}\right), \nn \\
&& (m_{\nu})_{31} =  (m_{\nu})_{13}, \nn \\
&& (m_{\nu})_{32} = (m_{\nu})_{23}, \nn \\
&&(m_{\nu})_{33} = -\frac{y_{\tau \tau}^{2}}{M_{N_3}}
-\frac{y_{\tau e}^{2}}{M_{N_1}}
-\frac{y_{\tau \mu}^{2}}{M_{N_2}}
+\frac{\tilde{y}_{\tau e}^{2}}{M_{N_1}}
+\frac{\tilde{y}_{\tau \mu}^{2}}{M_{N_2}}
-i\,2\left(\frac{y_{\tau e}\,\tilde{y}_{\tau e}}{M_{N_1}}
 +\frac{y_{\tau \mu}\,\tilde{y}_{\tau \mu}}{M_{N_2}}
 \right), \nn \\
 \label{mnuelements} \\ 
 &&m_{\nu} =
 \left(\begin{array}{ccc}
(m_{\nu})_{11} ~~&~~ (m_{\nu})_{12}
~~&~~(m_{\nu})_{13}\\
(m_{\nu})_{21}
~~&~~ (m_{\nu})_{22}
~~&~~(m_{\nu})_{23}\\
(m_{\nu})_{31} ~~&
~~ (m_{\nu})_{32} ~~&~~ (m_{\nu})_{33}\\
\end{array}\right) \,.
\label{neutrino-mass}
\end{eqnarray}
\section{Neutrino less double $\beta$ decay parameter $m_{\beta \beta}$}
\label{app:mbb}
Since the light neutrino mass matrix is Majorana in nature, it
is a complex symmetric matrix. A complex symmetric matrix $m_{\nu}$
can be diagonalised by a Unitary matrix $U_{\rm PMNS}$
(defined in Eq.~(\ref{mass-matrix})) in the following way,
\begin{eqnarray}
m_{\rm diag} &=& U_{\rm PMNS}^{\dagger} \,m_{\nu}\, U_{\rm PMNS}^{\ast}, \nn \\
\Rightarrow m_{\nu} &=& U_{\rm PMNS}\, m_{\rm diag}\, U_{\rm PMNS}^{T}.\nn
\end{eqnarray}
Now equating the $(i,\,j)^{\rm th}$ element from both sides of the
above equation, we get
\begin{eqnarray}
(m_{\nu})_{ij} = (U_{\rm PMNS})_{ik} \,
(m_{\rm diag})_{k k^{\prime}} \,
(U_{\rm PMNS}^{T})_{k^{\prime} j}.\nn
\end{eqnarray}
Since $m_{\rm dia}$ is a diagonal matrix, we can further
simplify $({m_{\nu}})_{ij}$ by using $m_{\rm diag} =
m_{k}\,\delta_{k k^{\prime}}$, where $m_k$ is the mass of
$k^{\rm th}$ light neutrino. Therefore $(m_{\nu})_{ij}$
takes the following form
\begin{eqnarray}
(m_{\nu})_{ij} = \sum_{k = 1}^{3} m_{k}\,(U_{\rm PMNS})_{ik}\,
(U_{\rm PMNS})_{jk}\,. \nn
 \end{eqnarray}
Above equation expresses the elements of light neutrino mass matrix
(Eq.~(\ref{mnuelements})) in terms of the light neutrino masses, the intergenerational
mixing angles and the phases. Taking $i=j=1$, we get the expression of the 
(1,1) element of $m_{\nu}$ i.e.
\begin{eqnarray}
(m_{\nu})_{11} = \sum_{k=1}^{3} m_{k}\,(U_{\rm PMNS})^2_{\,1\,k},\nn
\end{eqnarray}
which is related to the important parameter $m_{\beta \beta}$
of the neutrino less double $\beta$ decay \cite{DellOro:2016tmg} as,
\begin{eqnarray}
m_{\beta \beta} = \left|\sum_{k=1}^{3} m_{k}\,(U_{\rm PMNS})^2_{1\,k}\right|
=\left|(m_{\nu})_{11}\right|\,.
\label{beta-decay}
\end{eqnarray} 
\section{CP Asymmetric Parameter Calculation for Leptogenesis}
\label{App:cp_lepton}
The amount of lepton asymmetry generated 
in the out of equilibrium decay of the RH neutrino
$N_i$ is parametrised by the CP asymmetry
parameter ($\varepsilon_i$), which is defined as,
\begin{eqnarray}
\varepsilon_{i} = \frac{\sum_{j} \left[\Gamma (N_{i}
\rightarrow L_j\,\phi_h)
- \Gamma (N_{i} \rightarrow \bar{L}_j\,{\phi_h}^{\star})\right]}
{\sum_{j} \left[\Gamma (N_{i} \rightarrow L_j\,\phi_h)
+ \Gamma (N_{i} \rightarrow \bar{L}_j\,{\phi_h}^{\star})\right]}\,.
\label{cpasy}
\end{eqnarray}   
If we consider only the tree level decay process of $N_i$
(first diagram in Fig.\,\,\ref{feyn1-lepto}), there will
not be any CP violation. The nonzero CP asymmetry is generated
only by the interference between the tree level and the one
loop level diagrams. The expression of CP asymmetry
parameter ($\varepsilon_i$) is given by
\cite{Flanz:1996fb, Pilaftsis:1997jf, Pilaftsis:2003gt},
\begin{eqnarray}
\varepsilon_{i} = -\sum_{j\neq i} \frac{M_{N_i}}{M_{N_j}}
\frac{\Gamma_{j}}{M_{N_j}}
\left(\frac{V_{j}}{2} + S_{j} \right) \frac{{\rm Im}
\left[(\mathcal{M_{D}} \mathcal{M_{D}}^{\dagger})^{2}_{ij}\right]}
{(\mathcal{M_{D}} \mathcal{M_{D}}^{\dagger})_{ii}
\,(\mathcal{M_{D}} \mathcal{M_{D}}^{\dagger})_{jj}},
\label{epsai}
\end{eqnarray} 
where $V_{j}$ and $S_{j}$ are the contributions coming from
the vertex correction and the self energy correction diagrams
respectively (second and third diagrams in Fig.\,\ref{feyn1-lepto}).
The expressions of $V_{j}$ and $S_{j}$ have the
following forms \cite{Flanz:1996fb, Pilaftsis:1997jf,
Pilaftsis:2003gt},
\begin{eqnarray}
V_{j} &=& 2 \frac{M_{N_j}^{2}}{M_{N_i}^{2}} \left[ \left( 1 +
\frac{M_{N_j}^{2}}{M_{N_i}^{2}}\right) \log \left( 1 +
\frac{M_{N_i}^{2}}{M_{N_j}^{2}}\right) - 1\right], \nn \\
S_{j} &=& \frac{M_{N_j}^{2} \Delta M_{ij}^{2}}{(\Delta M_{ij}^{2})^{2}
+ M_{N_i}^{2} \Gamma_{j}^{2}}\,,
\label{vjsj}
\end{eqnarray}
with 
\begin{eqnarray}
\Delta M_{ij}^{2} = M_{N_j}^{2} - M_{N_i}^{2},
\end{eqnarray}
and $\Gamma_{j}$ denotes the tree level decay width of
the RH neutrino $N_j$ (neglecting subdominant one loop
corrections), which is given by
\begin{eqnarray}
\Gamma_{j} = \frac{M_{N_j}}{4 \pi v^{2}}
(\mathcal{M_{D}} \mathcal{M_{D}}^{\dagger})_{jj}\,\,.
\end{eqnarray}

Now, as mentioned in the beginning of this section, the
enhancement in the CP asymmetry factor (Eq.~(\ref{cpasy}))
occurs when two RH neutrinos are almost degenerate i.e.
$M_{N_2}-M_{N_1}\simeq \dfrac{\Gamma_1}{2}$. This is know as the
resonance condition. In the present scenario, we consider $M_{N_3}>
M_{N_2}\simeq M_{N_1}$. Therefore, resonance condition is satisfied only
for the two lightest RH neutrinos $N_2$ and $N_1$. Hence we can neglect
the contribution of $N_3$ in the CP asymmetry parameter (Eq.~(\ref{epsai}))
by considering the summation over only $N_1$ and $N_2$ (i.e. $j=1,\,2$). 
Using the resonance condition in Eq.~(\ref{vjsj}), one can easily notice that
$S_j\sim \mathcal{O}\left(\frac{M_{N_j}}{\Gamma_j}\right) >>1$ ($j=1,\,2)$
and 
\begin{eqnarray}
\varepsilon_i \simeq -\sum^2_{j\neq i,\,\,j=1} \frac{M_{N_i}}{M_{N_j}}
\frac{\Gamma_{j}}{M_{N_j}}\,S_{j}\,
\dfrac{{\rm Im}\left[(\mathcal{M_{D}}
\mathcal{M_{D}}^{\dagger})^{2}_{ij}\right]}
{(\mathcal{M_{D}} \mathcal{M_{D}}^{\dagger})_{ii}
\,(\mathcal{M_{D}} \mathcal{M_{D}}^{\dagger})_{jj}}\,\,,
\label{epsaiaprox}
\end{eqnarray}
where we have neglected the quantity $V_j$ which is, in the present
condition ($M_{N_2}-M_{N_1}\simeq \dfrac{\Gamma_1}{2}$), much smaller
compared to $S_j$. The resonance condition leads to
\begin{eqnarray}
\Delta M^2_{21} &=& M^2_{N_2} -M^2_{N_1} \,,\nn \\
&=& \dfrac{\Gamma_1}{2} \left(2\,M_{N_1}+\dfrac{\Gamma_1}{2}\right)\,,\nn\\
&\simeq& M_{N_1}\,\Gamma_1 + \mathcal{O}\left(\Gamma^2_1\right)\,.
\label{deltam12}
\end{eqnarray}
Using Eq.~(\ref{deltam12}) in Eq.~(\ref{vjsj}) we get,
\begin{eqnarray}
S_1 &\simeq& -\dfrac{M^2_{N_1}}{2\,M_{N_2}\,\Gamma_1}\,,\nn \\
S_2 &\simeq& \dfrac{M^2_{N_2}}{M_{N_1}}
\dfrac{\Gamma_1}{\Gamma^2_1+\Gamma^2_2}\,.
\end{eqnarray}
Now, substituting the expressions of $S_1$ and $S_2$ in
Eq.~(\ref{epsaiaprox}) and using
${\rm Im}\left[(\mathcal{M_{D}} \mathcal{M_{D}}^{\dagger})^{2}_{12}\right]=
-\,{\rm Im}\left[(\mathcal{M_{D}} \mathcal{M_{D}}^{\dagger})^{2}_{21}\right]$,
one obtains,
\begin{eqnarray}
\varepsilon_2 &\simeq& -\dfrac{1}{2}
\frac{{\rm Im}
\left[(\mathcal{M_{D}} \mathcal{M_{D}}^{\dagger})^{2}_{12}\right]}
{(\mathcal{M_{D}} \mathcal{M_{D}}^{\dagger})_{11}\,(\mathcal{M_{D}}
\mathcal{M_{D}}^{\dagger})_{22}} \,, 
\label{epsa2}\\
\varepsilon_1 &\simeq& -\dfrac{\Gamma_1\,\Gamma_2}{\Gamma^2_1+\Gamma^2_2}\,
\frac{{\rm Im}
\left[(\mathcal{M_{D}} \mathcal{M_{D}}^{\dagger})^{2}_{12}\right]}
{(\mathcal{M_{D}} \mathcal{M_{D}}^{\dagger})_{11}\,(\mathcal{M_{D}}
\mathcal{M_{D}}^{\dagger})_{22}} \,, 
\label{epsa1}\\
&\simeq& \dfrac{2\,\Gamma_1\,\Gamma_2}
{\Gamma^2_1+\Gamma^2_2}\,\varepsilon_2\,.
\label{epsa1-epsa2}
\end{eqnarray} 
\section{Expressions of decay widths of $h_2$, $h_1$ and $Z_{BL}$}
\label{App:AppendixB}
In the present work, we have considered the effect of electroweak
symmetry breaking on dark matter production.
After EWSB SM particles become massive and affect DM production, while before
EWSB those particles have no effect. To take this effect into account
we have defined an extra constant $C_{ASB}$.
In all the equations, the value of the constant $C_{ASB} = 0$ before the
EWSB and this is equal to unity i.e. $C_{ASB} = 1$ after the EWSB.
Also before the EWSB, there is no mixing between the SM and BSM Higgs
bosons, i.e. $\alpha=0$. The two vertices which are common
to all Higgs mediated diagrams are as follows,
\begin{eqnarray}
g_{h_{1} \phi_{DM}^{\dagger} \phi_{DM}} &=&
- \left(v\lambda_{Dh}\cos \alpha +
v_{BL}\lambda_{DH}\sin \alpha\right)\,, \nn \\
g_{h_{2} \phi_{DM}^{\dagger} \phi_{DM}} &=&
 \left(v\lambda_{Dh}\sin \alpha -
v_{BL}\lambda_{DH}\cos \alpha\right)\,.
\label{dm_vertex}
\end{eqnarray}
\underline{\bf Total decay width of $h_2$:}\\
\begin{itemize}

\item \underline{$h_2$ $\rightarrow VV$} ($V= W^\pm, Z$):
\newline
\begin{eqnarray}
g_{h_{2}VV} &=& -\dfrac{2M_{V}^{2}}{v}\,\sin\alpha\,,\nn \\
\Gamma(h_2 \rightarrow VV) &=&
\dfrac{C_{ASB}\,M_{h_2}^{3}\,g_{h_{2}VV}^{2}}
{64\,\pi M_{V}^{4}\,S_{V}}
\sqrt{1 - \dfrac{4 M_{V}^{2}}{M_{h_2}^{2}}}\,\,
\left(1 - \dfrac{4 M_{V}^{2}}{M_{h_2}^{2}}
+ \dfrac{12M_{V}^{4}}{M_{h_2}^{4}}\right)\,,
\end{eqnarray}
where $S_V=2\,(1)$ for $ZZ$\,($W^+W^-$) final state.

\item \underline{$h_{2}$ $\rightarrow$ $h_1$\,$h_1$}:
\newline
\begin{eqnarray}
&&\Gamma({h_{2}} \rightarrow {h_1}{h_1}) = 
\dfrac{g_{h_{1}h_{1}h_{2}}^{2}}{32\,\pi\,M_{h_2}}
\,\sqrt{1 - \dfrac{4M_{h_1}^{2}}{M_{h_2}^{2}}}\,.
\end{eqnarray}

\item \underline{$h_{2}$ $\rightarrow$ $\phi_{DM}^\dagger$
$\phi_{DM}$}:
\newline
\begin{eqnarray}
\label{h2dmdm}
\Gamma({h_{2}} \rightarrow \phi_{DM}^\dagger \phi_{DM}) &=& 
\dfrac{g_{h_{2}\phi_{DM}^\dagger\phi_{DM}}^{2}}{16\,\pi\,M_{h_2}}
\,\sqrt{1 - \dfrac{4 M_{DM}^{2}}{M_{h_2}^{2}}}\,.
\end{eqnarray}

\item \underline{$h_{2}$ $\rightarrow$ $f\bar{f}$}:
\newline
\begin{eqnarray}
g_{h_{2}ff} &=& \dfrac{M_{f}}{v}\,\sin\alpha \,,\nn \\
\Gamma(h_{2} \rightarrow {\rm f} \,\bar{\rm f})
&=& \dfrac{C_{ASB}\,\,n_{c}\, M_{h_{2}}\, g_{h_{2}ff}}{8\pi}
\,\left(1 - \dfrac{4 M_{f}^{2}}{M_{h_2}^{2}}\right)
^{{3}/{2}}\,,
\end{eqnarray}
$n_c$ is the color charge, for leptons it is $1$ and for
quarks it is  $3$.
\vskip 1cm
Total decay width of the extra Higgs $h_2$ in the present case
is,
\begin{eqnarray}
&&\Gamma_{h_2} = \sum_{V=W,Z}\Gamma(h_2 \rightarrow VV) +
\Gamma({h_{2}} \rightarrow {h_1}{h_1})
+ \Gamma({h_{2}} \rightarrow \phi_{DM} \phi_{DM})
+ \sum_{f}\Gamma(h_{2} \rightarrow {f} \,\bar{f})\,. \nn \\
\end{eqnarray}

\underline{\bf Total decay width of $h_1$}:\\
\begin{eqnarray}
\label{h1dmdm}
\Gamma({h_{1}} \rightarrow \phi_{DM}^{\dagger} \phi_{DM}) &=& 
\dfrac{C_{ASB}\,\,g_{h_{1}\phi_{DM}^{\dagger}\phi_{DM}}^{2}}{16\,\pi\,M_{h_1}}
\,\sqrt{1 - \dfrac{4 M_{DM}^{2}}{M_{h_1}^{2}}}\,.
\end{eqnarray}
Total decay width of SM-like Higgs boson is,
\begin{eqnarray}
\Gamma_{h_1} = \cos^{2}\alpha \,\Gamma_{\rm SM} +
\Gamma({h_{1}} \rightarrow \phi_{DM}^{\dagger} \phi_{DM})\,\,,
\end{eqnarray}
where $\Gamma_{\rm SM}$ is the total decay width of
SM Higgs boson.
\vskip 1cm
\underline{\bf Total decay width of $Z_{BL}$}:\\
\begin{eqnarray}
\Gamma(Z_{BL} \rightarrow f \bar{f}) &=&
\frac{M_{Z_{BL}}}{12 \pi} n_{c} (q_{f}\,g_{BL})^{2}
\left( 1 + \frac{2 M_{f}^{2}}{M_{Z_{BL}}^{2}}
\right) \sqrt{1 - \frac{4 M_{f}^{2}}{M_{Z_{BL}}^{2}}}\,,\nn\\
\Gamma(Z_{BL} \rightarrow \nu_{x} \bar{\nu_{x}})
&=& \frac{M_{Z_{BL}}}{24 \pi}  g_{BL}^{2}
\left( 1 - \frac{4 M_{\nu_{x}}^{2}}{M_{Z_{BL}}^{2}} \right)^{{3}/{2}}\,,\nn\\
\Gamma(Z_{BL} \rightarrow N_{x} \bar{N_{x}})
&=& \frac{M_{Z_{BL}}}{24 \pi}  g_{BL}^{2}
\left( 1 - \frac{4 M_{N_{x}}^{2}}{M_{Z_{BL}}^{2}} \right)^{{3}/{2}}\,,\nn \\
\Gamma(\zbl \rightarrow {\dm}^\dagger \dm) &=&
\dfrac{\gbl^2 \nbl^2 \mzbl}{48\,\pi}\left(1-\dfrac{4\,M^2_{DM}}
{\mzbl}\right)^{3/2} \,.
\label{zbldmdm}
\end{eqnarray}
\vskip 1cm
Total decay width of the extra neutral gauge boson $Z_{BL}$ is,
\begin{eqnarray}
\Gamma_{Z_{BL}} &=&\sum_{f}  \Gamma(Z_{BL} \rightarrow f \bar{f}) +
\Gamma(Z_{BL} \rightarrow \nu_{x} \bar{\nu_{x}}) + 
\Gamma(Z_{BL} \rightarrow N_{x} \bar{N_{x}})+
\Gamma(\zbl \rightarrow {\dm}^\dagger \dm)\,.\nn\\
\end{eqnarray}
\end{itemize}
\newpage
\section{Analytical Expression of relevant Cross sections}
\label{App:AppendixA}
Here we will give the expressions of cross sections
for all relevant processes which take part in the
FIMP DM production. 
\begin{itemize}

\item \underline{$h_{1}$\,$h_{1}$ $\rightarrow
\phi_{DM}^{\dagger} \,\phi_{DM}$}\,:

\begin{eqnarray}
&&g_{h_{1}h_{1}h_{1}} = -3\,[2\,v\lambda_{h}\cos^{3}\alpha
+ 2\,v_{BL}\,\lambda_{H}\sin^{3}\alpha +
\lambda_{hH}\sin\alpha\,\cos\alpha\,
(v\sin\alpha + v_{BL}\cos\alpha)],\nn \\
&&g_{h_{1}h_{1}h_{2}} = [6\,v\lambda_{h}\cos^{2}\alpha\sin\alpha -
6\,v_{BL}\lambda_{H}\sin^{2}\alpha\,\cos\alpha 
-(2-3\,\sin^{2}\alpha)\,v\,\lambda_{hH}\,\sin\alpha
\nn\\ 
&&~~~~~~~~~-(1-3\sin^{2}\alpha)v_{BL}\,
\lambda_{hH}\cos\alpha] \,,\label{h2h1h1} \\
&&g_{h_{1}h_{1}\phi_{DM}^{\dagger} \phi_{DM}} =
- (\lambda_{Dh} \cos^{2}\alpha + \lambda_{DH}
\sin^{2}\alpha) \,,\nn \\
&&M_{h_{1} h_{1}} = \left(\dfrac{C_{ASB}\,g_{h_{1}h_{1}h_{1}}\,
\,g_{h_{1}\phi_{DM}^{\dagger} \phi_{DM}}}{(s-M_{h_1}^{2}) +
i M_{h_1} \Gamma_{h_1}} + \dfrac{g_{h_{1}h_{1}h_{2}}
\,\,g_{h_{2}\phi_{DM}^{\dagger} \phi_{DM}}}{(s-M_{h_2}^{2}) +
i M_{h_2} \Gamma_{h_2}}\right) 
- g_{h_{1}h_{1}\phi_{DM}^{\dagger} \phi_{DM}}\,, \nn \\
&&\sigma_{h_{1}h_{1} \rightarrow \phi_{DM}^{\dagger} \phi_{DM}} =
\dfrac{1}{16 \pi s}\,\,\sqrt{\dfrac{s - 4M_{DM}^{2}}{s - 4M_{h_1}^{2}}}
\,\,\,|M_{h_{1}h_{1}}|^{2}\,.
\end{eqnarray}
\item \underline{$h_{2}$\,$h_{2}$ $\rightarrow
\phi_{DM}^{\dagger}\,\phi_{DM}$}\,:

\begin{eqnarray}
&&g_{h_{2}h_{2}h_{2}} = 3\,[2\,v\lambda_{h}\sin^{3}
\alpha - 2\,v_{BL}\lambda_{H}\cos^{3}\alpha +
\lambda_{hH}\sin\alpha\cos\alpha\,
(v\cos\alpha - v_{BL}\sin\alpha)],\nn \\
&&g_{h_{2}h_{2}h_{1}} = -[6\,v\lambda_{h}\sin^{2}
\alpha\cos\alpha + 6\,v_{BL}\lambda_{H}\cos^{2}\alpha\sin\alpha 
-(2-3\,\sin^{2}\alpha)v_{BL}\lambda_{hH}\sin\alpha
\,\,\nn \\
&&~~~~~~~~~+(1-3\sin^{2}\alpha)v\lambda_{hH}\cos\alpha]\,, \nn \\
&&g_{h_{2}h_{2}\phi_{DM}^{\dagger} \phi_{DM}} = -\,(\lambda_{Dh} \sin^{2}\alpha
+ \lambda_{DH} \cos^{2}\alpha) \,,\nn \\
&&M_{h_{2}h_{2}} = \left(C_{ASB}\,\,\dfrac{g_{h_{2}h_{2}h_{1}}\,
\,g_{h_{1}\phi_{DM}^{\dagger} \phi_{DM}}}{(s-M_{h_1}^{2}) +
i M_{h_1} \Gamma_{h_1}} + \dfrac{g_{h_{2}h_{2}h_{2}}
\,\,g_{h_{2}\phi_{DM}^{\dagger} \phi_{DM}}}{(s-M_{h_2}^{2}) +
i M_{h_2} \Gamma_{h_2}}\right) 
- g_{h_{2}h_{2}\phi_{DM}^{\dagger} \phi_{DM}}\,, \nn \\
&&\sigma_{h_{2}h_{2} \rightarrow \phi_{DM}^{\dagger} \phi_{DM}} =
\dfrac{1}{16 \pi s}\,\,\sqrt{\dfrac{s - 4M_{DM}^{2}}{s - 4M_{h_2}^{2}}}
\,\,\,|M_{h_{2}h_{2}}|^{2}\,.
\end{eqnarray}

\newpage
\item \underline{$h_{1}$\,$h_{2}$ $\rightarrow
\phi_{DM}^{\dagger}\,\phi_{DM}$}\,:

\begin{eqnarray}
&&g_{h_{1}h_{1}h_{2}} = [6\,v\lambda_{h}\cos^{2}\alpha\,\sin\alpha
- 6\,v_{BL}\,\lambda_{H}\sin^{2}\alpha\,\cos\alpha 
-(2-3\,\sin^{2}\alpha)\,v\,\lambda_{hH}\sin\alpha \nn \\
&&~~~~~~~~~
-(1-3\sin^{2}\alpha)v_{BL}\,\lambda_{hH}\cos\alpha] \nn \\
&&g_{h_{2}h_{2}h_{1}} = -[6\,v\lambda_{h}\sin^{2}\alpha\,\cos\alpha
+ 6\,v_{BL}\lambda_{H}\cos^{2}\alpha\sin\alpha 
-(2-3\,\sin^{2}\alpha)\,v_{BL}\,\lambda_{hH}\sin\alpha \nn \\
&&~~~~~~~~~ 
+(1-3\sin^{2}\alpha)v\lambda_{hH}\cos\alpha]\,, \nn \\
&&g_{h_{1}h_{2}\phi_{DM}^{\dagger} \phi_{DM}} = \,\sin\alpha \cos\alpha(\lambda_{Dh} -
\lambda_{DH}) \,,\nn \\
&&M_{h_{1}h_{2}} = -C_{ASB}\left(g_{h_{1}h_{2}\phi_{DM}^{\dagger} \phi_{DM}}
-\dfrac{g_{h_{2}h_{2}h_{1}}
\,\,g_{h_{2}\phi_{DM}^{\dagger} \phi_{DM}}}{(s-M_{h_2}^{2}) +
i M_{h_2} \Gamma_{h_2}}\right)
+ \dfrac{g_{h_{1}h_{1}h_{2}}\,
\,g_{h_{1}\phi_{DM}^{\dagger} \phi_{DM}}}{(s-M_{h_1}^{2}) +
i M_{h_1} \Gamma_{h_1}}, \nn \\
&&\sigma_{h_{1}h_{2} \rightarrow \phi_{DM}^{\dagger} \phi_{DM}} =
\dfrac{1}{16 \pi s}\,\,\sqrt{\dfrac{s(s - 4M_{DM}^{2})}
{(s-(M_{h_1}+M_{h_2})^{2})(s - (M_{h_2} - M_{h_1})^{2})}}
\,\,\,|M_{h_{1}h_{2}}|^{2}\,.
\end{eqnarray}

\item \underline{$W^{+}$\,$W^{-}$ $\rightarrow
\phi_{DM}^{\dagger}\,\phi_{DM}$}\,:
\begin{eqnarray}
g_{h_{1}WW} &=& \dfrac{2M_{W}^{2} \cos\alpha}{v}\,,\nn \\
g_{h_{2}WW} &=& -\dfrac{2M_{W}^{2} \sin\alpha}{v}, \nn \\
A_{WW} &=& C_{ASB}\left(\dfrac{g_{h_{1}WW}\,\,g_{h_{1}\phi_{DM}^{\dagger}
\phi_{DM}}}{(s-M_{h_1}^{2}) + i M_{h_1} \Gamma_{h_1}}
+ \dfrac{g_{h_{2}WW}\,g_{h_{2}\phi_{DM}^{\dagger} \phi_{DM}}}
{(s-M_{h_2}^{2}) + i M_{h_2} \Gamma_{h_2}}\right), \nn \\
M_{WW} &=& \dfrac{2}{9}\,\left(1 + \dfrac{(s - 2M_{W}^{2})^{2}}
{8M_{W}^{4}}\right)\,A_{WW}, \nn \\
\sigma_{WW \rightarrow \phi_{DM}^{\dagger} \phi_{DM}} &=& \dfrac{1}{16 \pi s}\,\,
\sqrt{\dfrac{s - 4M_{DM}^{2}}{s - 4M_{W}^{2}}} \,\,\,|M_{WW}|^{2}\,.
\end{eqnarray}

\item \underline{$Z\,Z \rightarrow
\phi_{DM}^{\dagger}\,\phi_{DM}$}\,:
\begin{eqnarray}
g_{h_{1}ZZ} &=& \dfrac{2M_{Z}^{2} \cos \alpha}{v}\,,\nn \\
g_{h_{2}ZZ} &=& -\dfrac{2M_{Z}^{2} \sin \alpha}{v}, \nn \\
A_{ZZ} &=& C_{ASB}\left(\dfrac{g_{h_{1}ZZ}\,g_{h_{1}\phi_{DM}^{\dagger}
\phi_{DM}}}{(s-M_{h_1}^{2}) + i M_{h_1} \Gamma_{h_1}}
+ \dfrac{g_{h_{2}ZZ}\,g_{h_{2}\phi_{DM}^{\dagger} \phi_{DM}}}
{(s-M_{h_2}^{2}) + i M_{h_2} \Gamma_{h_2}}\right), \nn \\
M_{ZZ} &=& \dfrac{2}{9}\,\,\left(1 + \dfrac{(s - 2M_{Z}^{2})^{2}}
{8M_{Z}^{4}}\right)\,A_{ZZ}, \nn \\
\sigma_{ZZ \rightarrow \phi_{DM}^{\dagger} \phi_{DM}}
&=& \dfrac{1}{16 \pi s}\,\,\sqrt{\dfrac{s - 4M_{DM}^{2}}
{s - 4M_{Z}^{2}}} \,\,\,|M_{ZZ}|^{2}\,.
\end{eqnarray}

\item \underline{$t\bar{t}$\, $\rightarrow
\phi_{DM}^\dagger \,\phi_{DM}$}\,:

\begin{eqnarray}
g_{h_{1}tt} &=& -\dfrac{M_{t}}{v} \cos\alpha \,, \nn \\
g_{h_{2}tt} &=& \dfrac{M_{t}}{v} \sin\alpha \,, \nn \\
g_{\zbl tt} &=& \dfrac{g_{BL}}{3} \,, \nn \\
M_{tt} &=& C_{ASB}\left(\dfrac{g_{h_{1}tt}\,
\,g_{h_{1}\phi_{DM}^{\dagger} \phi_{DM}}}{(s-M_{h_1}^{2}) +
i M_{h_1} \Gamma_{h_1}} + \dfrac{g_{h_{2}tt}
\,\,g_{h_{2}\phi_{DM}^{\dagger} \phi_{DM}}}{(s-M_{h_2}^{2}) +
i M_{h_2} \Gamma_{h_2}}\right), \nn \\
\sigma_{{t}{\bar{t}} \rightarrow \phi_{DM}^{\dagger} \phi_{DM}}^{h_{1} h_{2}} &=&
\dfrac{1}{32 \pi s\,n_{c}}\,\,(s - 4M_{t}^{2})\,\,
\sqrt{\dfrac{s - 4M_{DM}^{2}}{s - 4M_{t}^{2}}}
\,\,\,|M_{tt}|^{2}\,, \nn \\
\sigma_{{t}{\bar{t}} \rightarrow \phi_{DM}^{\dagger} \phi_{DM}}^{\zbl} &=&
\dfrac{g_{BL}^{2}\, n_{BL}^{2}}{64 \pi s\,n_{c}}\,\,
\sqrt{\dfrac{s - 4M_{DM}^{2}}{s - 4M_{t}^{2}}}\,\,
\frac{s\, (s - 4M_{DM}^{2})\, g_{\zbl t t}^{2}}
{(s - M^{2}_{\zbl})^{2}
+ \Gamma^{2}_{\zbl} M^{2}_{\zbl}}\,,\nn \\
\sigma_{{t}{\bar{t}} \rightarrow \phi_{DM}^{\dagger} \phi_{DM}} &=&
\sigma_{{t}{\bar{t}} \rightarrow \phi_{DM}^{\dagger} \phi_{DM}}^{h_1 h_2}
+ \sigma_{{t}{\bar{t}} \rightarrow \phi_{DM}^{\dagger} \phi_{DM}}^{\zbl}\,.
\end{eqnarray}

\item \underline{$N_{i}$\,$N_{i}$\,$\rightarrow
\phi_{DM}^\dagger\,\phi_{DM}\,\,(i=1,\, 2,\,3)$}\,:

\begin{eqnarray}
g_{h_{1}N_{i}N_{i}} &=& \frac{y_{N_i}\sin \alpha}{\sqrt{2}} \,, \nn \\
g_{h_{2}N_{i}N_{i}} &=& \frac{y_{N_i}\cos \alpha}{\sqrt{2}} \,, \nn \\
M_{N_{i}N_{i}} &=& \dfrac{C_{ASB}\,\,g_{h_{1}N_{i}N_{i}}\,
\,g_{h_{1}\phi_{DM}^{\dagger} \phi_{DM}}}{(s-M_{h_1}^{2}) +
i M_{h_1} \Gamma_{h_1}} + \dfrac{g_{h_{2}N_{i}N_{i}}
\,\,g_{h_{2}\phi_{DM}^{\dagger} \phi_{DM}}}{(s-M_{h_2}^{2}) +
i M_{h_2} \Gamma_{h_2}}\,, \nn \\
\sigma_{N_{i}N_{i} \rightarrow \phi_{DM}^{\dagger} \phi_{DM}}^{h_1 h_2} &=&
\dfrac{(s - 4\,M_{N_j}^{2})}{32 \pi s}\,
\sqrt{\dfrac{(s - 4M_{DM}^{2})}
{(s-4 M_{N_i}^{2})}}
\,\,\,|M_{N_{i}N_{i}}|^{2}\,,\nn \\
\sigma_{N_{i}N_{i} \rightarrow \phi_{DM}^{\dagger} \phi_{DM}}^{\zbl} &=&
\dfrac{g_{\mu\tau}^{4}n_{\mu\tau}^{2}}{192 \pi s} \,
\sqrt{\dfrac{s-4M_{DM}^{2}}{s-4M_{N_i}^{2}}}
\dfrac{(s-4M_{DM}^{2})(s-4M_{N_i}^{2})}{(s - M_{\zbl}^{2})^{2} +
\Gamma_{\zbl}^{2}M_{\zbl}^{2}}\,, \nn \\ 
\sigma_{N_{i}N_{i} \rightarrow \phi_{DM}^{\dagger} \phi_{DM}} &=&
\sigma_{N_{i}N_{i} \rightarrow \phi_{DM}^{\dagger} \phi_{DM}}^{h_1 h_2}
+\sigma_{N_{i}N_{i} \rightarrow \phi_{DM}^{\dagger} \phi_{DM}}^{\zbl}\,.
\end{eqnarray}

\item \underline{$\zbl\,\zbl \rightarrow
\phi_{DM}^{\dagger}\,\phi_{DM}$}\,:
\begin{eqnarray}
g_{h_{1} \zbl \zbl} &=& \dfrac{2M_{\zbl}^{2} \sin \alpha}{v}\,,\nn \\
g_{h_{2}\zbl \zbl} &=& -\dfrac{2M_{\zbl}^{2} \cos \alpha}{v}, \nn \\
g_{\zbl \zbl \dm^{\dagger} \dm} &=& 2\, g_{BL}^{2} n_{BL}^{2}\,,\nn \\
A_{\zbl \zbl} &=& C_{ASB}\left(\dfrac{g_{h_{1}\zbl \zbl}\,g_{h_{1}
\phi_{DM}^{\dagger} \phi_{DM}}}{(s-M_{h_1}^{2}) + i M_{h_1} \Gamma_{h_1}}
+ \dfrac{g_{h_{2}\zbl \zbl}\,g_{h_{2}\phi_{DM}^{\dagger} \phi_{DM}}}
{(s-M_{h_2}^{2}) + i M_{h_2} \Gamma_{h_2}} -  g_{\zbl \zbl \dm^{\dagger} \dm} \right), \nn \\
M_{\zbl \zbl} &=& \dfrac{2}{9}\,\,\left(1 + \dfrac{(s - 2M_{\zbl}^{2})^{2}}
{8M_{\zbl}^{4}}\right)\,A_{\zbl \zbl}, \nn \\
\sigma_{\zbl \zbl \rightarrow \phi_{DM}^{\dagger}\phi_{DM}}
&=& \dfrac{1}{16 \pi s}\,\,\sqrt{\dfrac{s - 4M_{DM}^{2}}
{s - 4M_{\zbl}^{2}}} \,\,\,|M_{\zbl \zbl}|^{2}\,.
\end{eqnarray}
\end{itemize}



\begin{thebibliography}{99}
\bibitem{Fukuda:1998mi} 
Y.~Fukuda {\it et al.} [Super-Kamiokande Collaboration],
``{\it Evidence for oscillation of atmospheric neutrinos}'',
Phys.\ Rev.\ Lett.\  {\bf 81}, 1562 (1998)
[hep-ex/9807003].
\bibitem{Ahmad:2002jz} 
Q.~R.~Ahmad {\it et al.} [SNO Collaboration],
``{\it Direct evidence for neutrino flavor transformation
from neutral current interactions in the Sudbury Neutrino Observatory}'',
Phys.\ Rev.\ Lett.\  {\bf 89}, 011301 (2002)
[nucl-ex/0204008].
\bibitem{Eguchi:2002dm} 
K.~Eguchi {\it et al.} [KamLAND Collaboration],
``{\it First results from KamLAND: Evidence for
reactor anti-neutrino disappearance}'',
Phys.\ Rev.\ Lett.\  {\bf 90}, 021802 (2003)
[hep-ex/0212021].
\bibitem{An:2015nua} 
  F.~P.~An {\it et al.} [Daya Bay Collaboration],
  ``{\it Measurement of the Reactor Antineutrino Flux and Spectrum at Daya Bay}'',
  Phys.\ Rev.\ Lett.\  {\bf 116}, no. 6, 061801 (2016)
  [arXiv:1508.04233 [hep-ex]].

\bibitem{RENO:2015ksa} 
  J.~H.~Choi {\it et al.} [RENO Collaboration],
  ``{\it Observation of Energy and Baseline Dependent Reactor
  Antineutrino Disappearance in the RENO Experiment}'',
  Phys.\ Rev.\ Lett.\  {\bf 116}, no. 21, 211801 (2016)
  [arXiv:1511.05849 [hep-ex]].
\bibitem{Abe:2014bwa} 
  Y.~Abe {\it et al.} [Double Chooz Collaboration],
  ``{\it Improved measurements of the neutrino mixing angle
  $\theta_{13}$ with the Double Chooz detector}'',
  JHEP {\bf 1410}, 086 (2014)
  Erratum: [JHEP {\bf 1502}, 074 (2015)]
  [arXiv:1406.7763 [hep-ex]].
\bibitem{Abe:2015awa} 
  K.~Abe {\it et al.} [T2K Collaboration],
  ``{\it Measurements of neutrino oscillation in appearance
  and disappearance channels by the T2K experiment with 6.6$\times$10$^{20}$ protons on target}'',
  Phys.\ Rev.\ D {\bf 91}, no. 7, 072010 (2015)
  [arXiv:1502.01550 [hep-ex]].

\bibitem{Salzgeber:2015gua} 
  M.~Ravonel Salzgeber [T2K Collaboration],
  ``{\it Anti-neutrino oscillations with T2K}'',
  arXiv:1508.06153 [hep-ex].
\bibitem{Adamson:2016tbq} 
  P.~Adamson {\it et al.} [NOvA Collaboration],
  ``{\it First measurement of electron neutrino appearance in NOvA}'',
  Phys.\ Rev.\ Lett.\  {\bf 116}, no. 15, 151806 (2016)
  [arXiv:1601.05022 [hep-ex]].
\bibitem{Adamson:2016xxw} 
  P.~Adamson {\it et al.} [NOvA Collaboration],
  ``{\it First measurement of muon-neutrino disappearance in NOvA}'',
  Phys.\ Rev.\ D {\bf 93}, no. 5, 051104 (2016)
  [arXiv:1601.05037 [hep-ex]].

\bibitem{xyz}
Talk on ``{\it Recent results from $T2K$ and Future Prospects}'',\\
https://indico.cern.ch/event/432527/contributions/2143636/


\bibitem{Gando:2012zm} 
  A.~Gando {\it et al.} [KamLAND-Zen Collaboration],
  ``{\it Limit on Neutrinoless $\beta\beta$ Decay of $^{136}$Xe from the First Phase of KamLAND-Zen and Comparison with the Positive Claim in $^{76}$Ge}'',
  Phys.\ Rev.\ Lett.\  {\bf 110}, no. 6, 062502 (2013)
  [arXiv:1211.3863 [hep-ex]].

\bibitem{Agostini:2013mzu} 
  M.~Agostini {\it et al.} [GERDA Collaboration],
  ``{\it Results on Neutrinoless Double-$\beta$ Decay
  of $^{76}$Ge from Phase I of the GERDA Experiment}'',
  Phys.\ Rev.\ Lett.\  {\bf 111}, no. 12, 122503 (2013)
  [arXiv:1307.4720 [nucl-ex]].

\bibitem{Albert:2014awa} 
  J.~B.~Albert {\it et al.} [EXO-200 Collaboration],
  ``{\it Search for Majorana neutrinos with the first two years of EXO-200 data}'',
  Nature {\bf 510}, 229 (2014)
  [arXiv:1402.6956 [nucl-ex]].

\bibitem{Asakura:2014lma} 
  K.~Asakura {\it et al.} [KamLAND-Zen Collaboration],
  ``{\it Results from KamLAND-Zen}'',
  AIP Conf.\ Proc.\  {\bf 1666}, 170003 (2015)
  [arXiv:1409.0077 [physics.ins-det]].


\bibitem{xyz1}
M. Agostini [GERDA Collaboration], ``{\it 	First
results from GERDA Phase II}'',
http://neutrino2016.iopconfs.org/programme.



\bibitem{KamLAND-Zen:2016pfg} 
A.~Gando {\it et al.} [KamLAND-Zen Collaboration],
``{\it Search for Majorana Neutrinos near the Inverted Mass Hierarchy Region with KamLAND-Zen}'',
Phys.\ Rev.\ Lett.\  {\bf 117}, no. 8, 082503 (2016)
Addendum: [Phys.\ Rev.\ Lett.\  {\bf 117}, no. 10, 109903 (2016)]
[arXiv:1605.02889 [hep-ex]].


\bibitem{Sofue:2000jx} 
Y.~Sofue and V.~Rubin,
``{\it Rotation curves of spiral galaxies}'',
Ann.\ Rev.\ Astron.\ Astrophys.\  {\bf 39}, 137 (2001)
[astro-ph/0010594].

\cite{Bartelmann:1999yn}
\bibitem{Bartelmann:1999yn} 
M.~Bartelmann and P.~Schneider,
``{\it Weak gravitational lensing}'',
Phys.\ Rept.\  {\bf 340}, 291 (2001)
[astro-ph/9912508].

\bibitem{Clowe:2003tk} 
D.~Clowe, A.~Gonzalez and M.~Markevitch,
``{\it Weak lensing mass reconstruction of the interacting
cluster 1E0657-558: Direct evidence for the existence of dark matter}'',
Astrophys.\ J.\  {\bf 604}, 596 (2004)
[astro-ph/0312273].

\bibitem{Biviano:1996bg} 
  A.~Biviano, P.~Katgert, A.~Mazure, M.~Moles, R.~denHartog, J.~Perea and P.~Focardi,
  ``{\it The eso nearby abell cluster survey. 3.
  Distribution and kinematics of emission-line galaxies}'',
  Astron.\ Astrophys.\  {\bf 321}, 84 (1997)
  [astro-ph/9610168].
\bibitem{Kahlhoefer:2013dca} 
  F.~Kahlhoefer, K.~Schmidt-Hoberg, M.~T.~Frandsen and S.~Sarkar,
  ``{\it Colliding clusters and dark matter self-interactions}'',
  Mon.\ Not.\ Roy.\ Astron.\ Soc.\  {\bf 437}, no. 3, 2865 (2014)
  [arXiv:1308.3419 [astro-ph.CO]].

\bibitem{Harvey:2015hha}
  D.~Harvey, R.~Massey, T.~Kitching, A.~Taylor and E.~Tittley,
  ``{\it The non-gravitational interactions of dark matter in colliding galaxy clusters}'',
  Science {\bf 347} (2015) 1462
  [arXiv:1503.07675 [astro-ph.CO]].


\bibitem{Hinshaw:2012aka} 
G.~Hinshaw {\it et al.} [WMAP Collaboration],
``{\it Nine-Year Wilkinson Microwave Anisotropy Probe (WMAP)
Observations: Cosmological Parameter Results}'',
Astrophys.\ J.\ Suppl.\  {\bf 208}, 19 (2013)
[arXiv:1212.5226 [astro-ph.CO]].

\bibitem{Ade:2015xua} 
P.~A.~R.~Ade {\it et al.} [Planck Collaboration],
``{\it Planck 2015 results. XIII. Cosmological parameters}'',
arXiv:1502.01589 [astro-ph.CO].

\bibitem{Aprile:2012zx} 
  E.~Aprile [XENON1T Collaboration],
  ``{\it The XENON1T Dark Matter Search Experiment}'',
  Springer Proc.\ Phys.\  {\bf 148}, 93 (2013)
  [arXiv:1206.6288 [astro-ph.IM]].

\bibitem{Akerib:2015rjg} 
  D.~S.~Akerib {\it et al.} [LUX Collaboration],
  ``{\it Improved Limits on Scattering of Weakly Interacting Massive Particles from Reanalysis of 2013 LUX Data}'',
  Phys.\ Rev.\ Lett.\  {\bf 116}, no. 16, 161301 (2016)
  [arXiv:1512.03506 [astro-ph.CO]].

\bibitem{Ahmed:2010wy} 
  Z.~Ahmed {\it et al.} [CDMS-II Collaboration],
  ``{\it Results from a Low-Energy Analysis of the CDMS II Germanium Data}'',
  Phys.\ Rev.\ Lett.\  {\bf 106}, 131302 (2011)
  [arXiv:1011.2482 [astro-ph.CO]].

\bibitem{Agnese:2014aze} 
  R.~Agnese {\it et al.} [SuperCDMS Collaboration],
  ``{\it Search for Low-Mass Weakly Interacting Massive
  Particles with SuperCDMS}'',
  Phys.\ Rev.\ Lett.\  {\bf 112}, no. 24, 241302 (2014)
  [arXiv:1402.7137 [hep-ex]].

\bibitem{Gondolo:1990dk} 
  P.~Gondolo and G.~Gelmini,
  ``{\it Cosmic abundances of stable particles: Improved analysis}''
  Nucl.\ Phys.\ B {\bf 360}, 145 (1991).
  
\bibitem{Jungman:1995df} 
  G.~Jungman, M.~Kamionkowski and K.~Griest,
  ``{Supersymmetric dark matter}'',
  Phys.\ Rept.\  {\bf 267}, 195 (1996)
  [hep-ph/9506380].

\bibitem{ArkaniHamed:2008qn} 
  N.~Arkani-Hamed, D.~P.~Finkbeiner, T.~R.~Slatyer and N.~Weiner,
  ``{\it A Theory of Dark Matter}'',
  Phys.\ Rev.\ D {\bf 79}, 015014 (2009)
  [arXiv:0810.0713 [hep-ph]].

\bibitem{Arcadi:2017kky} 
  G.~Arcadi, M.~Dutra, P.~Ghosh, M.~Lindner,
  Y.~Mambrini, M.~Pierre, S.~Profumo and F.~S.~Queiroz,
  ``{\it The Waning of the WIMP? A Review of Models, Searches, and Constraints}'''
  arXiv:1703.07364 [hep-ph].
\bibitem{Hall:2009bx} 
  L.~J.~Hall, K.~Jedamzik, J.~March-Russell and S.~M.~West,
  ``{\it Freeze-In Production of FIMP Dark Matter}''
  JHEP {\bf 1003}, 080 (2010)
  [arXiv:0911.1120 [hep-ph]].

\bibitem{Yaguna:2011qn} 
  C.~E.~Yaguna,
  ``{\it The Singlet Scalar as FIMP Dark Matter}''
  JHEP {\bf 1108}, 060 (2011)
  [arXiv:1105.1654 [hep-ph]].

\bibitem{Molinaro:2014lfa} 
  E.~Molinaro, C.~E.~Yaguna and O.~Zapata,
  ``{\it FIMP realization of the scotogenic model}''
  JCAP {\bf 1407}, 015 (2014)
  [arXiv:1405.1259 [hep-ph]].


\bibitem{Biswas:2015sva} 
  A.~Biswas, D.~Majumdar and P.~Roy,
  ``{\it Nonthermal two component dark matter model for
  Fermi-LAT $\gamma$-ray excess and $3.55$ keV X-ray line}''
  JHEP {\bf 1504}, 065 (2015)
  [arXiv:1501.02666 [hep-ph]].

\bibitem{Merle:2015oja} 
  A.~Merle and M.~Totzauer,
  ``{\it keV Sterile Neutrino Dark Matter from Singlet Scalar
  Decays: Basic Concepts and Subtle Features}''
  JCAP {\bf 1506}, 011 (2015)
  [arXiv:1502.01011 [hep-ph]].

\bibitem{Shakya:2015xnx} 
  B.~Shakya,
  ``{\it Sterile Neutrino Dark Matter from Freeze-In}''
  Mod.\ Phys.\ Lett.\ A {\bf 31}, no. 06, 1630005 (2016)
  [arXiv:1512.02751 [hep-ph]].

\bibitem{Biswas:2016bfo} 
  A.~Biswas and A.~Gupta,
  ``{\it Freeze-in Production of Sterile Neutrino Dark
  Matter in U(1)$_{\rm B-L}$ Model}''
  JCAP {\bf 1609}, no. 09, 044 (2016)
  [arXiv:1607.01469 [hep-ph]].

\bibitem{Konig:2016dzg} 
  J.~König, A.~Merle and M.~Totzauer,
  ``{\it keV Sterile Neutrino Dark Matter from Singlet
  Scalar Decays: The Most General Case}''
  JCAP {\bf 1611}, no. 11, 038 (2016)
  [arXiv:1609.01289 [hep-ph]].

\bibitem{Biswas:2016iyh} 
  A.~Biswas and A.~Gupta,
  ``{\it Calculation of Momentum Distribution Function of a Non-thermal Fermionic Dark Matter}'',
  JCAP03(2017)033
  [arXiv:1612.02793 [hep-ph]].

\bibitem{Biswas:2016yjr} 
  A.~Biswas, S.~Choubey and S.~Khan,
  ``{\it FIMP and Muon ($g-2$) in a U$(1)_{L_{\mu}-L_{\tau}}$ Model}'',
  JHEP {\bf 1702}, 123 (2017)
  [arXiv:1612.03067 [hep-ph]].


\bibitem{Olive:2016xmw} 
  C.~Patrignani {\it et al.} [Particle Data Group],
  ``{\it Review of Particle Physics}'',
  Chin.\ Phys.\ C {\bf 40}, no. 10, 100001 (2016).

\bibitem{Sakharov:1967dj} 
  A.~D.~Sakharov,
  ``{\it Violation of CP Invariance, c Asymmetry, and Baryon Asymmetry of the Universe}'',
  Pisma Zh.\ Eksp.\ Teor.\ Fiz.\  {\bf 5}, 32 (1967)
  [JETP Lett.\  {\bf 5}, 24 (1967)]
  [Sov.\ Phys.\ Usp.\  {\bf 34}, 392 (1991)]
  [Usp.\ Fiz.\ Nauk {\bf 161}, 61 (1991)].
\bibitem{Mohapatra:1980qe} 
  R.~N.~Mohapatra and R.~E.~Marshak,
  ``{\it Local B-L Symmetry of Electroweak Interactions, Majorana Neutrinos and Neutron Oscillations}'',
  Phys.\ Rev.\ Lett.\  {\bf 44}, 1316 (1980)
  Erratum: [Phys.\ Rev.\ Lett.\  {\bf 44}, 1643 (1980)].
\bibitem{Georgi:1981pg} 
  H.~M.~Georgi, S.~L.~Glashow and S.~Nussinov,
  ``{\it Unconventional Model of Neutrino Masses}'',
  Nucl.\ Phys.\ B {\bf 193}, 297 (1981).
  
\bibitem{Wetterich:1981bx} 
  C.~Wetterich,
  ``{\it Neutrino Masses and the Scale of B-L Violation}'',
  Nucl.\ Phys.\ B {\bf 187}, 343 (1981).

\bibitem{Lindner:2011it} 
  M.~Lindner, D.~Schmidt and T.~Schwetz,
  ``{\it Dark Matter and neutrino masses from global U(1)$_{B-L}$ symmetry breaking}'',
  Phys.\ Lett.\ B {\bf 705}, 324 (2011)
  [arXiv:1105.4626 [hep-ph]].

\bibitem{Okada:2010wd} 
  N.~Okada and O.~Seto,
  ``{\it Higgs portal dark matter in the minimal gauged $U(1)_{B-L}$ model}'',
  Phys.\ Rev.\ D {\bf 82}, 023507 (2010)
  [arXiv:1002.2525 [hep-ph]].

\bibitem{Okada:2012sg} 
  N.~Okada and Y.~Orikasa,
  ``{\it Dark matter in the classically conformal B-L model}'',
  Phys.\ Rev.\ D {\bf 85}, 115006 (2012)
  [arXiv:1202.1405 [hep-ph]].
  
\bibitem{Basso:2012ti} 
  L.~Basso, O.~Fischer and J.~J.~van der Bij,
  ``{\it Natural Z$^{\prime}$ model with an inverse seesaw mechanism and leptonic dark matter}'',
  Phys.\ Rev.\ D {\bf 87}, no. 3, 035015 (2013)
  [arXiv:1207.3250 [hep-ph]].

\bibitem{Basak:2013cga} 
  T.~Basak and T.~Mondal,
  ``{\it Constraining Minimal $U(1)_{B-L}$ model from Dark Matter Observations}'',
  Phys.\ Rev.\ D {\bf 89}, 063527 (2014)
  [arXiv:1308.0023 [hep-ph]].
  
\bibitem{Sanchez-Vega:2014rka} 
  B.~L.~Sánchez-Vega, J.~C.~Montero and E.~R.~Schmitz,
  ``{\it Complex Scalar DM in a B-L Model}'',
  Phys.\ Rev.\ D {\bf 90}, no. 5, 055022 (2014)
  [arXiv:1404.5973 [hep-ph]].
 
\bibitem{Basak:2014sza} 
  T.~Mondal and T.~Basak,
  ``{\it Class of Higgs-portal Dark Matter models
  in the light of gamma-ray excess
  from Galactic center}'',
  Phys.\ Lett.\ B {\bf 744}, 208 (2015)
  [arXiv:1405.4877 [hep-ph]].
 
\bibitem{Chatrchyan:2012oaa} 
  S.~Chatrchyan {\it et al.} [CMS Collaboration],
  ``{\it Search for heavy narrow dilepton resonances in $pp$ collisions at $\sqrt{s}=7$ TeV and $\sqrt{s}=8$ TeV}'',
  Phys.\ Lett.\ B {\bf 720}, 63 (2013)
  [arXiv:1212.6175 [hep-ex]].
    
\bibitem{Aad:2014cka} 
  G.~Aad {\it et al.} [ATLAS Collaboration],
  ``{\it Search for high-mass dilepton resonances
  in pp collisions at $\sqrt{s}=8$ TeV with the ATLAS detector}'',
  Phys.\ Rev.\ D {\bf 90}, no. 5, 052005 (2014)
  [arXiv:1405.4123 [hep-ex]].
   
\bibitem{Guo:2015lxa} 
  J.~Guo, Z.~Kang, P.~Ko and Y.~Orikasa,
  ``{\it Accidental dark matter: Case in the scale invariant local B-L model}'',
  Phys.\ Rev.\ D {\bf 91}, no. 11, 115017 (2015)
  [arXiv:1502.00508 [hep-ph]].


\bibitem{Rodejohann:2015lca} 
  W.~Rodejohann and C.~E.~Yaguna,
  ``{\it Scalar dark matter in the ${\rm B-L}$ model}'',
  JCAP {\bf 1512}, no. 12, 032 (2015)
  [arXiv:1509.04036 [hep-ph]].
  
\bibitem{Okada:2016gsh} 
  N.~Okada and S.~Okada,
  ``{\it $Z^\prime_{BL}$ portal dark matter and LHC Run-2 results}'',
  Phys.\ Rev.\ D {\bf 93}, no. 7, 075003 (2016)
  [arXiv:1601.07526 [hep-ph]].
 
\bibitem{Patra:2016ofq} 
  S.~Patra, W.~Rodejohann and C.~E.~Yaguna,
  ``{\it A new B-L model without right-handed neutrinos}'',
  JHEP {\bf 1609}, 076 (2016)
  [arXiv:1607.04029 [hep-ph]].
\bibitem{Okada:2016tci} 
  N.~Okada and S.~Okada,
  ``{\it $Z^\prime$-portal right-handed neutrino dark matter
  in the minimal U(1)$_X$ extended Standard Model}'',
  Phys.\ Rev.\ D {\bf 95}, no. 3, 035025 (2017)
  [arXiv:1611.02672 [hep-ph]].
   
\bibitem{Buchmuller:1992qc} 
  W.~Buchmuller and T.~Yanagida,
  ``{\it Baryogenesis and the scale of B-L breaking}'',
  Phys.\ Lett.\ B {\bf 302}, 240 (1993).
 
\bibitem{Buchmuller:1996pa} 
  W.~Buchmuller and M.~Plumacher,
  ``{\it Baryon asymmetry and neutrino mixing}'',
  Phys.\ Lett.\ B {\bf 389}, 73 (1996)
  [hep-ph/9608308].
  
\bibitem{Dulaney:2010dj} 
  T.~R.~Dulaney, P.~Fileviez Perez and M.~B.~Wise,
  ``{\it Dark Matter, Baryon Asymmetry, and Spontaneous B and L Breaking}'',
  Phys.\ Rev.\ D {\bf 83}, 023520 (2011)
  [arXiv:1005.0617 [hep-ph]].
\bibitem{Capozzi:2016rtj} 
F.~Capozzi, E.~Lisi, A.~Marrone, D.~Montanino and A.~Palazzo,
``{\it Neutrino masses and mixings: Status of known
and unknown $3\nu$ parameters}'',
Nucl.\ Phys.\ B {\bf 908}, 218 (2016),
[arXiv:1601.07777 [hep-ph]].

\bibitem{Manton:1983nd} 
  N.~S.~Manton,
  ``{\it Topology in the Weinberg-Salam Theory}'',
  Phys.\ Rev.\ D {\bf 28}, 2019 (1983).

\bibitem{Klinkhamer:1984di} 
  F.~R.~Klinkhamer and N.~S.~Manton,
  ``{\it A Saddle Point Solution in the Weinberg-Salam Theory}'',
  Phys.\ Rev.\ D {\bf 30}, 2212 (1984).

\bibitem{Kuzmin:1985mm} 
  V.~A.~Kuzmin, V.~A.~Rubakov and M.~E.~Shaposhnikov,
  ``{\it On the Anomalous Electroweak Baryon Number
  Nonconservation in the Early Universe}'',
  Phys.\ Lett.\  {\bf 155B}, 36 (1985).

  
\bibitem{Khlebnikov:1988sr} 
  S.~Y.~Khlebnikov and M.~E.~Shaposhnikov,
  ``{\it The Statistical Theory of Anomalous
  Fermion Number Nonconservation}'',
  Nucl.\ Phys.\ B {\bf 308}, 885 (1988).


\bibitem{Arcadi:2013aba} 
G.~Arcadi and L.~Covi,
``{\it Minimal Decaying Dark Matter and the LHC}'',
JCAP {\bf 1308}, 005 (2013)
[arXiv:1305.6587 [hep-ph]].


\bibitem{Biswas:2016ewm} 
  A.~Biswas, S.~Choubey and S.~Khan,
  ``{\it Galactic gamma ray excess and dark matter phenomenology
  in a $U(1)_{B-L}$ model}'',
  JHEP {\bf 1608}, 114 (2016),
  [arXiv:1604.06566 [hep-ph]].
 
\bibitem{Chakrabarty:2015yia} 
  N.~Chakrabarty, D.~K.~Ghosh, B.~Mukhopadhyaya and I.~Saha,
  ``{\it Dark matter, neutrino masses and high scale validity of an inert Higgs doublet model}'',
  Phys.\ Rev.\ D {\bf 92}, no. 1, 015002 (2015)
  [arXiv:1501.03700 [hep-ph]].

\bibitem{Adler:1969gk} 
S.~L.~Adler,
``{\it Axial vector vertex in spinor electrodynamics}'',
Phys.\ Rev.\  {\bf 177}, 2426 (1969).
\bibitem{Bardeen:1969md} 
W.~A.~Bardeen,
``{\it Anomalous Ward identities in spinor field theories}'',
Phys.\ Rev.\  {\bf 184}, 1848 (1969).
\bibitem{Delbourgo:1972xb} 
R.~Delbourgo and A.~Salam,
``{\it The gravitational correction to pcac}'',
Phys.\ Lett.\ B {\bf 40}, 381 (1972).
\bibitem{Eguchi:1976db} 
T.~Eguchi and P.~G.~O.~Freund,
``{\it Quantum Gravity and World Topology}'',
Phys.\ Rev.\ Lett.\  {\bf 37}, 1251 (1976).

\bibitem{Jarlskog:1985ht} 
  C.~Jarlskog,
``{\it Commutator of the Quark Mass Matrices
in the Standard Electroweak Model and a Measure
of Maximal CP Violation}'',
  Phys.\ Rev.\ Lett.\  {\bf 55}, 1039 (1985).

\bibitem{Abe:2017vif} 
  K.~Abe {\it et al.} [T2K Collaboration],
  ``{\it Measurement of neutrino and antineutrino
 oscillations by the T2K experiment including a new additional
 sample of $\nu_e$ interactions at the far detector}'',
 arXiv:1707.01048 [hep-ex].
\bibitem{DellOro:2016tmg} 
  S.~Dell'Oro, S.~Marcocci, M.~Viel and F.~Vissani,
  ``{\it Neutrinoless double beta decay: 2015 review}'',
  Adv.\ High Energy Phys.\  {\bf 2016}, 2162659 (2016)
  [arXiv:1601.07512 [hep-ph]].
  
\bibitem{Plumacher:1996kc} 
  M.~Plumacher,
  ``{\it Baryogenesis and lepton number violation}'',
  Z.\ Phys.\ C {\bf 74}, 549 (1997),
  [hep-ph/9604229].

\bibitem{Iso:2010mv} 
  S.~Iso, N.~Okada and Y.~Orikasa,
  ``{\it Resonant Leptogenesis in the Minimal B-L Extended
  Standard Model at TeV}'',
  Phys.\ Rev.\ D {\bf 83}, 093011 (2011),
  [arXiv:1011.4769 [hep-ph]].

\bibitem{Buchmuller:2002rq} 
  W.~Buchmuller, P.~Di Bari and M.~Plumacher,
  ``{\it Cosmic microwave background, matter - antimatter
  asymmetry and neutrino masses}''
  Nucl.\ Phys.\ B {\bf 643}, 367 (2002)
  Erratum: [Nucl.\ Phys.\ B {\bf 793}, 362 (2008)]
  [hep-ph/0205349].
 
  
\bibitem{Pilaftsis:1997jf} 
  A.~Pilaftsis,
  ``{\it CP violation and baryogenesis due to heavy Majorana neutrinos}'',
  Phys.\ Rev.\ D {\bf 56}, 5431 (1997),
  [hep-ph/9707235].
  
\bibitem{Pilaftsis:2003gt} 
  A.~Pilaftsis and T.~E.~J.~Underwood,
  ``{\it Resonant leptogenesis}'',
  Nucl.\ Phys.\ B {\bf 692}, 303 (2004),
  [hep-ph/0309342].
   
\bibitem{Heeck:2016oda} 
  J.~Heeck and D.~Teresi,
  ``{\it Leptogenesis and neutral gauge bosons}'',
  Phys.\ Rev.\ D {\bf 94}, no. 9, 095024 (2016)
  [arXiv:1609.03594 [hep-ph]].

\bibitem{Flanz:1996fb} 
  M.~Flanz, E.~A.~Paschos, U.~Sarkar and J.~Weiss,
  ``{\it Baryogenesis through mixing of heavy Majorana neutrinos}'',
  Phys.\ Lett.\ B {\bf 389}, 693 (1996),
  [hep-ph/9607310].

\bibitem{Carena:2004xs} 
  M.~Carena, A.~Daleo, B.~A.~Dobrescu and T.~M.~P.~Tait,
  ``{\it $Z^\prime$ gauge bosons at the Tevatron}'',
  Phys.\ Rev.\ D {\bf 70}, 093009 (2004)
  [hep-ph/0408098].

 
\bibitem{Cacciapaglia:2006pk} 
  G.~Cacciapaglia, C.~Csaki, G.~Marandella and A.~Strumia,
  ``{\it The Minimal Set of Electroweak Precision Parameters}'',
  Phys.\ Rev.\ D {\bf 74}, 033011 (2006)
  [hep-ph/0604111].
  
\bibitem{Schael:2013ita} 
  S.~Schael {\it et al.} [ALEPH and DELPHI and L3 and
  OPAL and LEP Electroweak Collaborations],
  ``{\it Electroweak Measurements in Electron-Positron
  Collisions at W-Boson-Pair Energies at LEP}'',
  Phys.\ Rept.\  {\bf 532}, 119 (2013)
  [arXiv:1302.3415 [hep-ex]].
  
\bibitem{Edsjo:1997bg} 
  J.~Edsjo and P.~Gondolo,
  ``{\it Neutralino relic density including coannihilations}'',
  Phys.\ Rev.\ D {\bf 56}, 1879 (1997)
  [hep-ph/9704361].

 
\bibitem{Biswas:2011td} 
  A.~Biswas and D.~Majumdar,
  ``{\it The Real Gauge Singlet Scalar Extension
  of Standard Model: A Possible Candidate of Cold Dark Matter}'',
  Pramana {\bf 80}, 539 (2013)
  [arXiv:1102.3024 [hep-ph]].

\bibitem{Semenov:2010qt} 
  A.~Semenov,
  ``{\it LanHEP - a package for automatic generation of Feynman
  rules from the Lagrangian. Updated version 3.1}'',
  arXiv:1005.1909 [hep-ph].
  
\end{thebibliography}
\end{document}